\documentclass[preprint]{revtex4-1}
\usepackage{graphicx} 
\usepackage{amsmath}
\usepackage{caption}
\usepackage{float}
\usepackage{mathptmx}
\usepackage{comment}
\usepackage{natbib}
\usepackage{enumitem}
\usepackage[colorlinks]{hyperref}
\usepackage{color, soul}
\usepackage{tabularx}
\usepackage{dcolumn}
\usepackage{bm}

\usepackage{caption}
\usepackage{subcaption}

\begin{document}

\title{ Turbulence closure modeling with data--driven techniques:\\Investigation of generalizable deep neural networks}

\author{Salar Taghizadeh\textsuperscript{1}}
\email{staghizadeh@tamu.edu}
\author{Freddie D. Witherden\textsuperscript{2}}
\author{Yassin A. Hassan\textsuperscript{1,3}}
\author{Sharath S. Girimaji\textsuperscript{1,2,4}}
\affiliation{\footnotesize $^1$J. Mike Walker ’66 Department of Mechanical Engineering, Texas A\&M University, College Station, TX 77843\looseness=-1}
\affiliation{$^2$Department of Ocean Engineering, Texas A\&M University, College Station, TX 77843}
\affiliation{$^3$Department of Nuclear Engineering, Texas A\&M University, College Station, TX 77843}
\affiliation{$^3$Department of Aerospace Engineering, Texas A\&M University, College Station, TX 77843}


\begin{abstract}
Generalizability of machine--learning (ML) based turbulence closures to accurately predict unseen practical flows remains an important challenge. It is well recognized that the neural network (NN) architecture and training protocol profoundly influence the generalizability characteristics. At the Reynolds--averaged Navier--Stokes (RANS) level, NN--based turbulence closure modeling is rendered difficult due to two important reasons: inherent complexity of the constitutive relation arising from flow--dependent non--linearity and bifurcations; and, inordinate difficulty in obtaining high--fidelity data covering the entire parameter space of interest. Thus, a predictive turbulence model must be robust enough to perform reasonably outside the domain of training. In this context, the objective of the work is to investigate the approximation capabilities of standard moderate--sized fully--connected NNs. We seek to systematically investigate the effects of: \emph{(i)} intrinsic complexity of the solution manifold; \emph{(ii)} sampling procedure (interpolation vs. extrapolation) and \emph{(iii)} optimization procedure. To overcome the data acquisition challenges, three proxy--physics turbulence surrogates of different degrees of complexity (yet significantly simpler than turbulence physics) are employed to generate the parameter--to--solution maps. Lacking a strong theoretical basis for finding the globally optimal NN architecture and hyperparameters in the presence of non--linearity and bifurcations, a `brute--force' parameter--space sweep is performed to determine a locally optimal solution. Even for this simple proxy--physics system, it is demonstrated that feed--forward NNs require more degrees of freedom than the original proxy--physics model to accurately approximate the true model even when trained with data over the entire parameter space (interpolation). Additionally, if deep fully--connected NNs are trained with data only from part of the parameter space (extrapolation), their approximation capability reduces considerably and it is not straightforward to find an optimal architecture. Overall, the findings provide a realistic perspective on the utility of ML turbulence closures for practical applications and identify areas for improvement.
\end{abstract}

\maketitle

\section{\label{sec:Intro}Introduction\protect}
Turbulent flows exhibit vastly different characteristics in different parameter regimes depending upon the mean strain--to--rotation rate ratio \cite{blaisdell1996simulation,girimaji2000pressure}, mean-to-turbulence time scale ratio \cite{speziale1991analytical, mishra2010pressure, mishra2017toward}, underlying flow instabilities \cite{speziale1996consistency}, large--scale unsteadiness, presence or absence of system--rotation or streamline--curvature \cite{speziale1991analytical}, body--force effects and flow geometry. In addition, due to inherent complexity of the turbulence phenomenon, closure models developed in one parameter regime cannot be presumed to be reasonable or even valid in other regimes. The strong dependence of flow statistics on the various physical parameters is one of the enduring challenges in the field of turbulence closure model development. The degree of difficulty of closure modeling depends upon the level of closure. In the Reynolds--averaged Navier--Stokes (RANS) method, the fundamental governing equations are averaged over all scales of motion leading to significant reduction in computational effort needed for performing a flow simulation. The reduction in computational burden comes at the cost of increased complexity of closure modeling. Ad hoc simplifications or assumptions are typically invoked to close various terms in the RANS equations. While RANS models may perform adequately in flows in which they are calibrated, they can be catastrophically wrong in other complex flows due to lack of generalizability. Despite inherent limitations, RANS is widely used in industrial applications involving complex flows due to ease of computations. At the other extreme of the closure spectrum, in the large--eddy simulation (LES) approach, dynamically important scales are resolved and only the small--scale motions are modeled. The small--scales are significantly easier to model as they embody most of the `universal’ aspects of turbulence and therefore, are more easily amenable to generalizability than their RANS counterpart. Thus, in LES, the relative simplicity and generalizability of subgrid closure models comes at the expense of significantly increased computational costs. The closure modeling challenges of scale resolving simulations (SRS) are of intermediate degree of difficulty as they resolve more scales than RANS but significantly lesser than LES \cite{girimaji2006partially, girimaji2006partially2}.

There is heightened expectation in recent times that ML techniques can be used to significantly improve RANS turbulence closure model performance in complex industrial flows. The rationale is that the shortcomings of the physics--based closures can be circumvented by appropriate data--based training of the models. Toward this end, many authors have used different ML methods to model the turbulence constitutive relation at all level of closure modeling, LES, SRS and RANS. ML--enhanced LES closures have been proposed in different studies \cite{sarghini2003neural, gamahara2017searching, maulik2017neural, maulik2018data, beck2019deep, zhou2019subgrid,  xie2020modeling, schoepplein2018application, reissmann2021application, maulik2019subgrid, wang2018investigations, maulik2019sub, xie2019artificial, xie2020artificial, xie2019modeling, park2021toward, wang2021artificial, subel2021data, pawar2020priori}. ML--enhanced RANS closures have also been proposed in numerous works using different ML algorithms including NNs~\cite{ling2016reynolds, zhang2019application, sotgiu2019towards, geneva2019quantifying, fang2020neural, jiang2020novel, yin2020feature, taghizadeh2020turbulence, jiang2021interpretable, zhou2021learning, yang2020improving, volpiani2021machine, maulik2021turbulent, zhu2019machine, peters2021s}, random forest~\cite{breiman2001random,wu2017priori,kaandorp2020data, wu2018physics, yang2020improving}, gene expression programming (GEP)~\cite{weatheritt2017development, weatheritt2017machine, akolekar2018development, zhao2020rans, lav2019framework, waschkowski2021multi} and deterministic symbolic regression models~\cite{schmelzer2020discovery, huijing2021data, beetham2020formulating, zhang2021customized}. A complete list of important contributions in this area can be found in recent review papers \cite{kutz2017deep, duraisamy2019turbulence, brunton2020machine, beck2021perspective, duraisamy2021perspectives}. ML methods for turbulence closure are not without their own challenges and shortcomings. There is no clear guidance on the optimal choice of architecture and hyperparameters or different elements of the training procedure. For instance, in order to develop a new algebraic Reynolds stress model using channel flow dataset, different NN architectures have been employed in literature. Zhang et al. \cite{zhang2019application} trained a NN with 4 hidden layers and 20 neurons per layer. Fang et al. \cite{fang2020neural} employed a NN with 5 hidden layers and 50 neurons per layer. Jiang et al. \cite{jiang2020novel} used a DNN with 9 hidden layers and varying number of neurons in each layer. For planner and periodic hill channel flow dataset, Sotgiu et al. \cite{sotgiu2019towards} used a NN with 8 hidden layers and 8 neurons in each layer. Lacking a formal procedure for NN selection and training protocol, ML--assisted turbulence closures can be as ad hoc as the traditional models and, more importantly, lack generalizability. In other areas, it has been shown that ML models developed with subject matter expertise has a better chance of succeeding~\cite{anand2020black}. Thus, it is of much interest to examine generalizability in turbulence--like systems that are much simpler to compute.

The earliest fundamental analyses of the approximation capabilities of NNs demonstrate that any continuous function on a bounded domain can be approximated to arbitrary accuracy with at least one hidden layer with sufficiently many neurons~\cite{cybenko1989approximation, hornik1989multilayer}. However, these results do not quantify the required sizes of NNs to achieve these rates. Approximation rates of NNs with different activation functions for various functions are presented in later studies~\cite{barron1993universal, leshno1993multilayer, maiorov1999lower, mhaskar1993approximation, mhaskar1996neural, yarotsky2017error, petersen2018optimal}. As the most efficient statistical learning methods, deep neural networks (DNNs) have been used to directly model the solution of PDEs~\cite{raissi2019physics, jin2021nsfnets, berner2020analysis, weinan2017deep, han2018solving, sirignano2018dgm} in which it is often plausible to bound the size of the involved NNs in a way that overcome the curse of dimensionality, i.e., the approximation rates do not grow exponentially with increasing input dimension. DNNs have also been used to approximate the parameter (solution) maps in parametric problems~\cite{dal2020data, hesthaven2018non, khoo2021solving, lee2020model, kutyniok2021theoretical, geist2021numerical}. In parametric PDE problems, parametric map connects the solution space of PDE to the parameters that describe mathematical and physical constraints of PDE, for instance shape of physical domain, boundary conditions. For linear parametric PDEs it has been theoretically shown that feed--forward NNs of sufficient depth and size with Rectified Linear Unit (ReLU) activation function are able to produce very efficient approximations. However, the results are not optimal and do not yield minimum depth and minimum number of neurons per layer, thus it is not clear whether deep NNs are indeed necessary~\cite{kutyniok2021theoretical}.

The practical learning problems consists of several aspects: \emph{(i)} capacity of NN architecture in describing the data; \emph{(ii)} data availability to describe the true model and \emph{(iii)} optimization procedure in finding the best fit~\cite{kutyniok2021theoretical}. In general, deep learning is a non--convex optimization problem and while a good approximation of a given unknown function by a NN may exist, it is unclear how that can be expeditiously determined in a practical application. Additionally, it is certainly feasible that the generated data might not contain sufficient information from the true model to assure that the optimization process will converge to the theoretically best approximation~\cite{geist2021numerical}. Therefore, it is necessary to numerically examine the existence of generalizable, reasonably--sized NNs from practical learning prospective. In this work we use supervised learning method, to train standard fully--connected NNs to fit the parameter--to--solution maps in the context of turbulence RANS closure modeling.

Turbulence has long been recognized as a complex flow phenomenon due to inherent chaotic tendencies, multi--scale interactions arising out of the non--linearity of the governing equations. As a consequence, a turbulence constitutive relation which relates Reynolds stresses to the mean--flow field statistics can exhibit multiple complex features:~\emph{(i)} strong history and inhomogeneity effects;~\emph{(ii)} non--linearity of unknown degree and~\emph{(iii)} multiple bifurcations in complex flows involving compressibility, system rotation, streamline curvature, additional body forces and complicated geometric features. Further, the RANS closures must satisfy key conservation laws, physical principles and mathematical constraints~\cite{taghizadeh2020turbulence}. Thus, generating a reasonable parameter--to--solution map for turbulence modeling at the RANS level is challenging. Even if the full parameter space is identified, obtaining data from direct numerical simulations (DNS) or experiments could be prohibitively expensive and indeed infeasible.

At the current state of development, data--driven methods work best for interpolation, while application to parameter space outside the training domain is fraught with uncertainty. When a NN--based turbulence model is used for computing flows of engineering interest, parameter regimes outside the training domain will likely be encountered. Therefore, it will be useful to assess the extrapolation capabilities of the NN--based models.

To circumvent data sampling limitations, we propose to employ simplified proxy--physics turbulence surrogates to generate parameter--to--solution maps. Using this data, we systematically examine three main challenges of practical learning problems in the context of turbulence closure modeling:

\begin{enumerate}
 \item \textbf{\emph{Effect of the intrinsic complexity:}} The effect of the complexity of the solution manifold on the performance of the NNs is investigated by employing three proxy--physics surrogates of different degrees of non--linearity and bifurcation characteristics. The turbulence surrogates mimic some of the key features of turbulence and provide sufficiently many training/testing data at low computational expense. It must be iterated that the actual turbulence constitutive relationship can be considerably more complex due to history (transient) effects, flow inhomogeneity and multiple bifurcations, which are not considered in the proxy--physics models.
 \item \textbf{\emph{Effects of sampling procedure (interpolation vs. extrapolation):}} The proxy--model approach allows for training data to be generated over all of the parameter space or select regimes to investigate the effects of sampling. Based on underlying physics of these proxy--physics models, training and testing datasets are segmented into three different scenarios to assess generalizability (interpolation vs. extrapolation) characteristics.
 \item \textbf{\emph{Effects of the optimization procedure:}} Architecture and the hyperparameters of the DNNs can significantly affect their performance and generalizability characteristics. In this study we examine the existence of generalizable NNs by performing a systematic search in state--space of hyperparameters and network--size for different data sampling scenarios. We seek to establish the optimal choice of hyperparameters and DNN neurons of required to achieve a predetermined level of accuracy.
\end{enumerate}

To enable a reasonably rigorous analysis, we \emph{(i)} restrict our consideration to statistically two--dimensional homogeneous turbulence, which represents the most elementary non--trivial flows of interest; and \emph{(ii)} utilize data generated from proxy--physics turbulence surrogates that incorporate some of the key aspects of real homogeneous flows. We consider using standard fully--connected NNs to reproduce the results of the simplified proxy--physics models based around a cubic polynomial. Such a model represents a gross oversimplification of the true physical dynamics of turbulence. Hence, if a moderate--sized NN, given an arbitrarily large amount of training data is unable to adequately reproduce this model, it would be unreasonable to expect NN--based ML models to perform well in unseen turbulent flows. Further, the lessons learned for this simpler system can yield valuable insight into the closure modeling of the more complex turbulence phenomenon. 

The various NN hyperparameters and training elements considered in this study are: number of hidden layers or depth of network, number of neurons in each layer or width of network, type of loss function, type of activation function for neurons, type of training optimizer, learning rare, batch size and regularization coefficient. The organization in the rest of the paper is as follows. The selected proxy--physics turbulence surrogates are explained in Sec.~\ref{sec:Method}. The NN hyperparameters and training elements are described in Sec.~\ref{sec:ML}. Section~\ref{sec:DataG} details how data is generated with the proxy--physics turbulence surrogates in different regimes of two--dimensional homogeneous turbulence. The results from different architectures and training methods are presented in Sec.~\ref{sec:Res}. We conclude with a summary of findings and recommendations in Sec.~\ref{sec:Conc}.

\section{\label{sec:Method}Proxy--Physics Methodology\protect}
In a statistically steady two--dimensional homogeneous incompressible turbulent flow field without body forces, only two parameters govern the physical behavior – normalized mean strain ($s_{ij}$) and rotation rates ($r_{ij}$),
\begin{equation}  
  s_{ij}=\frac{k}{\epsilon}S_{ij},\quad
  r_{ij}=\frac{k}{\epsilon}R_{ij},
\label{eq:Strain}
\end{equation}
where 
\begin{equation}  
  S_{ij}=\frac{1}{2}(\frac{\partial U_i}{\partial x_j}+\frac{\partial U_j}{\partial x_i}),\quad
  R_{ij}=\frac{1}{2}(\frac{\partial U_i}{\partial x_j}-\frac{\partial U_j}{\partial x_i}).
\label{eq:Strain2}
\end{equation}

Here $k$ and $\epsilon$ are respectively turbulent kinetic energy and turbulent dissipation. This set of flows do not exhibit any large--scale instabilities, coherent structures, complex geometrical features or statistical unsteadiness. Despite the apparent simplicity, homogeneous two--dimensional turbulent flows embody multiple complex phenomena that are strongly dependent upon the parameter values. The flow goes from hyperbolic to rectilinear to elliptic streamline geometry with increasing mean rotation rate. Depending upon the mean flow to turbulence strain rate ratio, the flow physics can range from the rapid distortion limit to decaying anisotropic turbulence. 

Using representation theory, the four--term expansion of the anisotropy tensor in terms of the normalized strain rate ($s$) and normalized rotation rate ($r$) tensors for two--dimensional mean flows can be expressed as \cite{girimaji1996fully},  
\begin{equation}  
b_{ij} = G_1(s_{ij})+G_2(s_{ik}r_{kj}-r_{ik}s_{kj})+G_3(s_{ik}s_{kj}-\frac{1}{3}\delta_{ij}s_{mn}s_{nm})+G_4(r_{ik}r_{kj}-\frac{1}{3}\delta_{ij}r_{mn}r_{nm}),
\label{eq:Anis.}
\end{equation}
where scalar coefficients $G_1-G_4$ are constitutive closure coefficients (CCC) \cite{taghizadeh2020turbulence} that must be modeled. They are functions of scalar invariant of the strain and rotation--rate tensors ($\eta_1=s_{ij} s_{ij},\eta_2=r_{ij} r_{ij}$) and other flow quantities such as - turbulent kinetic energy ($k$), turbulence frequency ($\omega$) and the coefficients of the pressure--strain correlation model.

As mentioned in the Introduction, generalizability of the NN models in this simple system is a prerequisite to generalizability in actual turbulent flows. Further, due to the fact that proxy--physics turbulence surrogates capture many key statistical features of turbulence, this study will yield much valuable insight into RANS ML modeling. Complete knowledge of the proxy--physics solution enables precise error assessment incurred during generalization to test flows. In this study, the proxy--physics data in different parts of the domain are used to train NNs of different architectures and hyperparameters. The networks are then tested in other parameter regimes to examine generalizability. 

\subsection{\label{sec:Surrogates} Proxy--Physics Surrogates}
As stated before the main objective of this work is to look at the challenges posed by non--linearity and bifurcation effects of turbulence, therefore three proxy--physics surrogates of different non--linearity and bifurcation characteristics are employed to generate the parameter--to--solution maps: two algebraic Reynolds stress models (ARSM) and one non--linear constitutive relationship. 

\textbf{\emph{Algebraic Reynolds Stress Model.}} ARSM does reasonably well in capturing key flow physics in different parts of the flow regime \cite{gatski1993explicit, girimaji1996fully}. The model requires the solution of a cubic equation and appropriate root must be chosen in different flow regimes to yield the correct behavior \cite{girimaji1996fully}. In this study, we use three--term self--consistent, nonsingular and fully explicit algebraic Reynolds stress model (EARSM) proposed by Girimaji \cite{girimaji1996fully} with two pressure--strain correlation models, Launder, Reece and Rodi (LRR) \cite{launder1975progress} and and Speziale, Sarkar and Gatski (SSG) \cite{speziale1991modelling} to generate stress--strain (constitutive) relationship datasets for different normalized strain and rotation rates.

An implicit algebraic equation for the anisotropy tensor can be obtained in the weak equilibrium limit of turbulence using the following simplification \cite{girimaji1996fully},
\begin{equation}  
\frac{Db_{ij}}{Dt} = \frac{\partial b_{ij}}{\partial t} + U_k\frac{\partial b_{ij}}{\partial x_k} \approx 0,
\label{eq:Weak Equi.}
\end{equation}
where $D/Dt$ is the substantial derivative following the mean flow. The weak--equilibrium assumption is valid for many flows wherein the timescale of anisotropy evolution is rapid compared to the timescales of mean flow, turbulent kinetic energy, and dissipation rate \cite{girimaji2001lower, gomez2014explicit}. Using Eqs.~\eqref{eq:Anis.} and \eqref{eq:Weak Equi.}, the non--linear algebraic Reynolds stress equation with the three--term model ($G_1-G_3$) can be written as the following cubic fixed--point equation for $G_1$ \cite{girimaji1996fully},
\begin{equation}  
(\eta_1 L_1^1)^2G_1^3 - (2\eta_1L_1^0L_1^1)G_1^2 + \Big[(L_1^0)^2 + \eta_1L_1^1L_2 - \frac{2}{3}\eta_1(L_3)^2 + 2\eta_2(L_4)^2\Big]G_1 -L_1^0L_2 = 0.
\label{eq:F_P Equi.}
\end{equation}
Then $G_2$ and $G_3$ can be expressed as \cite{girimaji1996fully}, 
\begin{equation}  
  G_2=\frac{-L_4G_1}{L_1^0-\eta_1L_1^1G_1},\quad
  G_3=\frac{2L_3G_1}{L_1^0-\eta_1L_1^1G_1}.
\label{eq:2Gs}
\end{equation}
The closure coefficients in Eqs.~\eqref{eq:F_P Equi.} and \eqref{eq:2Gs} are as follows:
\begin{equation}  
  L_1^0=\frac{C_1^0}{2}-1,\quad
  L_1^1=C_1^1+2,\quad
  L_2=\frac{C_2}{2}-\frac{2}{3}, \quad
  L_3=\frac{C_3}{2}-1,\quad
  L_4=\frac{C_4}{2}-1,
\label{eq:Coeff}
\end{equation}
where the $C$'s are numerical constants of the pressure--strain correlation models. The numerical constants for LRR pressure--strain correlation model, linear in the anisotropy tensor ($C_1^1=0$), and SSG, quasilinear in the anisotropy tensor ($C_1^1\neq0$), are given in Table~\ref{tab:1}. 

\begin{table}
  \caption{Coefficients in the LRR and SSG models}
  \label{tab:1}
  \begin{tabularx}{0.8\textwidth}{XXXXXX}
    \hline\hline
    Model & $C_1^0$ & $C_1^1$ & $C_2$ & $C_3$ & $C_4$\\
    \hline
    LRR & 3.0 & 0 & 0.8 & 1.75 & 1.31 \\
    SSG & 3.4 & 1.8 & 0.36 & 1.25 & 0.4 \\
    \hline\hline
  \end{tabularx}
\end{table}

The cubic relation in Eq.~\eqref{eq:F_P Equi.} has multiple real and complex roots and the selection of the appropriate solution for $G_1$ is not straightforward. By considering two physical selection criteria; \emph{(i)} continuity of $G_1$; and \emph{(ii)} $G_1 \leq 0$, Girimaji \cite{girimaji1996fully} derived a fully explicit solution of the cubic Eq.~\eqref{eq:F_P Equi.} for scalar coefficient $G_1$ as follows:
\begin{equation}
G_1 =
\left\{
	\begin{array}{ll}
		\frac{L_1^0L_2}{(L_1^0)^2+2\eta_2(L_4)^2} & \mbox{for  } \eta_1=0, \\
		\frac{L_1^0L_2}{(L_1^0)^2-\frac{2}{3}\eta_1(L_3)^2+2\eta_2(L_4)^2} & \mbox{for  } L_1^1=0, \\
        -\frac{p}{3}+(-\frac{b}{2}+\sqrt{D})^{\frac{1}{3}}+(-\frac{b}{2}-\sqrt{D})^{\frac{1}{3}} & \mbox{for  } D > 0, \\	
        -\frac{p}{3}+2\sqrt{\frac{-a}{3}}cos(\frac{\theta}{3}) & \mbox{for  } D < 0, b<0, \\	        
        -\frac{p}{3}+2\sqrt{\frac{-a}{3}}cos(\frac{\theta}{3}+\frac{2\pi}{3}) & \mbox{for  } D < 0, b>0.
	\end{array}
\right.
\label{eq:Sol}
\end{equation}
Here the discriminant $D$ of the cubic Eq.~\eqref{eq:F_P Equi.} is calculated as:
\begin{equation}  
  D=\frac{b^2}{4}+\frac{a^3}{27},
\label{eq:D}
\end{equation}
other parameters of Eqs.~\eqref{eq:Sol} and \eqref{eq:D} are defined as below:
\begin{equation}  
\begin{split}
  a=(q-\frac{p^2}{3}),&\quad
  b=\frac{1}{27}(2p^3-9pq+27r),\\
  p=-\frac{2L_1^0}{\eta_1L_1^1}, \quad
  q=\frac{1}{(\eta_1L_1^1)^2}&\Big[(L_1^0)^2+\eta_1L_1^1L_2-\frac{2}{3}\eta_1(L_3)^2+2\eta_2(L_4)^2\Big],\\
  r=-\frac{L_1^0L_2}{(\eta_1L_1^1)^2},& \quad
  cos(\theta)=\frac{-b/2}{\sqrt{-a^3/27}}.
\label{eq:OtherParameter}
\end{split}
\end{equation}
The first two cases in Eq.~\eqref{eq:Sol} are special limiting cases of the last three and it can be shown that the limiting behavior can be easily calculated from the general expressions \cite{girimaji1996fully}. After the coefficient $G_1$ is calculated, other CCC can also be obtained using Eq.~\eqref{eq:2Gs} in the entire parameter space.

\begin{figure}
\centering
\includegraphics[width=0.5\textwidth]{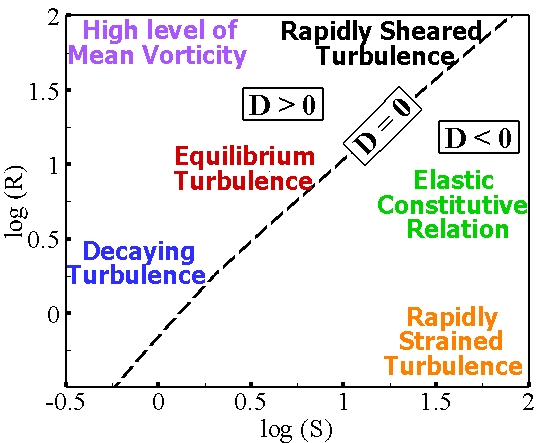}
\caption{\label{fig_1} Different states of turbulence of ARSM}
\end{figure}

The goal of this study is to use proxy--physics turbulence surrogates to provide some turbulence subject matter expertise. For this reason, it is necessary to ensure that the selected proxy model adequately incorporates some of the known features of turbulence. Different turbulence states covered by the ARSM turbulence surrogate are depicted for the considered parameter space in Fig.~\ref{fig_1} ($S \equiv \sqrt{s_{ij} s_{ij}} \equiv \sqrt{\eta_1}$ and $ R \equiv \sqrt{r_{ij} r_{ij}} \equiv \sqrt{\eta_2}$). At rapidly strained turbulence state, strain rate dominates over rotation rate and the discriminant $D$ (Eq.~\eqref{eq:D}) of the cubic fixed--point equation for $G_1$ is negative. The realizability violations occur at this rapidly strained region due to the governing elastic constitutive relationship. Rotation rate dominates over strain rate at high values of mean vorticity state. Other important turbulence states are also represented in this figure. Hyperbolic streamline flows occur when $S>>R$; and the streamlines are elliptic when $S<<R$. When $S\approx R$ rectilinear shear flows are seen \cite{mishra2013intercomponent}.

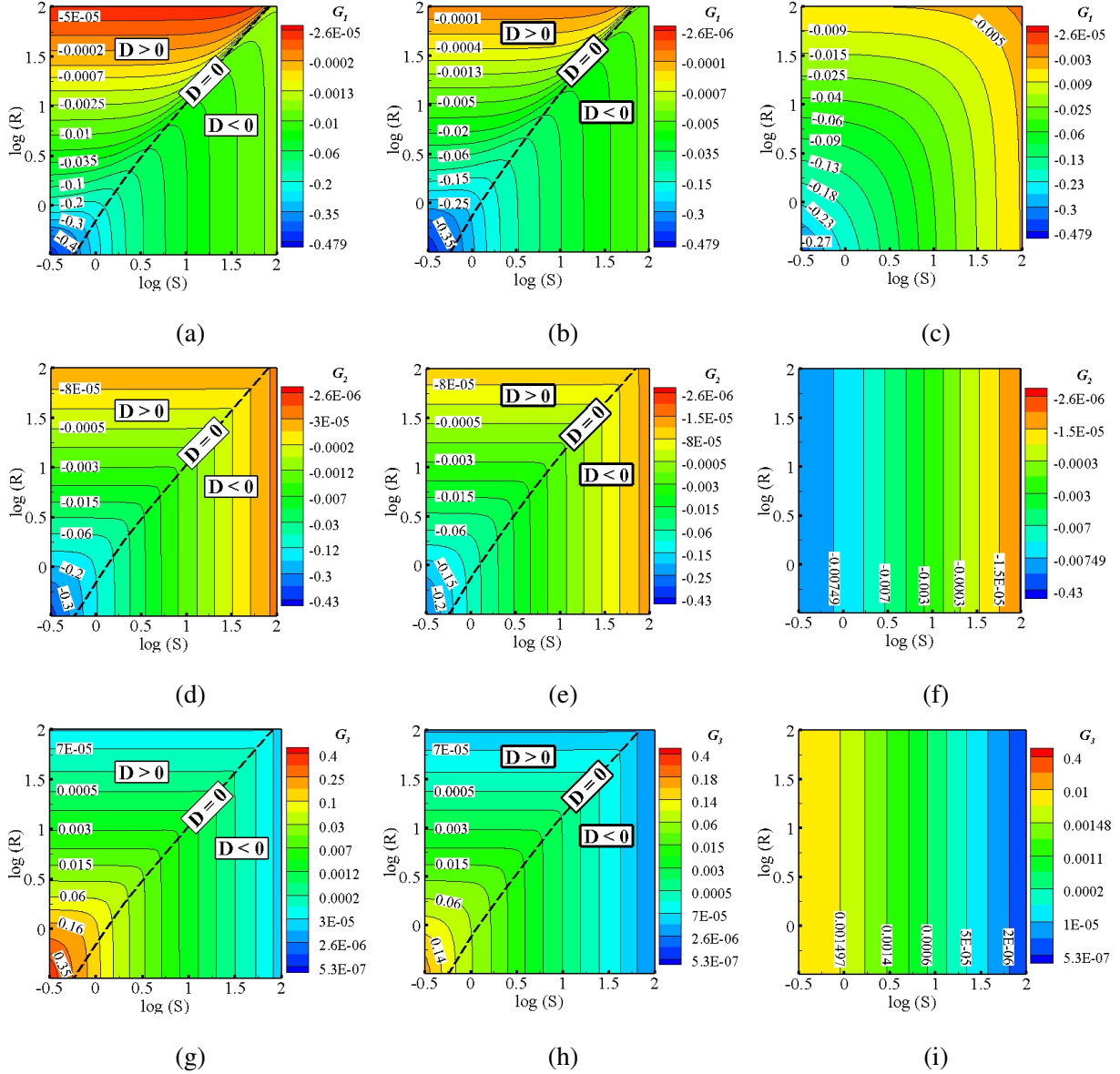
\begin{figure}
        \centering
 		\begin{subfigure}[b]{0.325\textwidth}
               \includegraphics[width=\textwidth,trim=2 2 2 2,clip]{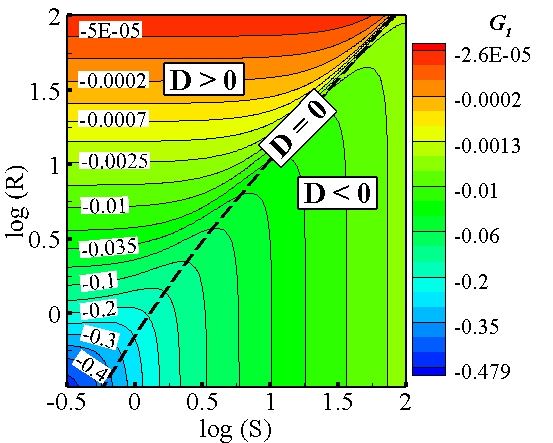}\begin{picture}(0,0)\put(-138,0){}\end{picture}
               \caption{}
        \end{subfigure}                                 
        \begin{subfigure}[b]{0.325\textwidth}
                \includegraphics[width=\textwidth,trim=1 1 1 2,clip]{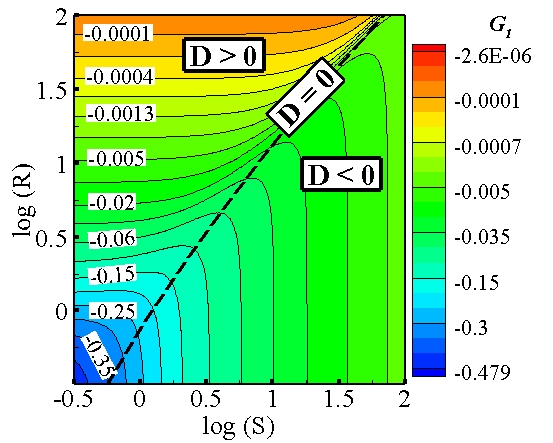}\begin{picture}(0,0)\put(-138,0){}\end{picture}
               \caption{} 
        \end{subfigure}
        \begin{subfigure}[b]{0.325\textwidth}
                \includegraphics[width=\textwidth,trim=1 1 1 2,clip]{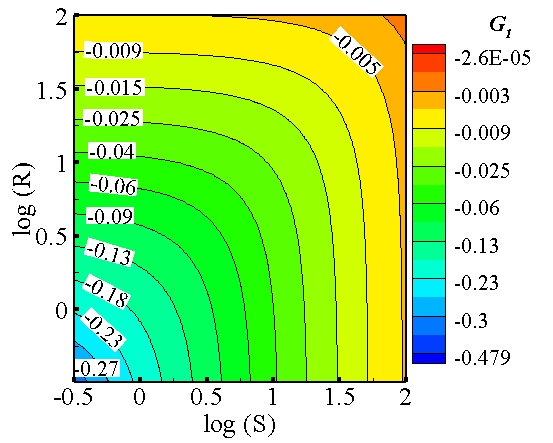}\begin{picture}(0,0)\put(-138,0){}\end{picture}
               \caption{}
        \end{subfigure}        
        \begin{subfigure}[b]{0.325\textwidth}
               \includegraphics[width=\textwidth,trim=2 2 2 2,clip]{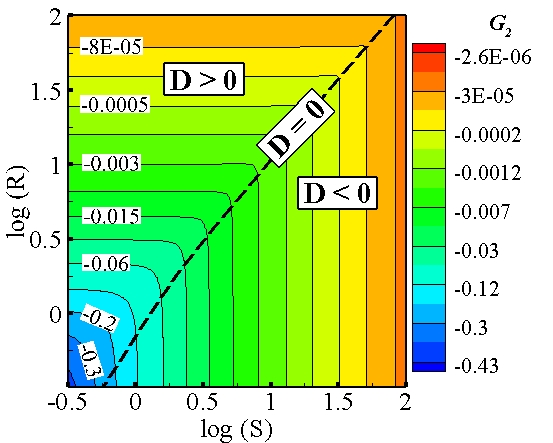}\begin{picture}(0,0)\put(-138,0){}\end{picture}
                \caption{}
        \end{subfigure}                                 
        \begin{subfigure}[b]{0.325\textwidth}
                \includegraphics[width=\textwidth,trim=1 1 1 2,clip]{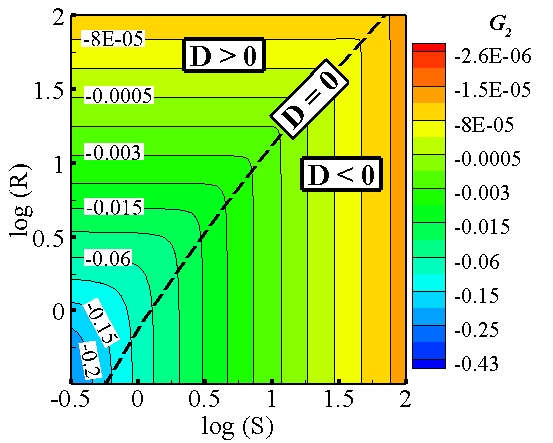}\begin{picture}(0,0)\put(-138,0){}\end{picture}
               \caption{}
        \end{subfigure}
        \begin{subfigure}[b]{0.325\textwidth}
                \includegraphics[width=\textwidth,trim=1 1 1 2,clip]{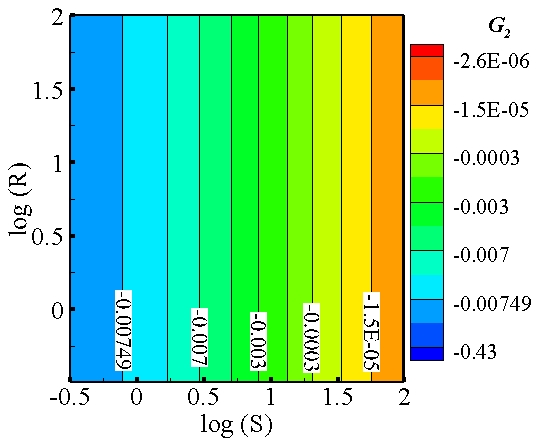}\begin{picture}(0,0)\put(-138,0){}\end{picture}
               \caption{}
        \end{subfigure}  
        \begin{subfigure}[b]{0.325\textwidth}
               \includegraphics[width=\textwidth,trim=2 2 2 2,clip]{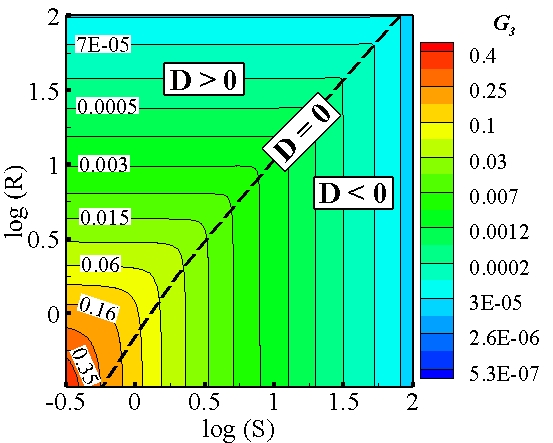}\begin{picture}(0,0)\put(-138,0){}\end{picture}
                \caption{}
        \end{subfigure}                                 
        \begin{subfigure}[b]{0.325\textwidth}
                \includegraphics[width=\textwidth,trim=1 1 1 2,clip]{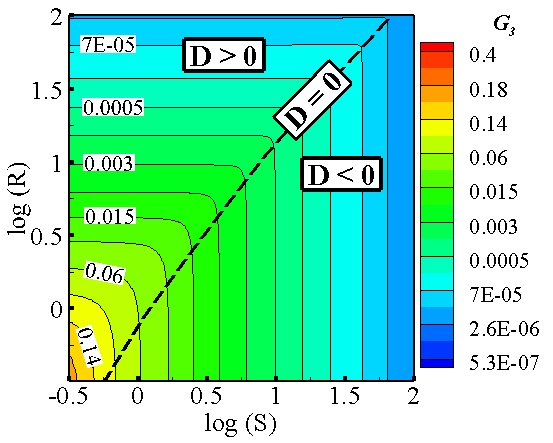}\begin{picture}(0,0)\put(-138,0){}\end{picture}
               \caption{}
        \end{subfigure}
        \begin{subfigure}[b]{0.325\textwidth}
                \includegraphics[width=\textwidth,trim=1 1 1 2,clip]{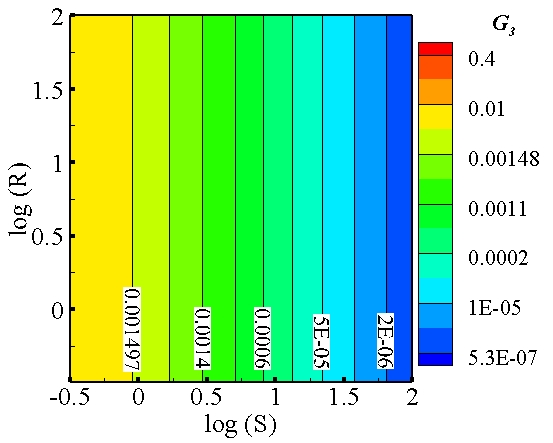}\begin{picture}(0,0)\put(-138,0){}\end{picture}
               \caption{}
        \end{subfigure}          
        \caption{\label{fig_2} Parameter--to--solution maps, (a), (d) and (g) ARSM with SSG, (b), (e) and (h) ARSM with LRR, (c), (f) and (i) SZL} 
        \end{figure}

The parameter--to--solution maps obtained from ARSM with two pressure--strain correlation models are shown in Fig.~\ref{fig_2}. It is shown that the coefficient $G_1$ is a continuous function across bifurcation line $D=0$ in ARSM model. Therefore, $G_1$, i.e., the effective turbulent viscosity is well defined in the entire parameter space wherein different important turbulence states are covered. Fig.~\ref{fig_2} shows well defined values for $G_2$ and $G_3$ coefficients in the entire parameter space. The corresponding CCC contours for ARSM with both the pressure--strain correlation models have similar shapes. However, CCC values are in wider ranges with SSG model. In particular, the magnitude of the $G_1$ is approximately zero, $10^{-5}$ for larger values of parameters, while it is in the order of $10^{-1}$ in decaying turbulence state. As mentioned earlier, small values of $G_1$ for large $S$ is a consequence of elastic constitutive behaviour and is needed for preserving realizability. 

\textbf{\emph{Non--linear Constitutive Relationship.}} In addition to the three--term ARSM surrogate proposed by Girimaji \cite{girimaji1996fully} with different pressure--strain correlation models, the four--term non--linear constitutive relationship proposed by Shih et al.~\cite{shih1993realizable}(Shih, Zhu and Lumley -- SZL model) is also considered as a proxy--physics turbulence surrogate to generate the parameter--to--solution map in this study. In SZL model, the CCC are expressed as \cite{shih1993realizable, craft1996development}, 
\begin{equation}  
\begin{split}
  G_1&=-C_{\mu}=\frac{-\frac{2}{3}}{1.25+\sqrt{2\eta_1}+0.9\sqrt{2\eta_2}}, \quad G_2=\frac{-\frac{15}{2}}{1000+(\sqrt{2\eta_1})^3}, \\
  G_3&=\frac{\frac{3}{2}}{1000+(\sqrt{2\eta_1})^3}, \quad
  \hspace{2.75cm}G_4=\frac{\frac{-19}{2}}{1000+(\sqrt{2\eta_1})^3}.
\label{eq:4Gs}
\end{split}
\end{equation}
It should be noted that the CCC expressions proposed in SZL model are simpler compared to ARSM model counterparts. The non--linear SZL model does not have bifurcation and the continuous function of $G_1$ is obtained with one single relation as given by Eq.~\eqref{eq:4Gs} in the entire parameter space. However, in ARSM two different relationships are employed. For the considered parameter space, the third relation in Eq.~\eqref{eq:Sol} is used in the rotation--dominated region $D>0$ and the fifth relation is used in the strain--dominated region $D<0$. Therefore, the parameter--to--solution map generated by SZL model has less complexity level compared to ARSM, Fig.~\ref{fig_2}. In SZL model, $G_2-G_4$ coefficients are only functions of the invariant of strain rate ($\eta_1$), therefore contour plots show vertical lines in the parameter space. Overall, the physics underlying CCC contours is a reasonable facsimile of real turbulent flows at different complexity levels of proxy--physics turbulence surrogates.

Therefore, three proxy--physics turbulence surrogates of different degrees of complexity can be used to examine the challenges posed by non--linearity and bifurcation effects of the solution manifold: \emph{(i)} SZL model, non--linear constitutive relation with no bifurcation in the parameter space; \emph{(ii)} ARSM with LRR model, mildly non--linear constitutive relation with bifurcation in the regime of interest; and \emph{(iii)} ARSM with SSG model, moderately non--linear constitutive model with bifurcation in the parameter space. The objective of this study is to investigate if a reasonably--sized, fully--connected NN, given an arbitrarily large amount of training data can simulate the simple surrogates of turbulence. If the NN--based ML models can not perform well with the simplified descriptions, it is unlikely to perform successfully in real unseen turbulent flows.

\section{\label{sec:ML} Machine Learning\protect}
\textbf{\emph{Previous investigations.}} ML is a general term to describe a class of algorithms which uses data to generate models. The selection of modeling strategy depends on the type of problem. Recently, DNNs have been widely used for modeling turbulent flows. Ling et al. \cite{ling2016reynolds} trained a DNN with 8 hidden layers, and 30 neurons per hidden layer with available high--fidelity data: duct and channel flow, perpendicular and inclined jet in cross--flow, flow around a square cylinder and flow through a converging--diverging flow. Their results showed that incorporating the Galilean invariance property in the DNN architecture can improve the predictive capability of ML turbulence models. Zhang et al. \cite{zhang2019application} trained a NN with 4 hidden layers and 20 neurons per layer in order to develop a model to predict the Reynolds stress of a channel flow at different Reynolds numbers. They obtained well behaved models by introducing regularization in their training algorithm. Using same channel flow datasets, Fang et al. \cite{fang2020neural} trained a NN with 5 hidden layers and 50 neurons per layer in order to develop an improved Reynolds stress tensor model. Jiang et al. \cite{jiang2020novel} used DNS of channel flow at different Reynolds number in order to developed a new algebraic Reynolds stress model by training a DNN with 9 hidden layers and varying number of neurons in each layer as, 12, 18, 21, 27, 32, 35, 30, 28, and 27, respectively. Sotgiu et al. \cite{sotgiu2019towards} developed a Reynolds stress constitutive model by training a NN with 8 hidden layers and 8 neurons. They used planner channel and periodic hill channel flow datasets for training the ML algorithm.
Geneva and Zabaras \cite{geneva2019quantifying} trained a NN with 5 hidden layers and tapered the number of neurons in the last two hidden layers to prevent weights from being too small and improve training performance. The required training data was generated by performing LES simulations of different flows: converging–diverging channel, periodic hills, square duct, square cylinder and tandem cylinders. The importance of judicious choice of network architecture has been recognized in the field of turbulence modeling. However, there have been no studies in literature to examine if the inherent complexity of turbulence, sampling and training procedure pose any further challenges.

\textbf{\emph{Objective.}} Determining a suitable network architecture and training hyperparameters of a DNN is an empirical task and it has been shown that there is a strong positive correlation between the final performance of the trained NNs and experience of the user in optimizing the hyperparameters~\cite{anand2020black}. There is no rigorous guidance on the right choice of architecture and different elements of the training procedure in developing ML--assisted turbulence models. The objective of this study is to examine the existence of generalizable, reasonably--sized, fully--connected NNs under full and partial availability of the training datasets. Adopted proxy–physics turbulence surrogates are used to easily generate the training data for all flows in the parameter space. Based on underlying physics of the simplified proxy--physics models, training and testing datasets are segmented at three different scenarios to assess the effect of sampling procedure on optimization capabilities and generalizability characteristics (interpolation vs. extrapolation) of the DNNs of different architectures and hyperparameters. 

\begin{figure}
\centering
\includegraphics[width=0.5\textwidth]{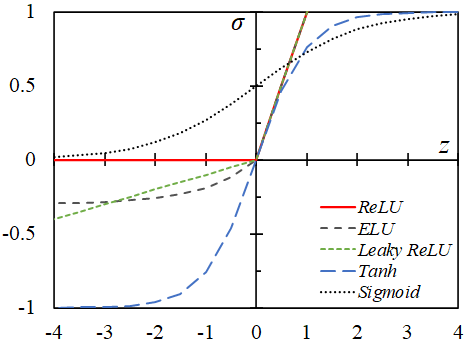}
\caption{\label{fig_3} Profiles of different activation functions}
\end{figure}

\textbf{\emph{Deep neural networks.}} A DNN is a class of model which transforms the input parameters (features) through several layers of units (neurons). Each unit (neuron) is connected by affine linear maps between units in successive layers and then with non--linear (scalar) activation functions within units. Different activation functions such as, rectified linear unit (ReLU), leaky ReLU, exponential linear unit (ELU), Sigmoid and hyperbolic tangent \cite{goodfellow2016deep} are shown in Fig.~\ref{fig_3}. The ReLU function which is the popular choice in the ML literature is defined as:
\begin{equation}  
  \sigma(z)=max(z, 0),
\label{eq:Relu}
\end{equation}
where, $z$ is the input to a neuron. In this study, we examine the effects of employing different activation functions on training and performance of the NNs. 

\begin{figure}
\centering
\includegraphics[width=0.7\textwidth]{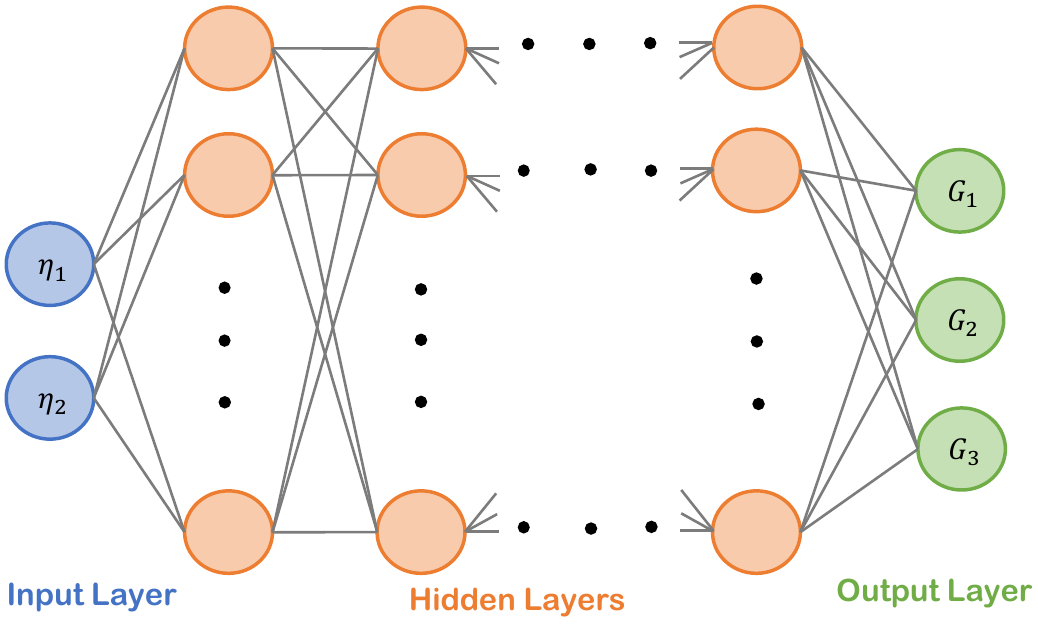}
\caption{\label{fig_4} Illustration of a deep feed--forward fully--connected NN}
\end{figure}

It should be noted that DNNs consist of simple functions and their efficiency emanates from the interactions between large number of hidden layers \cite{goodfellow2016deep}. Due to their flexible architecture and superior performance in modeling non--linear and complicated relationships with high--dimensional data, they have become a popular subset of ML approaches. Although a variety of network structures, such as convolutional neural networks (CNN), recurrent neural networks (RNN), or long short--term memory (LSTM) networks have been proposed in the ML literature \cite{goodfellow2016deep}, we will restrict ourselves to fully--connected architectures. Fig.~\ref{fig_4} shows the schematic of a deep feed--forward NN (also termed as a multi--layer perceptron (MLP)) which is trained using back--propagation with gradient descent method. The input layer, the hidden layers and the output layer are also shown in this figure.

\textbf{\emph{Loss function.}} The DNNs are trained by minimizing a loss (objective) function, which measures the difference between the predicted output of the model and labels (ground truth data). It has been shown that the success of ML models depends on the formulation of the loss function used for optimization of the model coefficients (weights and biases of neurons) during the ML training process \cite{taghizadeh2020turbulence}. The type of loss function is also problem specific and need to be selected properly. The root mean squared error (RMSE) is the commonly used loss function: 
\begin{equation}  
  RMSE=\sqrt{\frac{\sum_{i=1}^{N}(y_i-\hat{y}_i)^2}{N}},
\label{eq:RMSE}
\end{equation}
where $\hat{y}_i$ denotes the ML prediction and $y_i$ is the true labeled data; $N$ is the number of training data. Alternatively, the mean absolute percentage error (MAPE) is also considered, 
\begin{equation}  
  MAPE=\frac{1}{N}\sum_{i=1}^{N}{\frac{\lvert y_i-\hat{y}_i \rvert}{\lvert y_i \rvert}},
\label{eq:MAPE}
\end{equation}
The RMSE loss works well in most cases, while the MAPE loss is better for the case where the output has a large range of function values \cite{cai2021deepm}. In order to find the appropriate type of loss function, the models trained with MAPE and RMSE are compared in this study. 

Adding a regularization term in the loss function formulation is one of the common ways of avoiding overfitting. Overfitting occurs when the expressivity of the ML model is too large for the complexity of the function it approximates. Introducing a regularization term in the loss function formulation shrinks the model coefficients towards zero, decrease the complexity of the model and hence significantly reduces the variance \cite{james2013introduction}. The $L_1$ norm (Lasso Regression) and $L_2$ norm (Ridge Regression) are the two common regularization methods \cite{james2013introduction}, 
\begin{equation}  
\begin{split}
  MAPE (y, y_i) + \lambda \sum_{i=1}^{N}{\lvert \omega_i \rvert}, \quad  L_1 - norm\\
  MAPE (y, y_i) + \lambda \sum_{i=1}^{N}{\omega_i^2}, \quad  L_2 - norm
\label{eq:Reg}
\end{split}
\end{equation}
where $\omega$ denotes the weight of each neuron and $\lambda$ is a positive hyperparameter to determine the strength of regularization. For this work, to control the overfitting in some experiments, first a comparison between two regularization methods, $L_1$ and $L_2$ norms are made and then best performing method is applied in all the hidden layers during the training of the NNs.

\textbf{\emph{Network hyperparameters.}} Architecture and all training hyperparameters of the NNs need to be suitably specified in order to build a robust and effective model that can generalize to unseen datasets. For instance, hyperparameters include size of the networks (number of layers or depth, width or breadth of each layer), formulation and type of the loss function, type of regularization and optimization algorithm, type of activation function (act), batch size (bs), learning rate (lr), type of initialization and etc. Hyperparameter optimization is an empirical task and grid search is usually adopted, i.e., many networks with several different combinations of interval values of each hyperparameter are trained and compared based on their accuracy and generalization ability. 

In principle, all of the hyperparameters in the ML algorithm can be varied and they might have significant impact on the model performance. However, the predictive capability and generalization of a NN is mostly controlled by its architecture; the depth and breadth of network. In this study we perform a systematic search in state--space of hyperparameters and network--size to train NNs that are efficient and ensure a low generalization error in interpolation and extrapolation cases. We consider a matrix of network sizes by varying the depth for four values of 1, 3, 5, 7 and the width for four values of 3, 5, 7, 15. Other hyperparameters examined in this study are shown in Table~\ref{tab:2}. 
\begin{table}
  \caption{Different values of the considered hyperparameters}
  \label{tab:2}
  \begin{tabularx}{0.8\textwidth}{XX}
    \hline\hline
    Hyperparameter & Value\\
    \hline
    Activation function (act) & ReLU, ELU, leaky ReLU, Tanh, Sigmoid   \\
    Learning rate (lr) & $1\times 10^{-6}$, $1\times 10^{-5}$, $1\times 10^{-4}$, $1\times 10^{-3}$, $1\times 10^{-2}$  \\
    Batch size (bs) & 25, 50, 100, 1000 \\
    Optimization algorithm (opt) & Adam \cite{kingma2014adam}, RMSProp\\
    Regularization coefficient $(\lambda)$ & 0.01, 0.1, 0.2\\
    Initialization function & Xavier normal \cite{glorot2010understanding} \\   
    \hline\hline
  \end{tabularx}
\end{table}

\section{\label{sec:DataG} Data Generation with proxy--physics turbulence surrogates \protect}
The data needed for training the ML algorithm is generated using the proxy--physics surrogates as discussed in Sec.~\ref{sec:Method}. The number of data points that are non--uniformly extracted for each input parameter ($\eta_1$, $\eta_2$) in the range [$10^{-1},10^4$] is 150. Therefore, the total number of data points is 22,500 in the entire parameter space. In this work, different investigations are conducted in order to address the challenges in finding optimum NN architecture and hyperparameters when training is performed as follows: 

\begin{enumerate}[noitemsep]
  \item {Fully available data in the entire parameter space from ARSM and SZL models } 
  \item {Partially available data only in one part of parameter space from ARSM model}
  \item {Partially available data only in a narrow band of parameter space from ARSM model} 
\end{enumerate}
These three investigations lead to important inferences about existence of the moderate--sized NN approximation solutions for data manifolds with different complexity levels and their generalizability (interpolation vs. extrapolation) characteristics. We generate the training datasets in three different scenarios as follows.

\begin{figure}
        \centering
 		\captionsetup{justification=centering} 
 		\begin{subfigure}[b]{0.325\textwidth}
               \includegraphics[width=\textwidth,trim=2 2 2 2,clip]{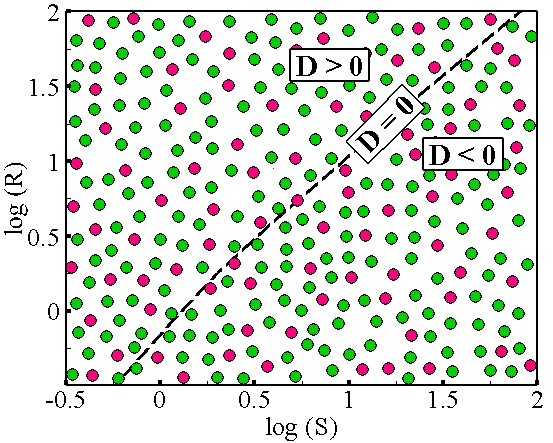}\begin{picture}(0,0)\put(-138,0){}\end{picture}
                \caption{}\label{fig:5a}
        \end{subfigure}                                 
        \begin{subfigure}[b]{0.325\textwidth}
                \includegraphics[width=\textwidth,trim=1 1 1 2,clip]{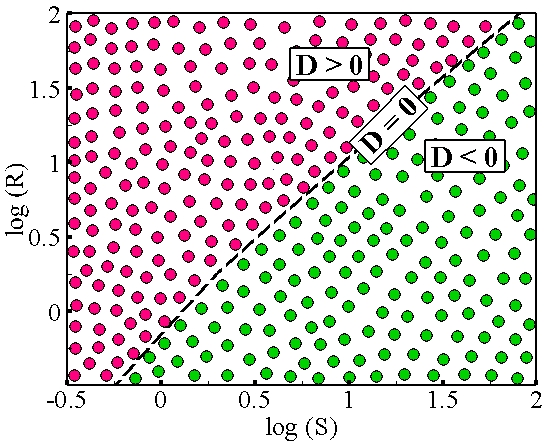}\begin{picture}(0,0)\put(-138,0){}\end{picture}
               \caption{} \label{fig:5b}
        \end{subfigure}
        \begin{subfigure}[b]{0.325\textwidth}
                \includegraphics[width=\textwidth,trim=1 1 1 2,clip]{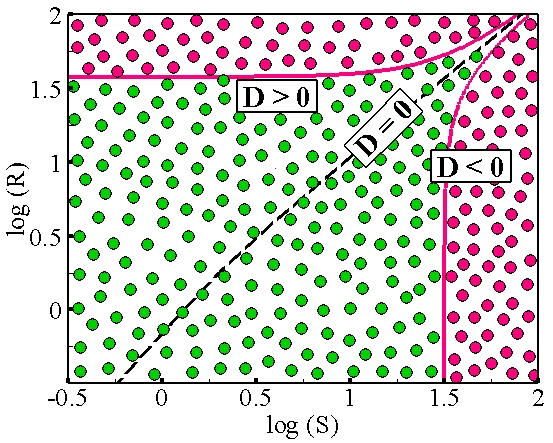}\begin{picture}(0,0)\put(-138,0){}\end{picture}
               \caption{} \label{fig:5c}
        \end{subfigure}        
        \caption{\label{fig_5} Training (green circles) and testing (red circles) datasets for (a) Case--1, (b) Case--2, (c) Case--3} 
        \end{figure}
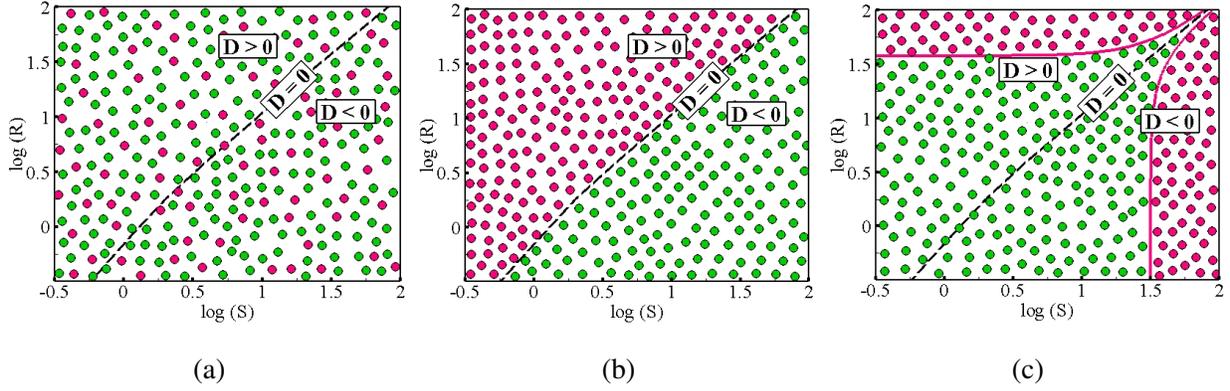
\subsection{\label{sec:Study-1} Training data over the entire parameter space}
This case represents the ideal scenario in which the training data fully describe the true model within the parameter space. Thus there is no physics that is unseen by the ML model which is trained in supervised manner. Sufficient data is gathered over strain, shear and rotational flow regimes. It should be noted that providing the real turbulence data in the entire parameter space, requires expensive DNS over a wide range of flows. In this case, generalizability is expected to be trivially straightforward. The optimum choice of hyperparameters and necessary number of neurons of DNNs needed to have a sufficiently accurate approximation are investigated. As one of the objectives of this study is to examine the effects of underlying physics of the solution manifold (non–linearity and bifurcation effects) on training and performance of the ML models, three parameter--to--solution maps generated by proxy--physics turbulence surrogates discussed in Sec.~\ref{sec:Method} are considered for this ideal case. For all the experiments in this case, the generated data points in the parameters space are randomly split into 75\% for training (Training Data) and 25\% for validation and testing the model (Testing Data). The testing data is used for final evaluation of the models. Fig.~\ref{fig:5a} represents randomly distributed data in the entire parameter space for this case.

\subsection{\label{sec:Study-2} Training data only in the strain--dominated region (D < 0)}
In this case, the training data is partially available and restricted to a subset of the parameter space. Therefore, part of the physics in the data manifold is seen by the NNs. Data points in the strain--dominated region $D<0$ (data generated with the fifth relation in Eq.~\eqref{eq:Sol}) are used for training and data points in the rotation--dominated region $D>0$ (data generated with third relation in Eq.~\eqref{eq:Sol}) are used for testing. Fig.~\ref{fig:5b} illustrates the segmented training and testing datasets for this case. This represents an important generalizability challenge (extrapolation) as training and testing are performed in parameter regimes of distinctly different turbulence physics. Existence of generalizable DNNs for this case is investigated for parameter--to--solution maps generated by ARSM with SSG and LRR pressure--strain correlation models.

\subsection{\label{sec:Study-3} Limited training data in the shear--dominated region}
In many instances training data is partially available only in a very narrow region of the parameter space. In this regard, we examine the existence of a generalizble NN trained with the limited dataset that covers shear--driven turbulence physics. In this case, the parameter--to--solution map generated by ARSM with SSG proxy--physics model is considered. As it can be seen from ARSM equation, Eq.~\eqref{eq:Sol}, shear flow represents the bifurcation region between strain and rotation flows. An arbitrary zone near the bifurcation line $D=0$ is defined as, $-1000<(\eta_1-0.7\eta_2)<1000$. The data points inside this region are used for training and data points outside this region are used for testing. The segmented regions are represented in Fig.~\ref{fig:5c}. 
\begin{table}
  \caption{Selected hyperparameters}
  \label{tab:3}
  \begin{tabularx}{0.8\textwidth}{XX}
    \hline\hline
    Hyperparameter & Value\\
    \hline
    Learning rate (lr) & $1\times 10^{-3} \sim 1\times10^{-6}$  \\
    Batch size (bs) & 50 \\
    Optimization algorithm (opt) & Adam\\
    Initialization function & Xavier normal \cite{glorot2010understanding} \\    
    \hline\hline
  \end{tabularx}
\end{table}
\section{\label{sec:Res} Results\protect}
\subsection{\label{sec:Result_1} RMSE loss function vs. MAPE loss function}
It has been shown that the appropriate formulation of loss function and optimal choice of the flow statistics contributing to the loss function impact the success of the ML trained turbulence models \cite{taghizadeh2020turbulence}. As mentioned in Sec.~\ref{sec:ML} we examine the performance of the ML models trained with different loss functions in this study. For this analysis the parameter--to--solution map generated by ARSM with SSG proxy--physics model is considered. First, the randomly generated dataset in the entire parameter space is segmented as 75\% for training and 25\% for testing of the models. Then, we train two DNNs with RMSE and MAPE loss functions without any regularization. The selected fixed network architecture for both of the cases has 7 hidden layers with 7 computation neurons in each layer. In this case, the type of activation function is ReLU and all other hyperparameters of the models are as shown in Table~\ref{tab:3}. As shown in the table, a variable learning rate is employed for training the models. 

\begin{figure}
        \centering
 		\begin{subfigure}[b]{0.325\textwidth}
               \includegraphics[width=\textwidth,trim=2 2 2 2,clip]{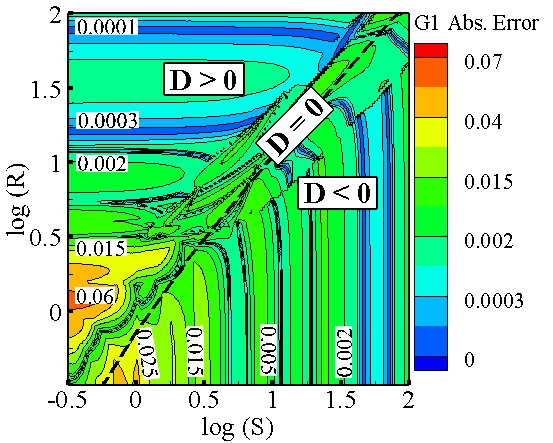}\begin{picture}(0,0)\put(-138,0){}\end{picture}
               \caption{}
        \end{subfigure}                                 
        \begin{subfigure}[b]{0.325\textwidth}
                \includegraphics[width=\textwidth,trim=1 1 1 2,clip]{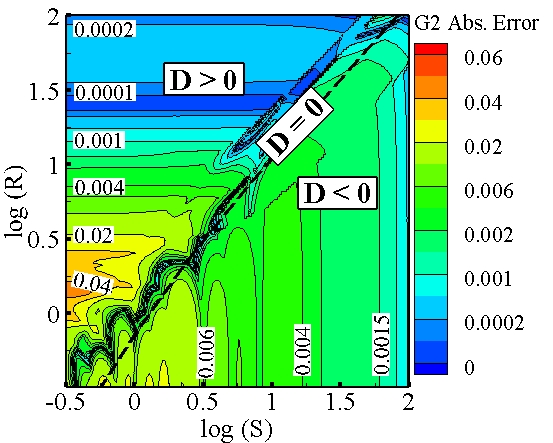}\begin{picture}(0,0)\put(-138,0){}\end{picture}
               \caption{} 
        \end{subfigure}
        \begin{subfigure}[b]{0.325\textwidth}
                \includegraphics[width=\textwidth,trim=1 1 1 2,clip]{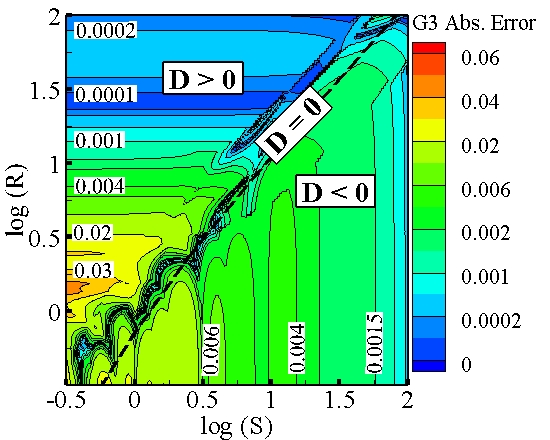}\begin{picture}(0,0)\put(-138,0){}\end{picture}
               \caption{}
        \end{subfigure}    
 		\begin{subfigure}[b]{0.325\textwidth}
               \includegraphics[width=\textwidth,trim=2 2 2 2,clip]{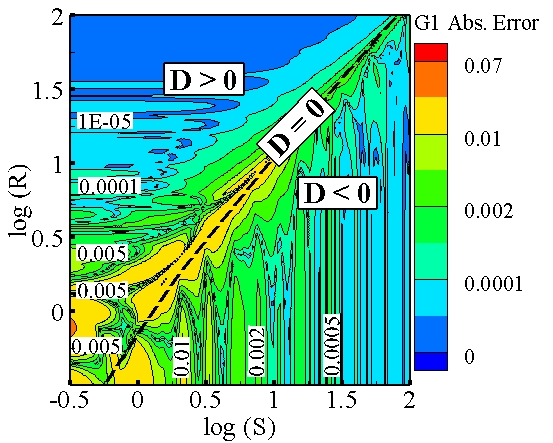}\begin{picture}(0,0)\put(-138,0){}\end{picture}
                \caption{}
        \end{subfigure}                                 
        \begin{subfigure}[b]{0.325\textwidth}
                \includegraphics[width=\textwidth,trim=1 1 1 2,clip]{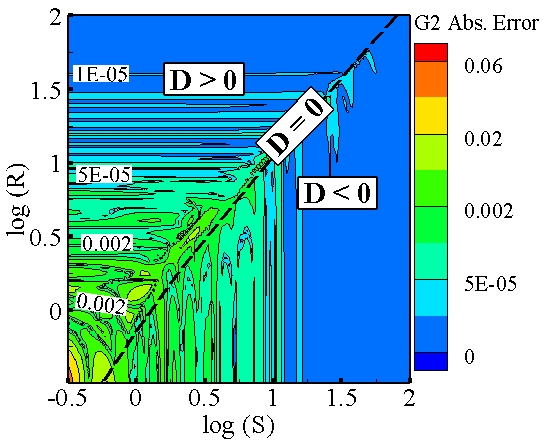}\begin{picture}(0,0)\put(-138,0){}\end{picture}
               \caption{} 
        \end{subfigure}
        \begin{subfigure}[b]{0.325\textwidth}
                \includegraphics[width=\textwidth,trim=1 1 1 2,clip]{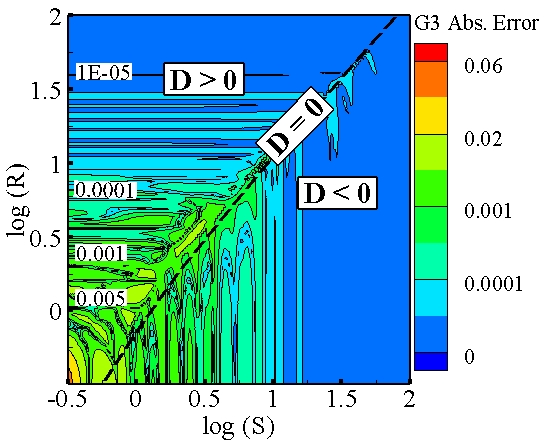}\begin{picture}(0,0)\put(-138,0){}\end{picture}
               \caption{} 
        \end{subfigure}    
        \caption{\label{fig_6} Absolute error contours of CCC for ML models trained with different loss functions, (a)-(c) RMSE, (d)-(f) MAPE} 
        \end{figure}

\begin{figure}
        \centering
 		\captionsetup{justification=centering} 
 		\begin{subfigure}[b]{0.325\textwidth}
               \includegraphics[width=\textwidth,trim=2 2 2 2,clip]{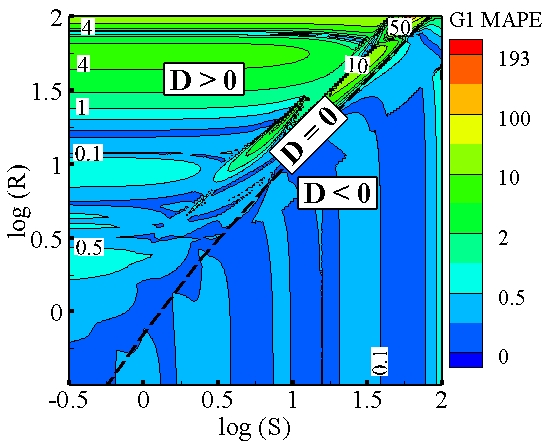}\begin{picture}(0,0)\put(-138,0){}\end{picture}
               \caption{}
        \end{subfigure}                                 
        \begin{subfigure}[b]{0.325\textwidth}
                \includegraphics[width=\textwidth,trim=1 1 1 2,clip]{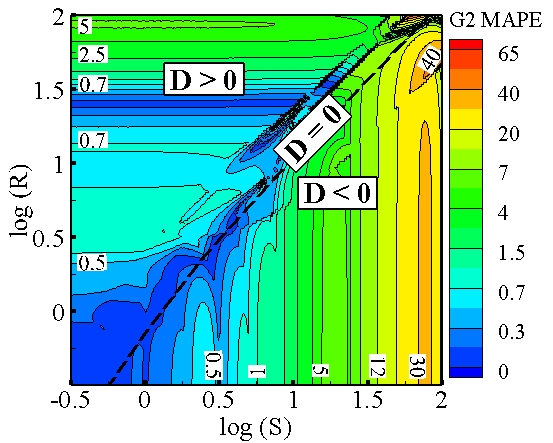}\begin{picture}(0,0)\put(-138,0){}\end{picture}
               \caption{} 
        \end{subfigure}
        \begin{subfigure}[b]{0.325\textwidth}
                \includegraphics[width=\textwidth,trim=1 1 1 2,clip]{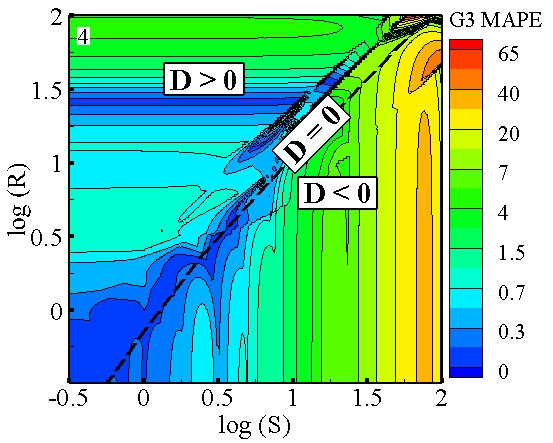}\begin{picture}(0,0)\put(-138,0){}\end{picture}
               \caption{} 
        \end{subfigure}    
 		\begin{subfigure}[b]{0.325\textwidth}
               \includegraphics[width=\textwidth,trim=2 2 2 2,clip]{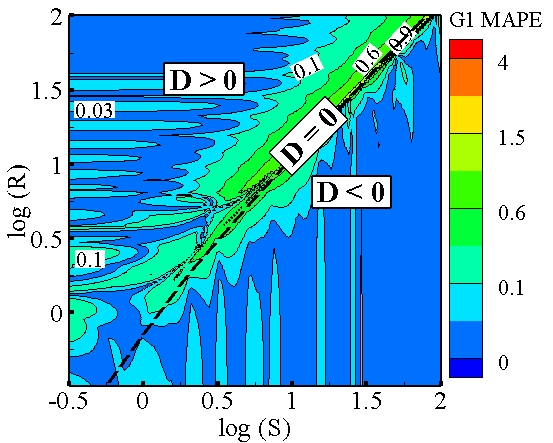}\begin{picture}(0,0)\put(-138,0){}\end{picture}
                \caption{}
        \end{subfigure}                                 
        \begin{subfigure}[b]{0.325\textwidth}
                \includegraphics[width=\textwidth,trim=1 1 1 2,clip]{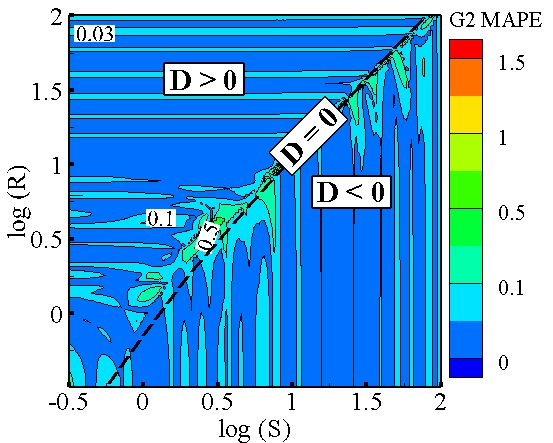}\begin{picture}(0,0)\put(-138,0){}\end{picture}
               \caption{} 
        \end{subfigure}
        \begin{subfigure}[b]{0.325\textwidth}
                \includegraphics[width=\textwidth,trim=1 1 1 2,clip]{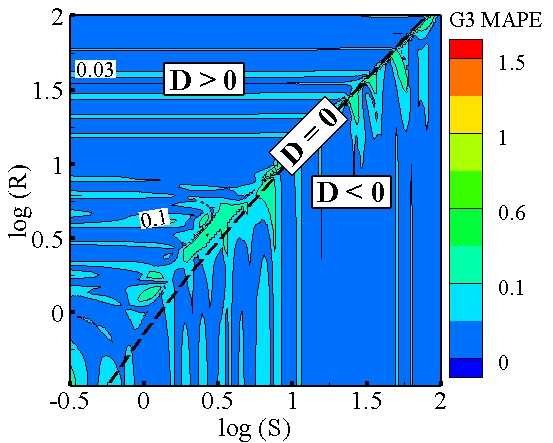}\begin{picture}(0,0)\put(-138,0){}\end{picture}
               \caption{} 
        \end{subfigure}    
        \caption{\label{fig_7} MAPE contours of CCC for ML models trained with different loss functions, (a)-(c) RMSE, (d)-(f) MAPE} 
        \end{figure}

The performance of the models trained with different loss functions are reported with both MAPE and RMSE error metrics on training and testing datasets in each column of Table~\ref{tab:4}. It can be seen that the network with MAPE loss function has smaller training and testing errors compared to the network with RMSE loss function. As mentioned in Sec.~\ref{sec:Method}, the CCC have a large range of values over the parameter space. When the RMSE is used as the loss function, the training and back--propagation process are mostly dominated by the CCC with larger magnitudes, as the small value coefficients contribute very little to the NN optimization process. Local error contours of the models trained with different loss functions are shown in Figs.~\ref{fig_6} and~\ref{fig_7}. It can be seen that the ML algorithm trained with MAPE has smaller local errors in different turbulence physics regions in the entire parameter space. 

Comparing different error metrics, MAPE vs. RMSE for final performance of the models in Table~\ref{tab:4} and absolute error vs. MAPE for local error contours in Figs.~\ref{fig_6} and~\ref{fig_7}, clearly shows that MAPE metric has easily interpretable presentation of final model performance in this study. Therefore, MAPE is selected as the appropriate type of loss function and evaluation metric for the remainder of the analysis.
\begin{table}
  \caption{Performance of models trained with different loss functions}
  \label{tab:4}
  \begin{tabularx}{0.95\textwidth}{XXXXX}
    \hline\hline
    & \multicolumn{4}{c}{Errors with various evaluation metrics} \\
    ML model & RMSE--training & RMSE--testing & MAPE--training & MAPE--testing  \\
    \hline
    RMSE--loss func. & $9.28\times10^{-3}$ & $9.34\times10^{-3}$ & 4.26 & 4.24 \\
    MAPE--loss func. & $4.03\times10^{-3}$ & $3.90\times10^{-3}$ & 0.043 & 0.044 \\
    \hline\hline
  \end{tabularx}
\end{table}
        
\subsection{\label{sec:Result_2} Case--1: Training data over the entire parameter space}
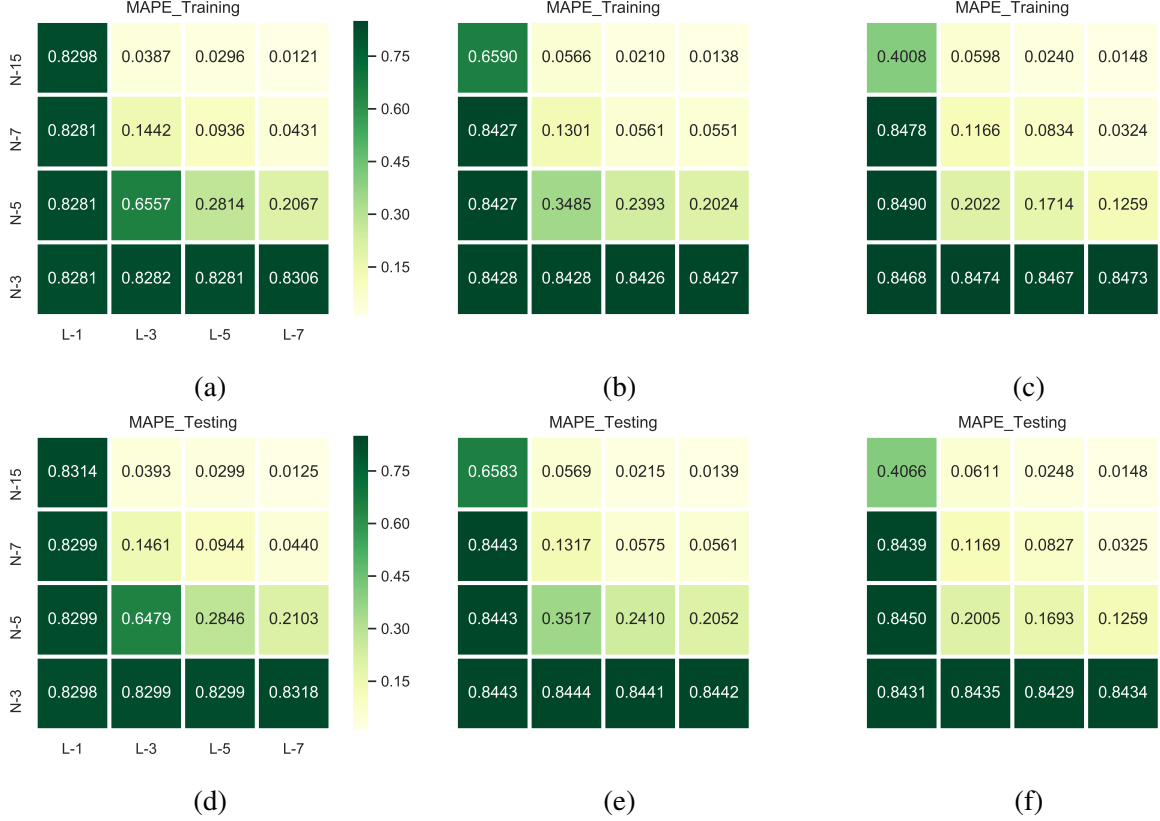
\begin{figure}
        \centering
 		\begin{subfigure}[b]{0.325\textwidth}
               \includegraphics[width=\textwidth,trim=85 18 50 20,clip]{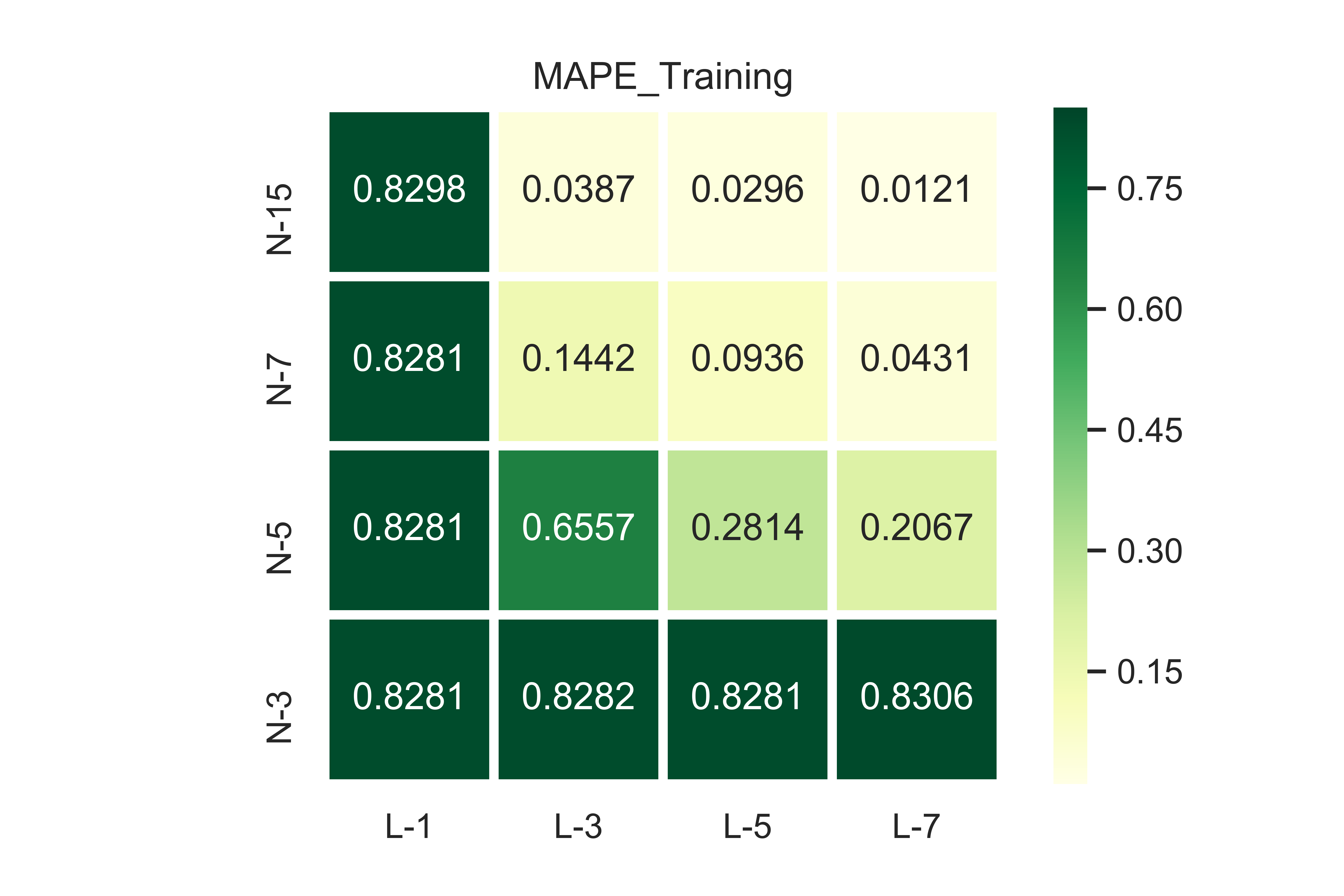}\begin{picture}(0,0)\put(-138,0){}\end{picture}
               \caption{}
        \end{subfigure}                                 
        \begin{subfigure}[b]{0.325\textwidth}
                \includegraphics[width=\textwidth,trim=85 18 50 20,clip]{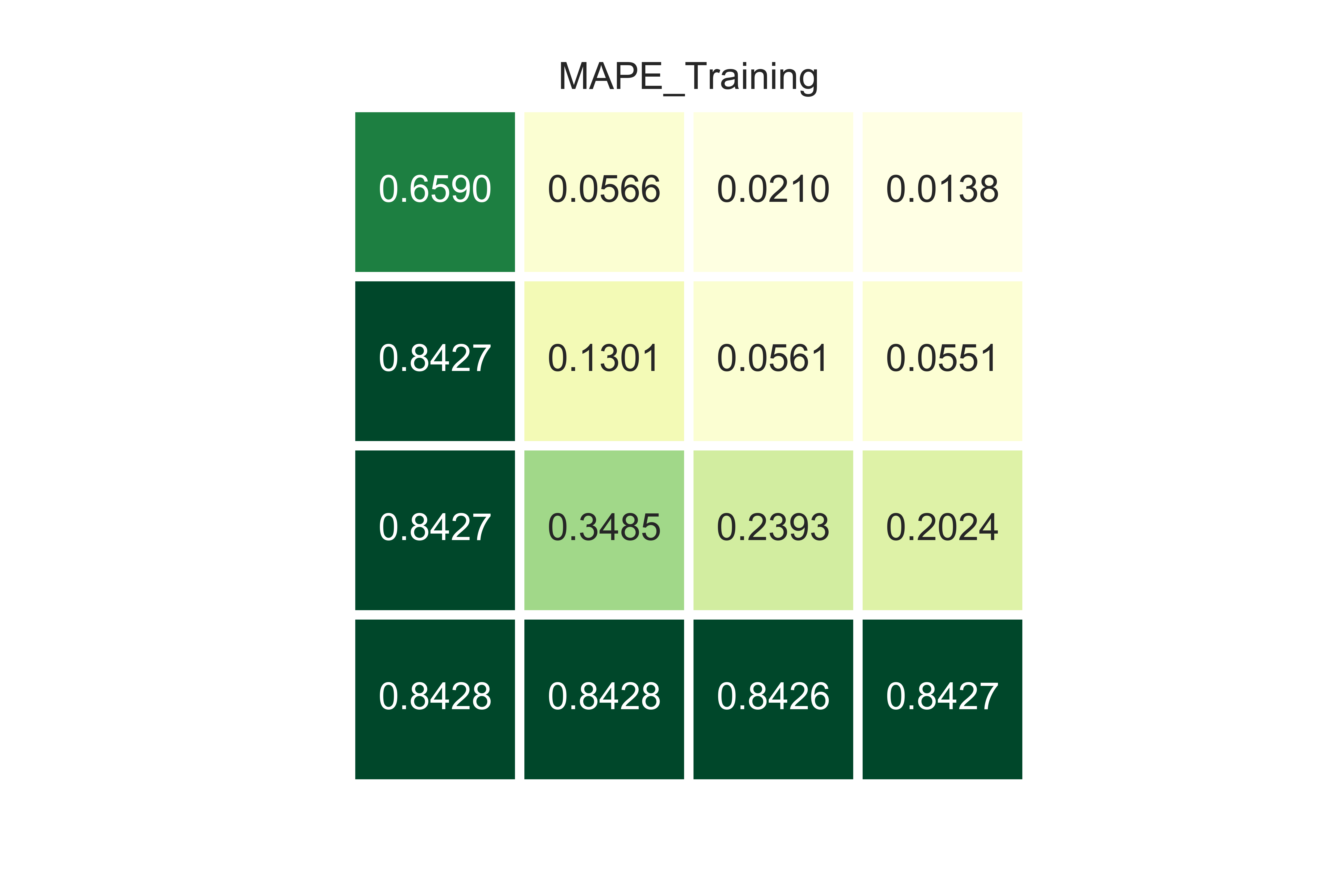}\begin{picture}(0,0)\put(-138,0){}\end{picture}
               \caption{} 
        \end{subfigure}
        \begin{subfigure}[b]{0.325\textwidth}
                \includegraphics[width=\textwidth,trim=85 18 50 20,clip]{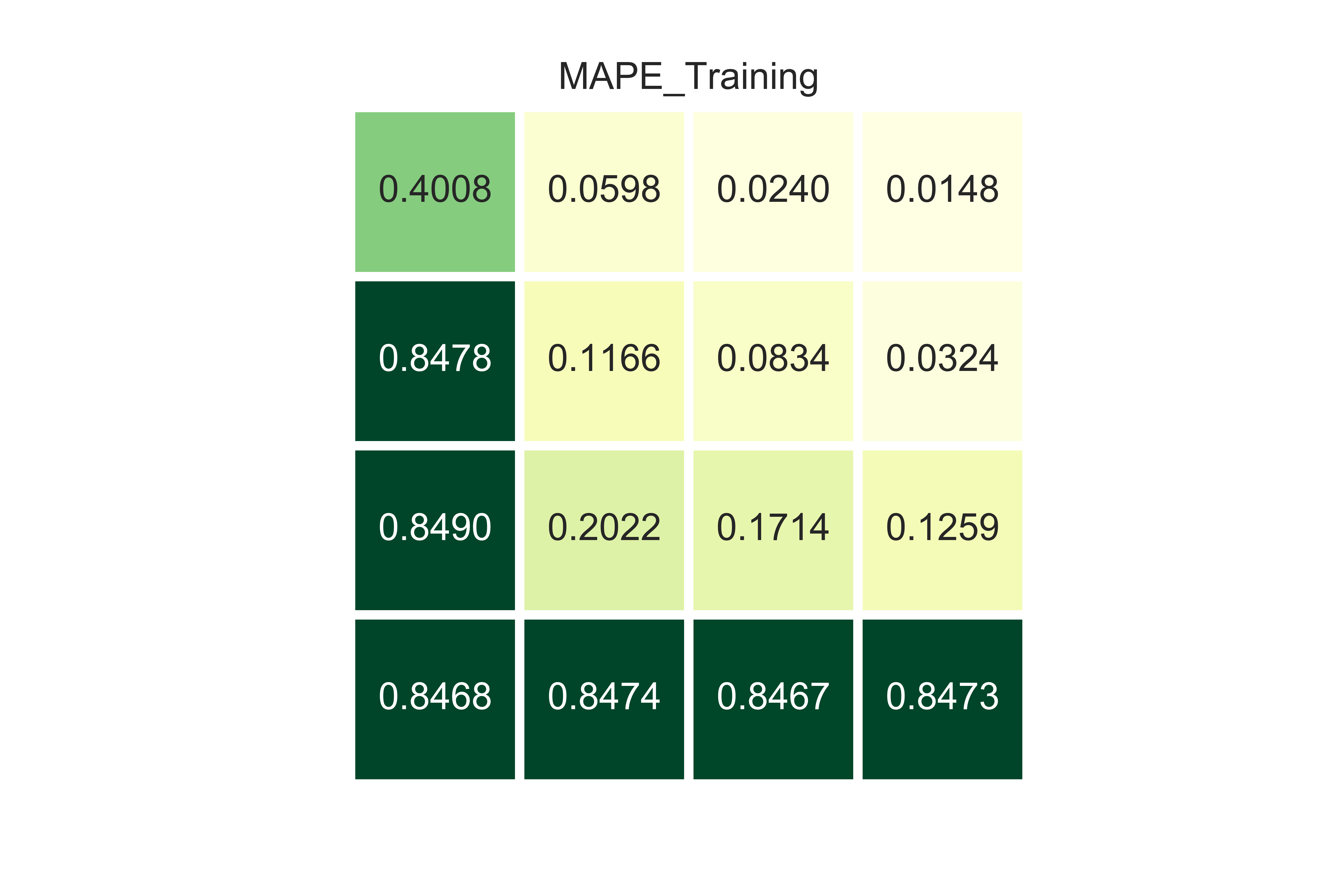}\begin{picture}(0,0)\put(-138,0){}\end{picture}
               \caption{}
        \end{subfigure}    
 		\begin{subfigure}[b]{0.325\textwidth}
               \includegraphics[width=\textwidth,trim=85 18 50 20,clip]{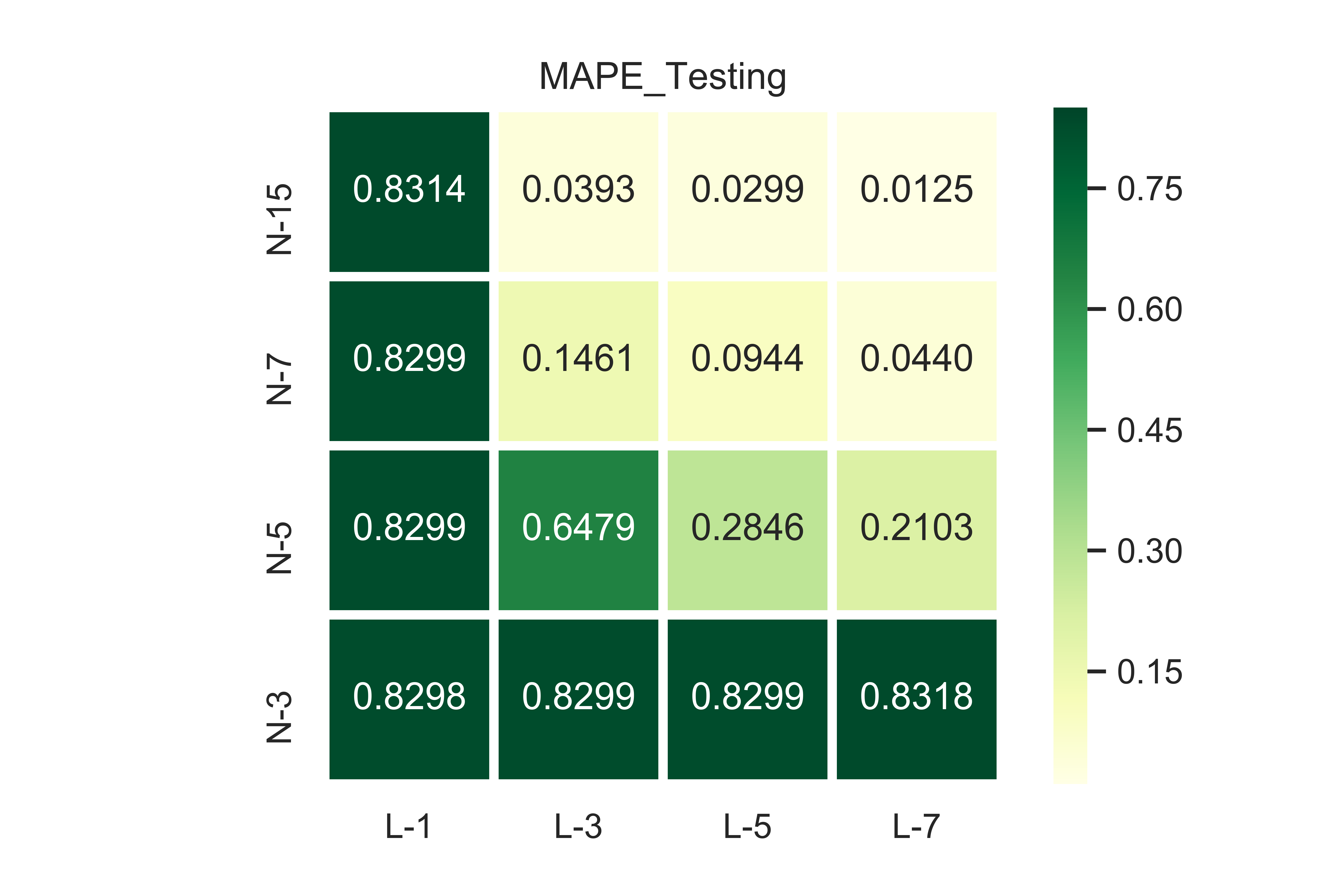}\begin{picture}(0,0)\put(-138,0){}\end{picture}
                \caption{}
        \end{subfigure}                                 
        \begin{subfigure}[b]{0.325\textwidth}
                \includegraphics[width=\textwidth,trim=85 18 50 20,clip]{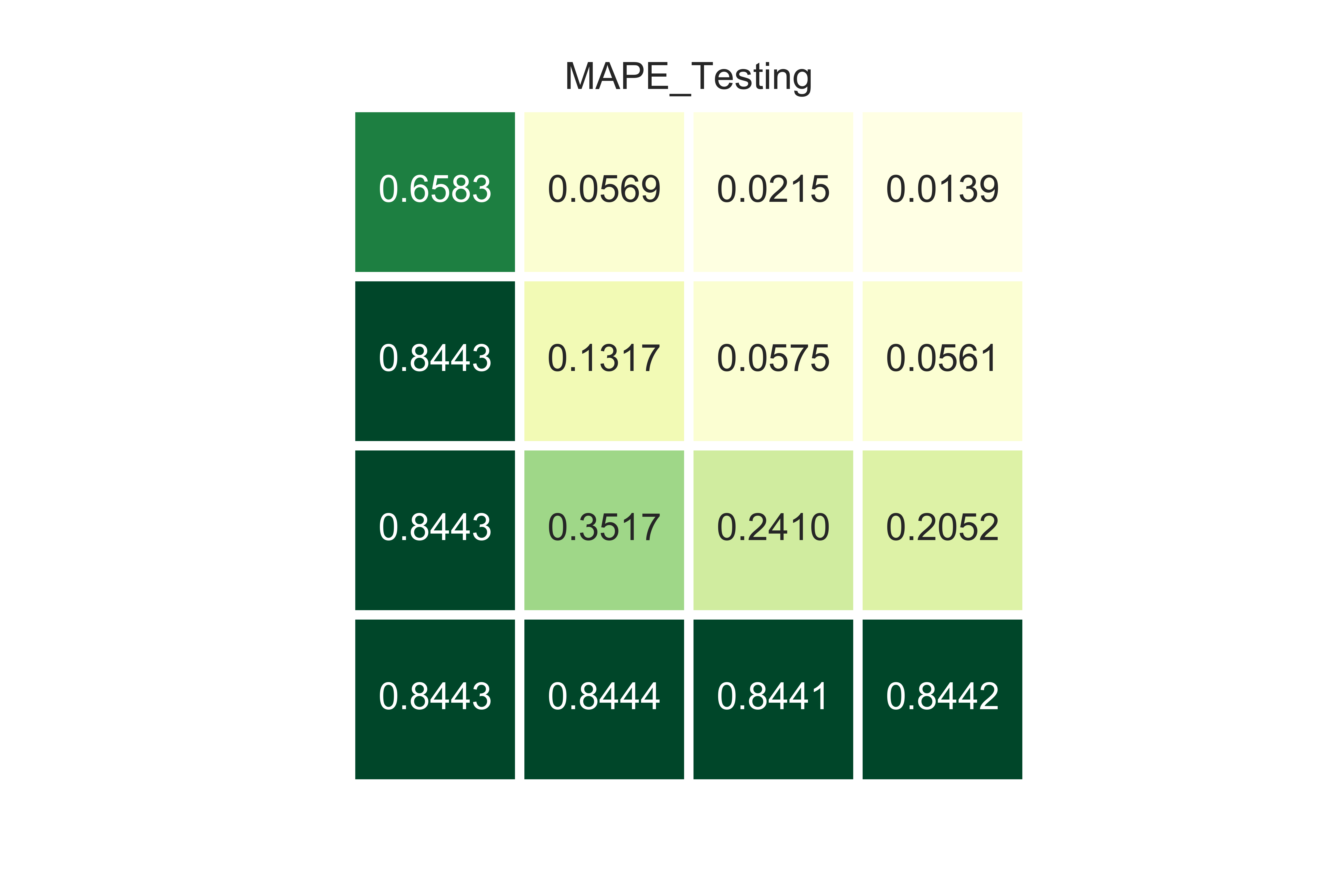}\begin{picture}(0,0)\put(-138,0){}\end{picture}
               \caption{} 
        \end{subfigure}
        \begin{subfigure}[b]{0.325\textwidth}
                \includegraphics[width=\textwidth,trim=85 18 50 20,clip]{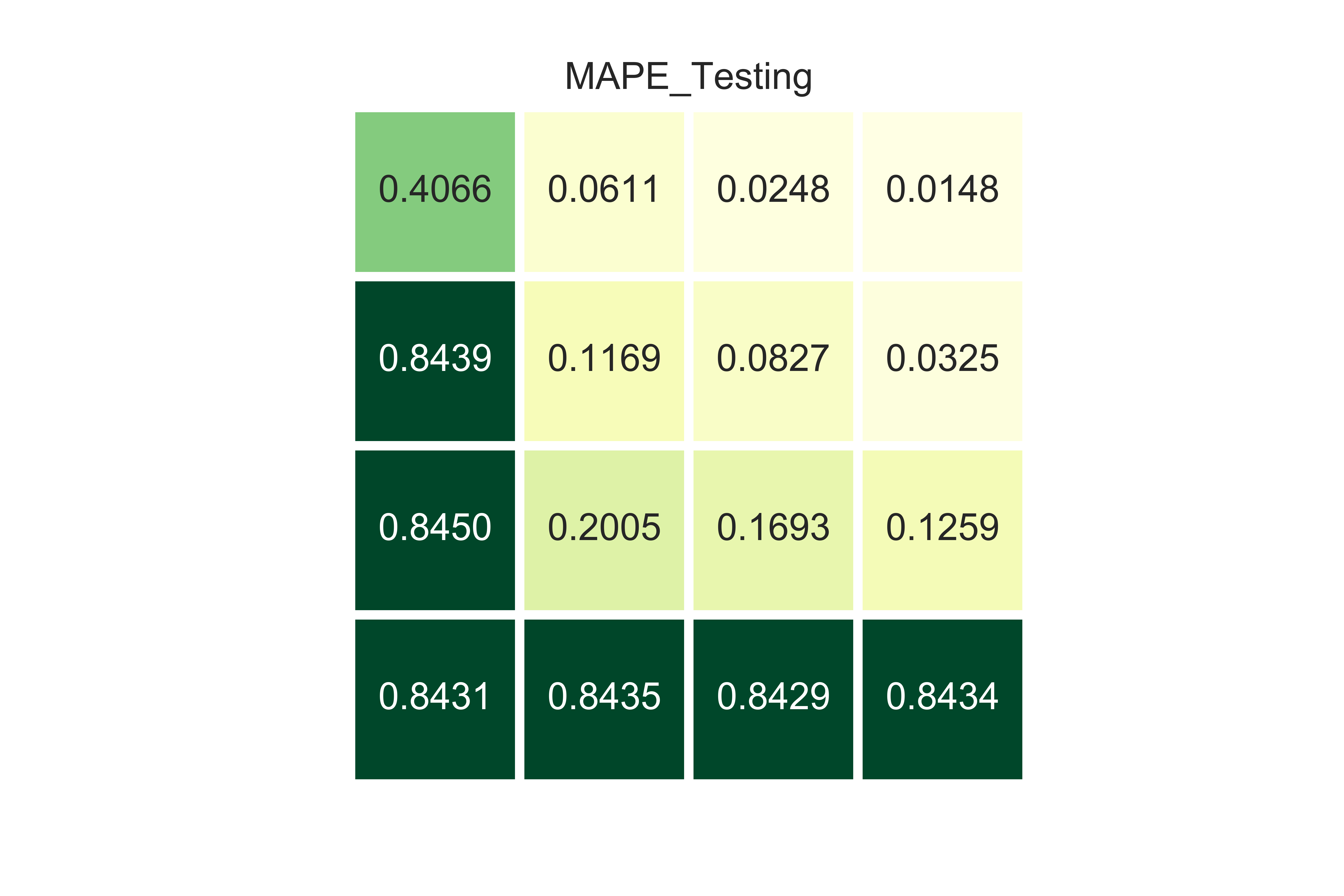}\begin{picture}(0,0)\put(-138,0){}\end{picture}
               \caption{} 
        \end{subfigure}    
        \caption{\label{fig_8} Training and testing MAPE for Case--1, (a) and (d) SSG, (b) and (e) LRR, (c) and (f) SZL} 
        \end{figure}

In this experiment, three parameter--to--solution maps generated by SZL, ARSM with SSG and LRR proxy--physics turbulence surrogates are considered. The datasets are randomly distributed for training the NNs, i.e., 75\% for training and 25\% for testing the models. Using the generated datasets, networks with different architectures are trained. The type of activation function for all computation neurons is ReLU and all other hyperparameters are fixed as shown in Table~\ref{tab:3} for this case. Fig.~\ref{fig_8} demonstrates the training and testing MAPE errors of all the 16 network architectures trained with different data manifolds. In this figure number of hidden layers and number of neurons in each layer are shown with horizontal and vertical axes, respectively. Although for this case, $L_2$ regularization is not used during ML training, all the 16 networks have similar performance on both the training and testing datasets and models test quite well without over/under fitting. It can be seen that shallow NNs (networks with one layer) and DNNs with small width (networks with three neurons in each layer) have the worst training and testing errors for all the data manifolds in the interpolation case. Additionally, as the number of the layers and neurons in each layer increase, the training and testing errors decrease and for this case NN with largest degrees of freedom (7L--15N) has the lowest training and testing errors. This is a surprising observation that the NN-based models require very large networks (more degrees of freedom than true proxy--physics model) to reduce errors to a reasonable level even for this interpolation case. Therefore, DNNs are not efficient approximations for the non--linear solution manifolds created by simplified proxy--physics turbulence surrogates.

\begin{figure}
        \centering
 		\captionsetup{justification=centering} 
 		\begin{subfigure}[b]{0.325\textwidth}
               \includegraphics[width=\textwidth,trim=2 2 2 2,clip]{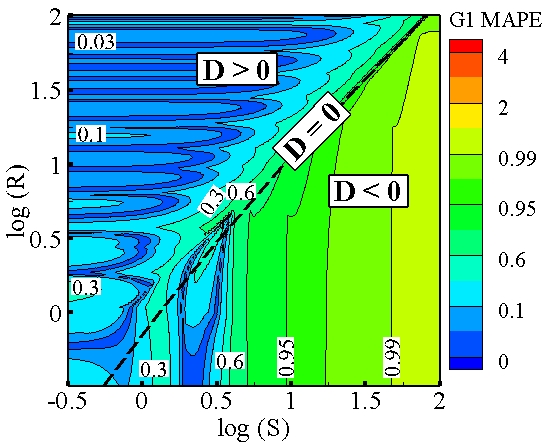}\begin{picture}(0,0)\put(-138,0){}\end{picture}
                \caption{}
        \end{subfigure}                                 
        \begin{subfigure}[b]{0.325\textwidth}
                \includegraphics[width=\textwidth,trim=1 1 1 2,clip]{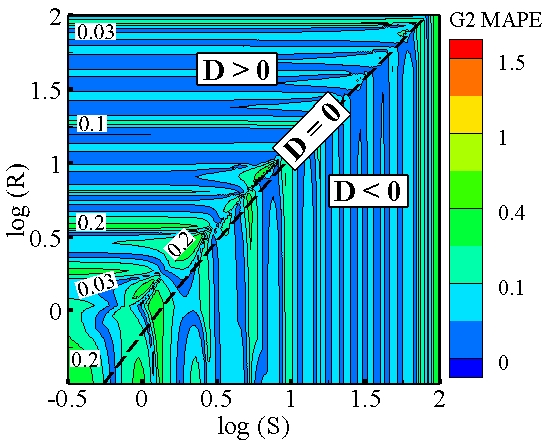}\begin{picture}(0,0)\put(-138,0){}\end{picture}
               \caption{} 
        \end{subfigure}
        \begin{subfigure}[b]{0.325\textwidth}
                \includegraphics[width=\textwidth,trim=1 1 1 2,clip]{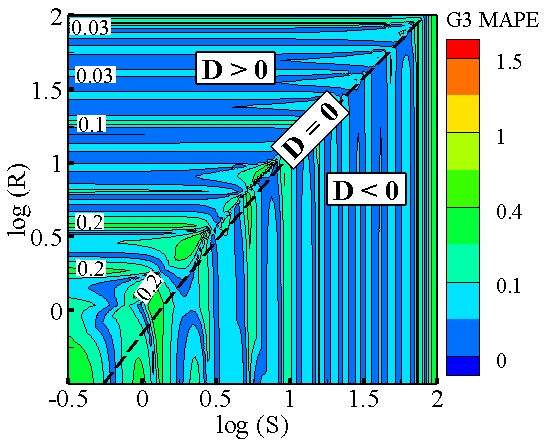}\begin{picture}(0,0)\put(-138,0){}\end{picture}
               \caption{} 
        \end{subfigure}    
 		\begin{subfigure}[b]{0.325\textwidth}
               \includegraphics[width=\textwidth,trim=2 2 2 2,clip]{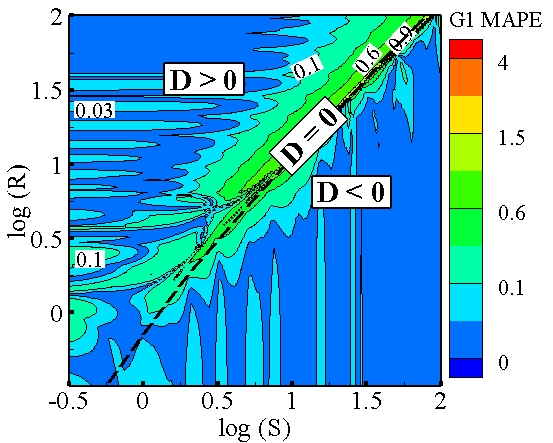}\begin{picture}(0,0)\put(-138,0){}\end{picture}
                \caption{}
        \end{subfigure}                                 
        \begin{subfigure}[b]{0.325\textwidth}
                \includegraphics[width=\textwidth,trim=1 1 1 2,clip]{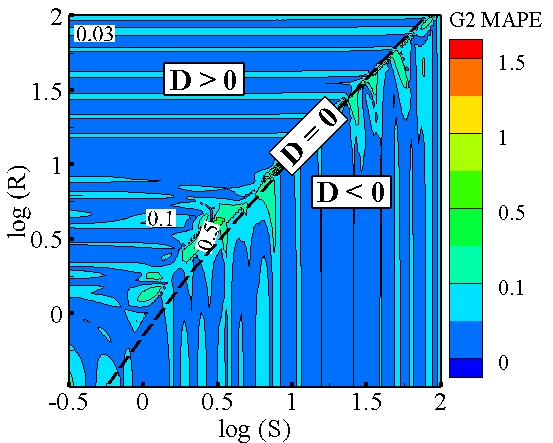}\begin{picture}(0,0)\put(-138,0){}\end{picture}
               \caption{} 
        \end{subfigure}
        \begin{subfigure}[b]{0.325\textwidth}
                \includegraphics[width=\textwidth,trim=1 1 1 2,clip]{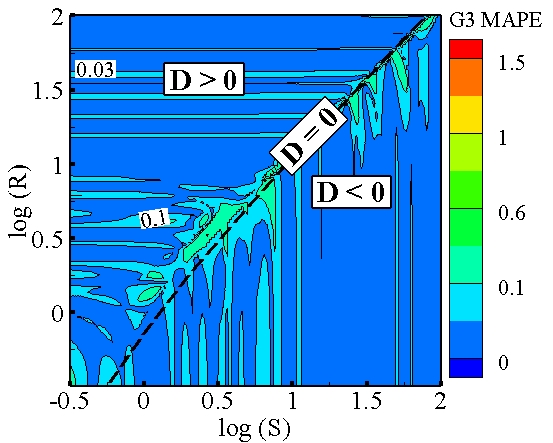}\begin{picture}(0,0)\put(-138,0){}\end{picture}
               \caption{} 
        \end{subfigure}           
 		\begin{subfigure}[b]{0.325\textwidth}
               \includegraphics[width=\textwidth,trim=2 2 2 2,clip]{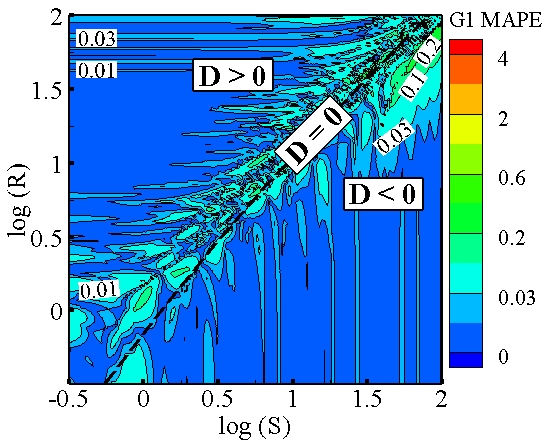}\begin{picture}(0,0)\put(-138,0){}\end{picture}
                \caption{}
        \end{subfigure}                                 
        \begin{subfigure}[b]{0.325\textwidth}
                \includegraphics[width=\textwidth,trim=1 1 1 2,clip]{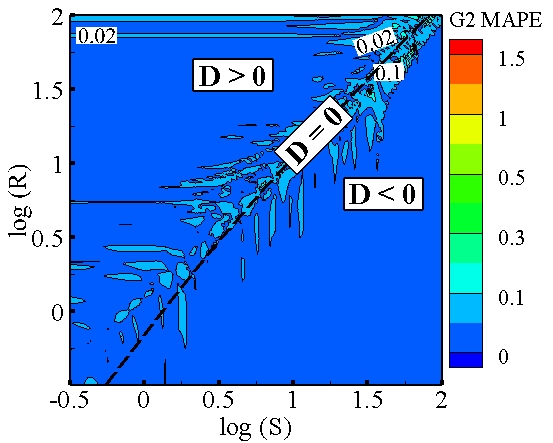}\begin{picture}(0,0)\put(-138,0){}\end{picture}
               \caption{} 
        \end{subfigure}
        \begin{subfigure}[b]{0.325\textwidth}
                \includegraphics[width=\textwidth,trim=1 1 1 2,clip]{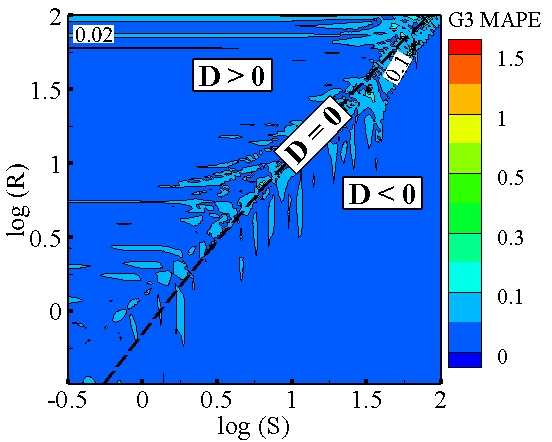}\begin{picture}(0,0)\put(-138,0){}\end{picture}
               \caption{} 
        \end{subfigure}           
        \caption{\label{fig_9} MAPE contours for Case--1, (a)-(c) 7L--5N, (d)-(f) 7L--7N, (g)-(i) 7L--15N} 
        \end{figure}

The approximation capabilities of the large NNs with the data manifolds of different complexity levels are comparable in this case. However, for the networks with reasonable degrees of freedom (total number of neurons less than 25), the errors are smallest in data manifold with no bifurcation (SZL model). These networks have bigger errors in data manifold with moderate non--linearity (ARSM with SSG model) compared to data manifold with mild non--linearity (ARSM with LRR model). Local MAPE contours in the entire parameter space for the networks with 7 hidden layers and different neurons trained with data manifold generated by ARSM and SSG model are shown in Fig.~\ref{fig_9}. It is evident that by increasing the number of neurons in each layer, error decreases in the entire parameter space and for the network with large architecture (7L--15N) the maximum error occurs mainly near the bifurcation line $D=0$ where the physics of turbulence undergoes rapid change. It should be noted that ReLU activation function is considered for all the NNs so far.

\begin{figure}
        \centering
 		\begin{subfigure}[b]{0.495\textwidth}
               \includegraphics[width=\textwidth,trim=85 18 50 20,clip]{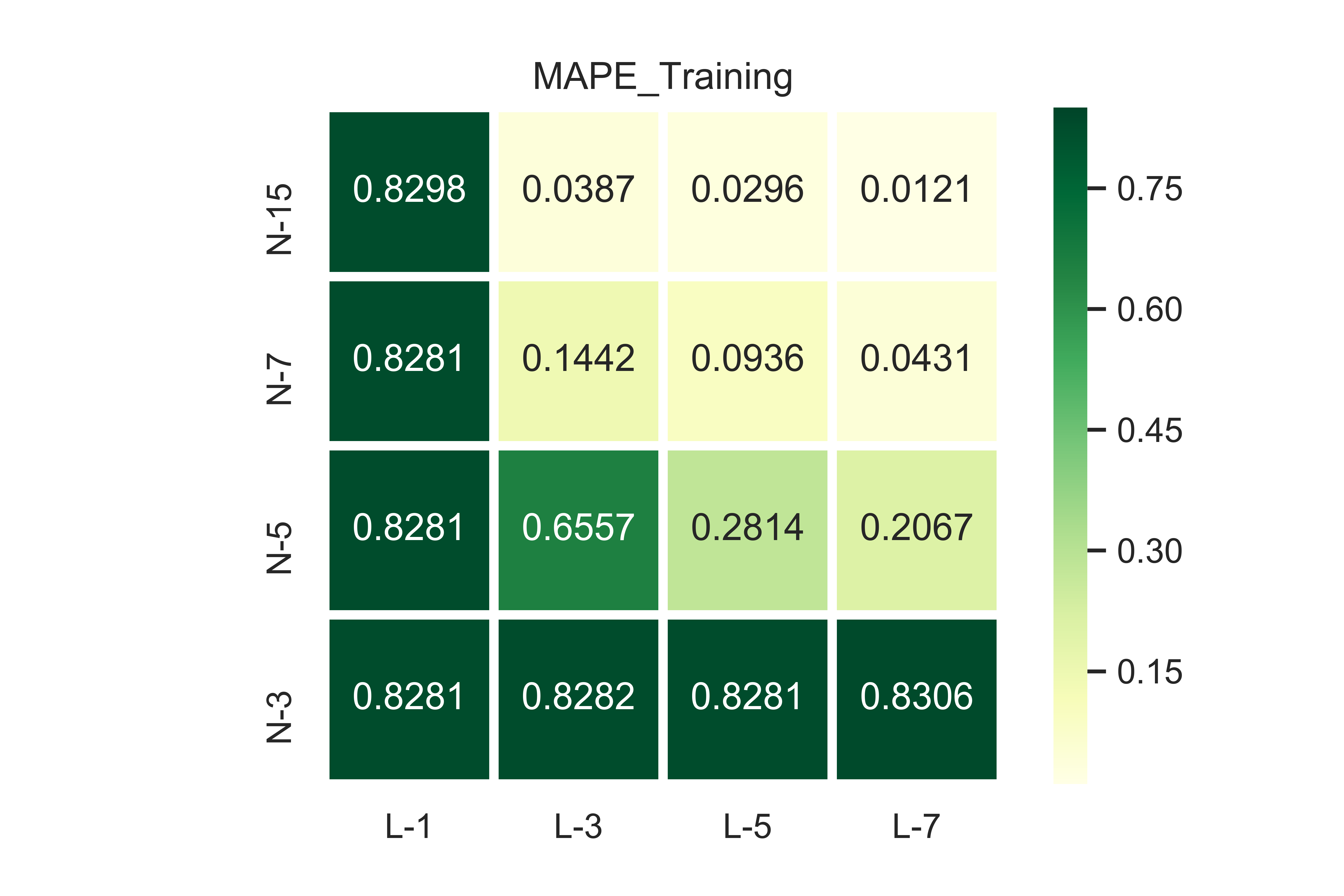}\begin{picture}(0,0)\put(-138,0){}\end{picture}
               \caption{}
        \end{subfigure}                                 
        \begin{subfigure}[b]{0.495\textwidth}
                \includegraphics[width=\textwidth,trim=85 18 50 20,clip]{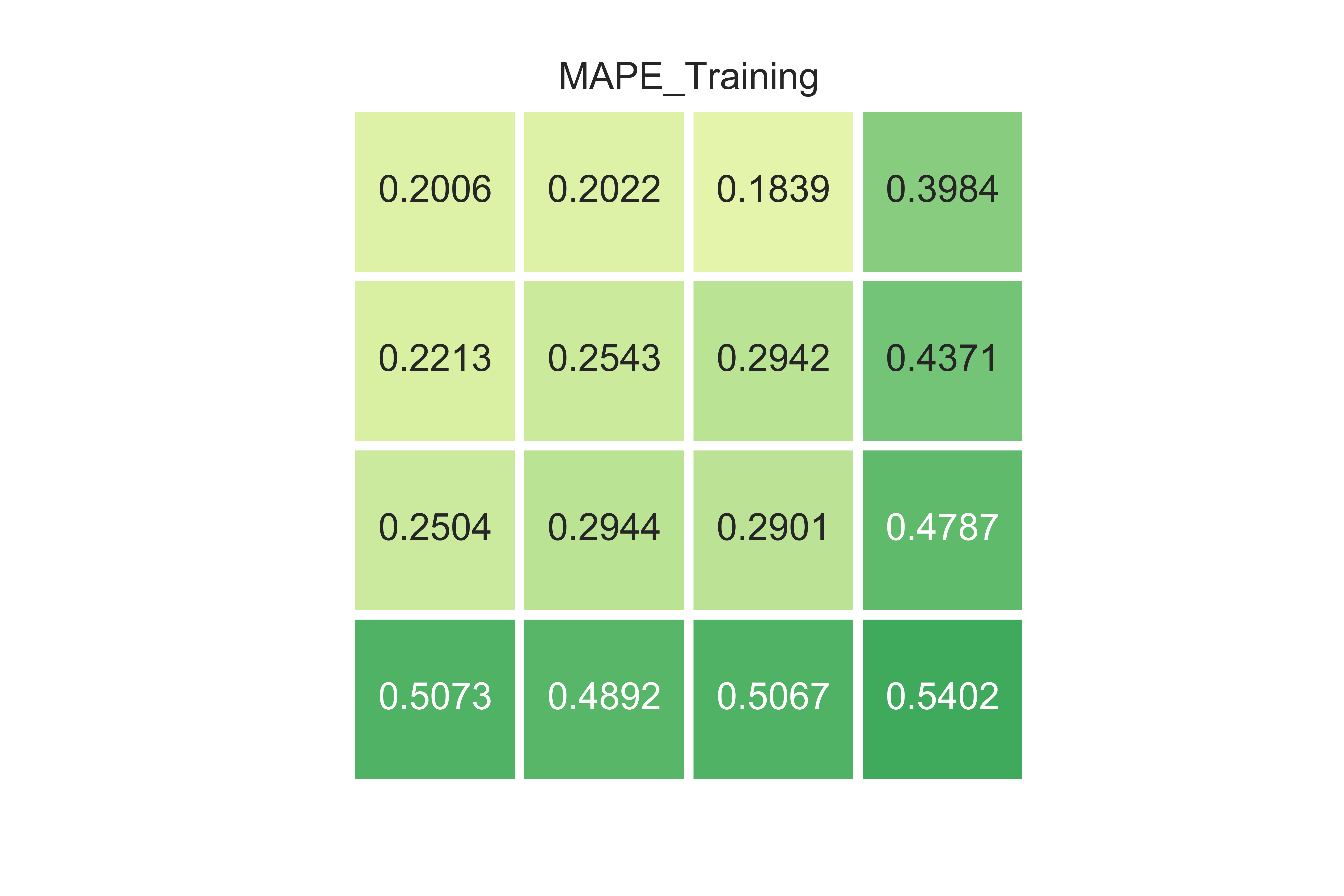}\begin{picture}(0,0)\put(-138,0){}\end{picture}
               \caption{} 
        \end{subfigure}
 		\begin{subfigure}[b]{0.495\textwidth}
               \includegraphics[width=\textwidth,trim=85 18 50 20,clip]{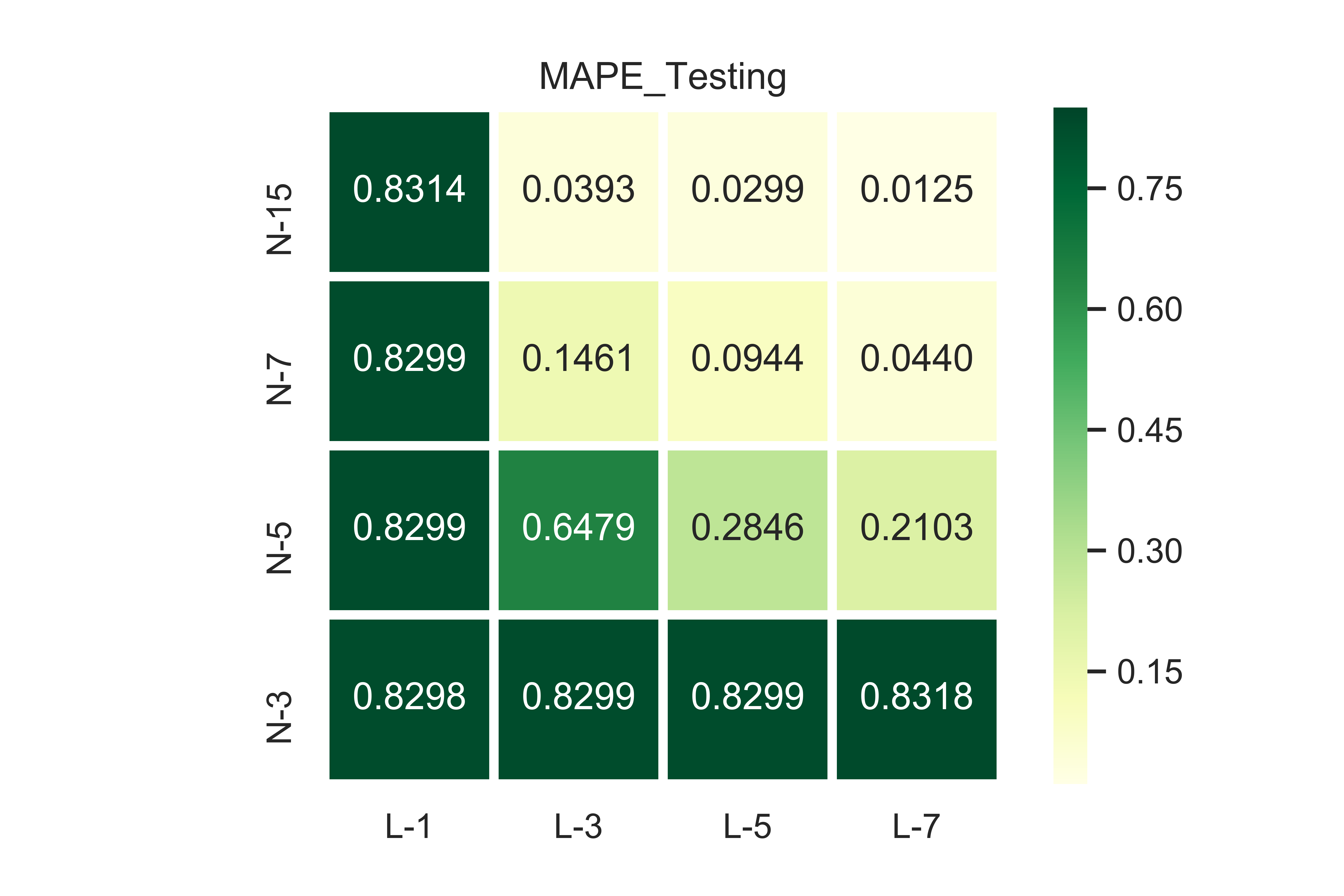}\begin{picture}(0,0)\put(-138,0){}\end{picture}
                \caption{}
        \end{subfigure}                                 
        \begin{subfigure}[b]{0.495\textwidth}
                \includegraphics[width=\textwidth,trim=85 18 50 20,clip]{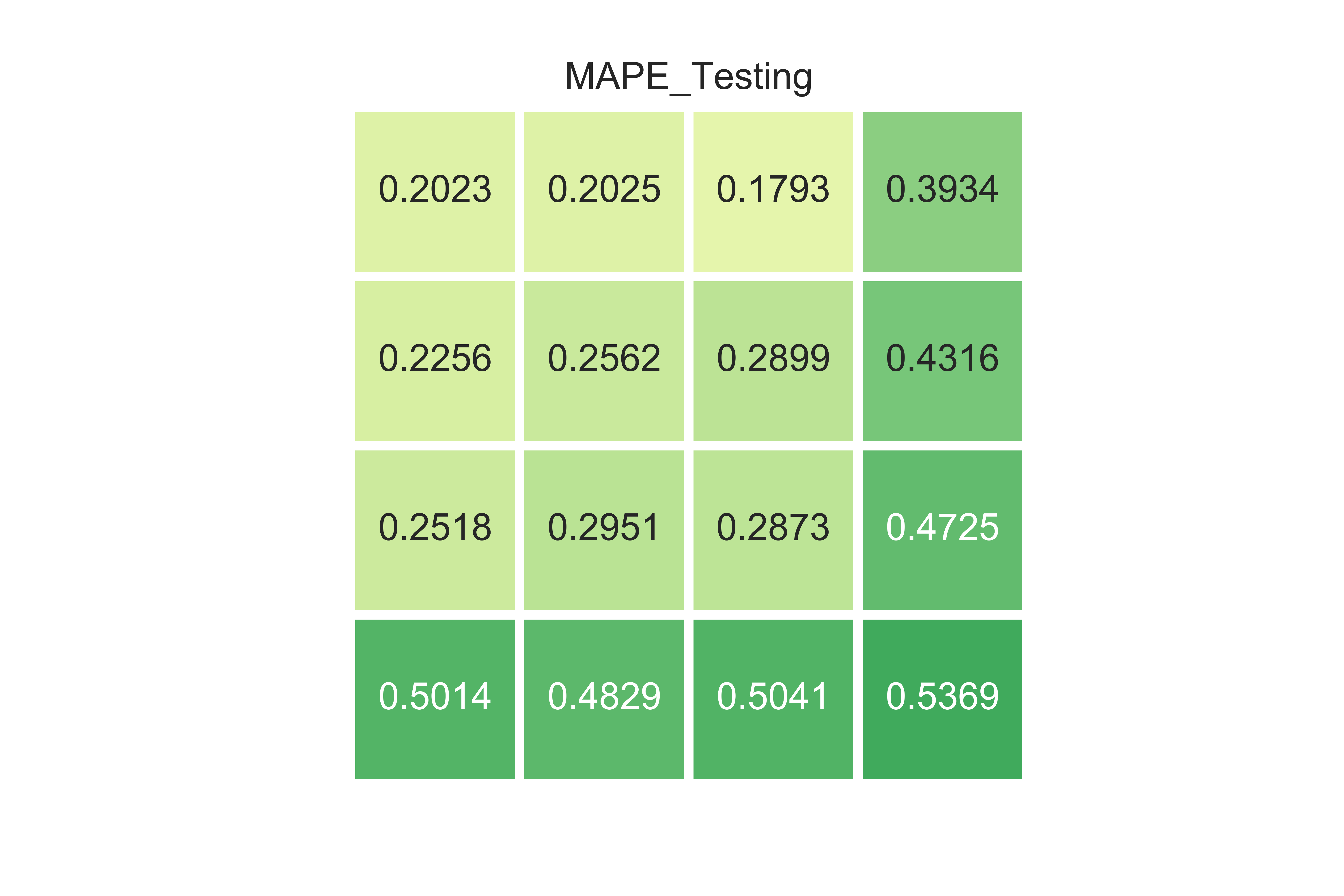}\begin{picture}(0,0)\put(-138,0){}\end{picture}
               \caption{} 
        \end{subfigure}
        \caption{\label{fig_10} Training and testing MAPE for Case--1, (a) and (c) ReLU, (b) and (d) Sigmoid} 
        \end{figure}
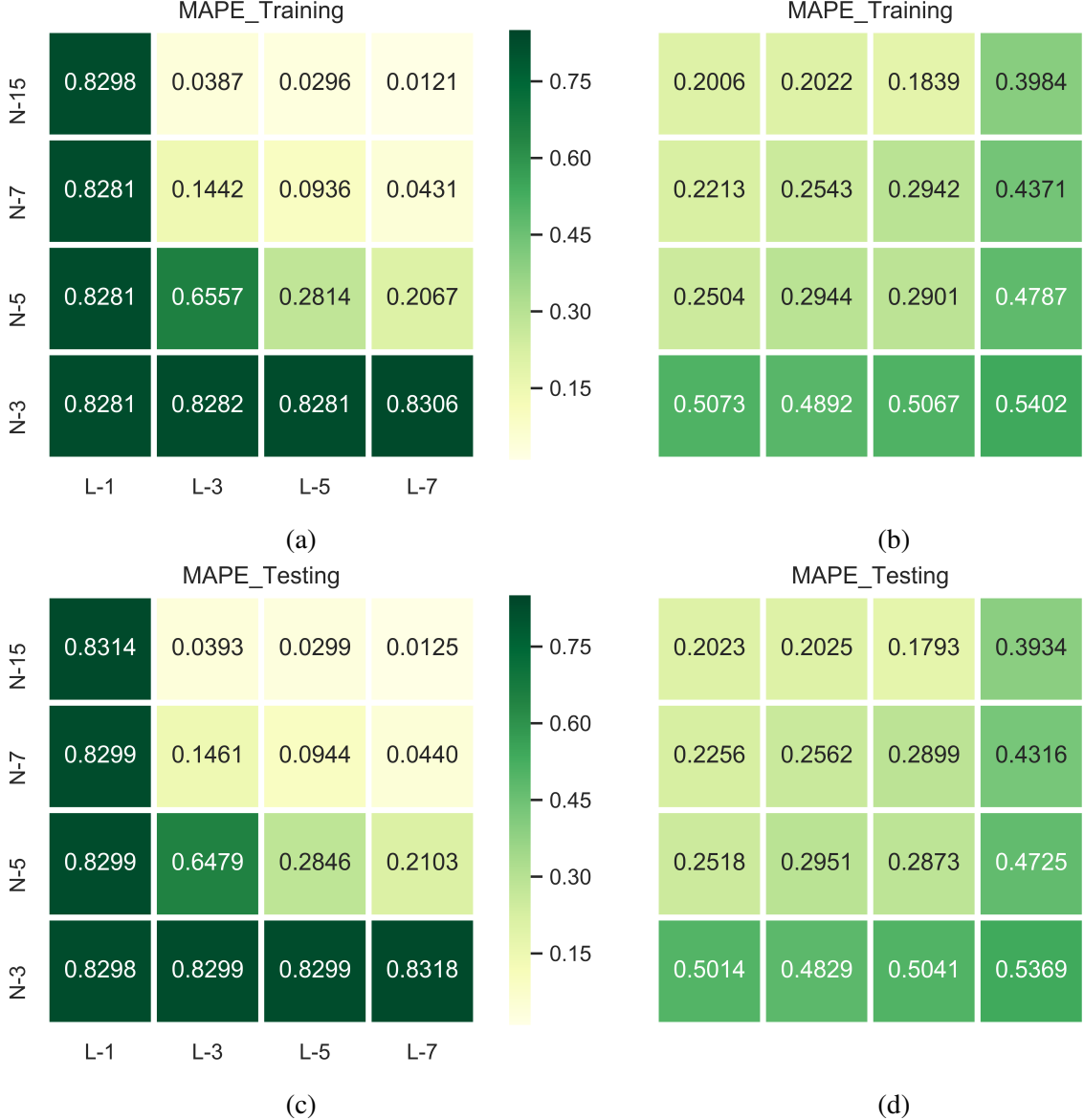

For Case--1, the effects of using ReLU and Sigmoid activation functions on approximation capability of the NNs with different architectures are further investigated. In this experiment all the hyperparameters are kept fixed as shown in Table~\ref{tab:3} and the data manifold generated by ARSM with SSG model is considered. The training and testing MAPE errors of all the 16 network architectures for both activation functions are illustrated in Fig.~\ref{fig_10}. It is evident that the performance of the NNs with the considered activation functions are significantly different. For both activation functions, networks with small width have large training and testing errors. But small width networks with Sigmoid activation function have smaller training and testing errors. Unlike the networks with ReLU activation function, the training and testing errors does not decrease as the number of the layers increases in networks with Sigmoid activation function and these networks outperform in shallow with large width architectures. It should be noted that Sigmoid activation function involves expensive operations (exponentials, etc.) compared to ReLU which is simply thresholded at zero (Eq.~\eqref{eq:Relu}). As illustrated in Fig.~\ref{fig_3}, Sigmoid activation function takes an input ($z$) and outputs a value ($\sigma$) in the range between 0 and 1. It is known that when a neuron with Sigmoid activation function saturates at either tail of 0 and 1, it `kills' the gradient and the information is not transferred through the neuron \cite{goodfellow2016deep}. Hence the saturation of the neurons with Sigmoid activation function could negatively affect the learning of the large networks. On the other hand, a large gradients for ReLU neurons in shallow and small width networks could unfavorably update the weights during the back--propagation process. Therefore, the ReLU neurons can irreversibly `die' during training since they can get knocked off the data manifold \cite{goodfellow2016deep}.
\subsection{\label{sec:Result_2-1} Case--2: Training data only in the strain--dominated region (D < 0)}
\begin{figure}
        \centering
 		\captionsetup{justification=centering} 
 		\begin{subfigure}[b]{0.325\textwidth}
               \includegraphics[width=\textwidth,trim=2 2 2 2,clip]{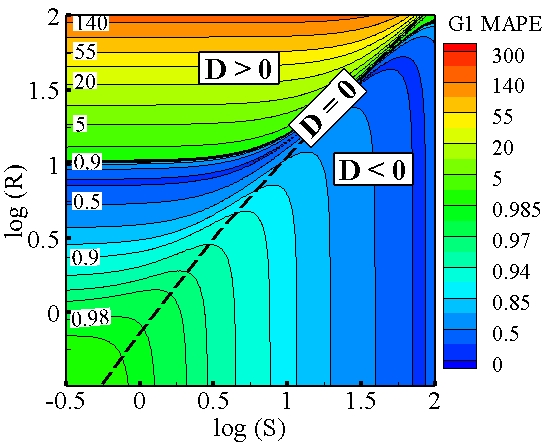}\begin{picture}(0,0)\put(-138,0){}\end{picture}
                \caption{}
        \end{subfigure}                                 
        \begin{subfigure}[b]{0.325\textwidth}
                \includegraphics[width=\textwidth,trim=1 1 1 2,clip]{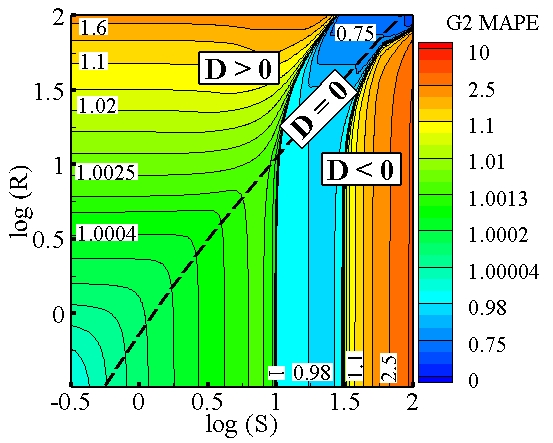}\begin{picture}(0,0)\put(-138,0){}\end{picture}
               \caption{} 
        \end{subfigure}
        \begin{subfigure}[b]{0.325\textwidth}
                \includegraphics[width=\textwidth,trim=1 1 1 2,clip]{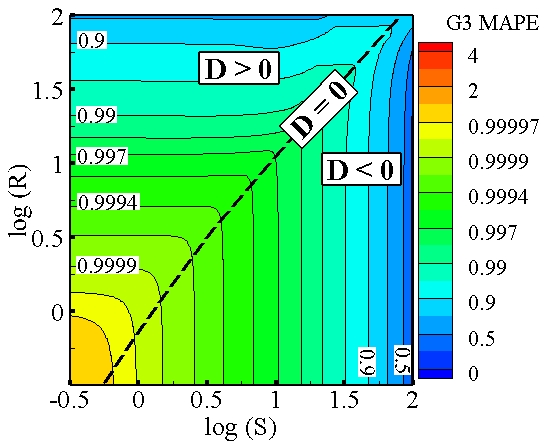}\begin{picture}(0,0)\put(-138,0){}\end{picture}
               \caption{} 
        \end{subfigure}    
 		\begin{subfigure}[b]{0.325\textwidth}
               \includegraphics[width=\textwidth,trim=2 2 2 2,clip]{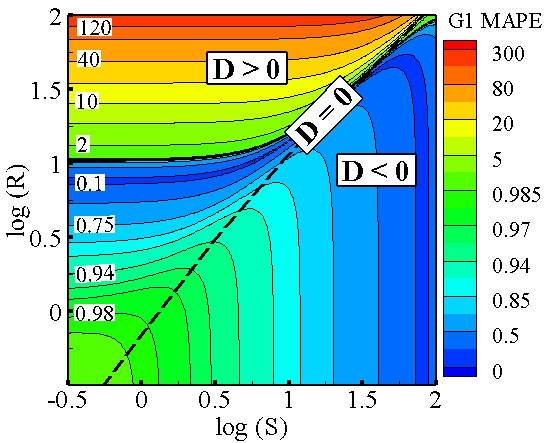}\begin{picture}(0,0)\put(-138,0){}\end{picture}
                \caption{}
        \end{subfigure}                                 
        \begin{subfigure}[b]{0.325\textwidth}
                \includegraphics[width=\textwidth,trim=1 1 1 2,clip]{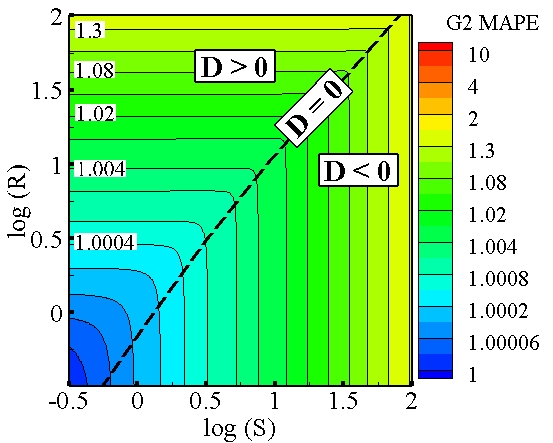}\begin{picture}(0,0)\put(-138,0){}\end{picture}
               \caption{} 
        \end{subfigure}
        \begin{subfigure}[b]{0.325\textwidth}
                \includegraphics[width=\textwidth,trim=1 1 1 2,clip]{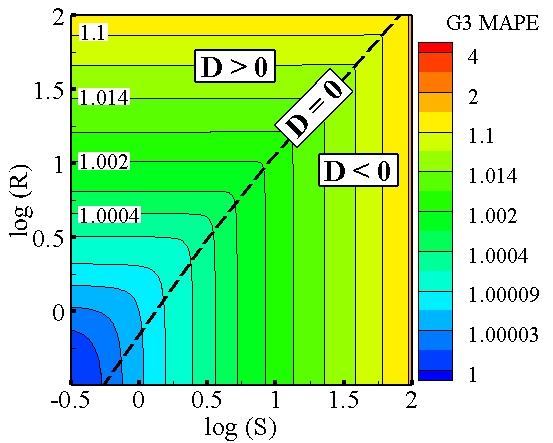}\begin{picture}(0,0)\put(-138,0){}\end{picture}
               \caption{} 
        \end{subfigure}    
 		\begin{subfigure}[b]{0.325\textwidth}
               \includegraphics[width=\textwidth,trim=2 2 2 2,clip]{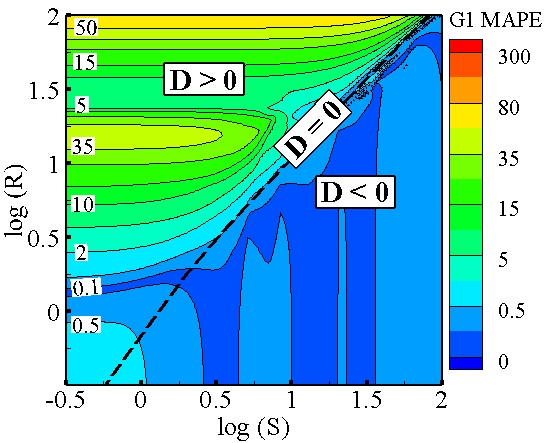}\begin{picture}(0,0)\put(-138,0){}\end{picture}
                \caption{}
        \end{subfigure}                                 
        \begin{subfigure}[b]{0.325\textwidth}
                \includegraphics[width=\textwidth,trim=1 1 1 2,clip]{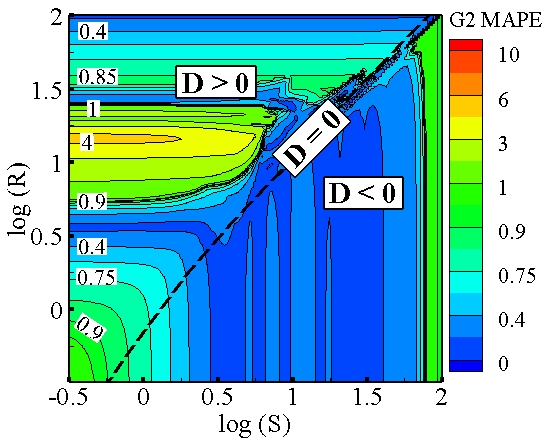}\begin{picture}(0,0)\put(-138,0){}\end{picture}
               \caption{} 
        \end{subfigure}
        \begin{subfigure}[b]{0.325\textwidth}
                \includegraphics[width=\textwidth,trim=1 1 1 2,clip]{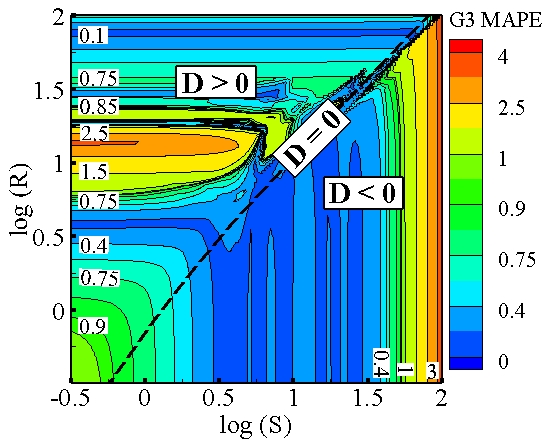}\begin{picture}(0,0)\put(-138,0){}\end{picture}
               \caption{} 
        \end{subfigure}           
        \caption{\label{fig_12} MAPE contours for Case--2, (a)-(c) 7L--5N, (d)-(f) 7L--7N, (g)-(i) 7L--15N} 
        \end{figure}

In this experiment the existence of generalizable NNs is examined when training data is partially available only in one side of parameter space (extrapolation). The data points in the strain--dominated region $D < 0$ are used for training and the data points in the rotation--dominated region $D > 0$ are used for testing of the models. For this analysis, ReLU activation function and $L_2$ norm regularization with $\lambda = 0.1$ are used. All other hyperparameters are fixed as shown in Table~\ref{tab:3}. A justification for the choice of hyperparameters for this case is given in appendix, Sec.~\ref{sec:App}. Fig.~\ref{fig_11} shows the training and testing errors of the 16 network architectures trained with partially available data generated by ARSM with SSG and LRR models. Unlike the interpolation case (Case--1), the ML models trained with limited (biased) data show overfitting for large networks. It is seen from Fig.~\ref{fig_11} that for both the data manifolds, the NNs with one and three hidden layers have the worst performance in training dataset. As the number of hidden layers increases to five, the capability of the NNs in capturing the non--linear relationship between the input parameters and CCC labels increases. Although, the training error reduces with increasing number of hidden layers from three to five, the testing error increases. By further increasing the number of hidden layers and neurons the performance of the model in training and testing datasets oscillates. Comparing the performance of the networks trained with different data manifolds in this case, shows that similar to the interpolation case, NNs trained with moderately non--linear data manifold (ARSM with SSG model) have bigger errors than NNs trained with mildly non--linear data manifold (ARSM with LRR model). However, unlike the interpolation case, finding a generalizable NN with a reasonable size is not straightforward when training data partially describe the true proxy--physics surrogate in the parameter space (only the strain--dominated region in this case). Local MAPE contours of DNNs with 7 hidden layers and different neurons trained with moderately non--linear data manifold are compared in Fig.~\ref{fig_12}. It is seen that DNNs with large number of neurons has relatively smaller errors in both training and testing regions of parameter space for all the CCC. 

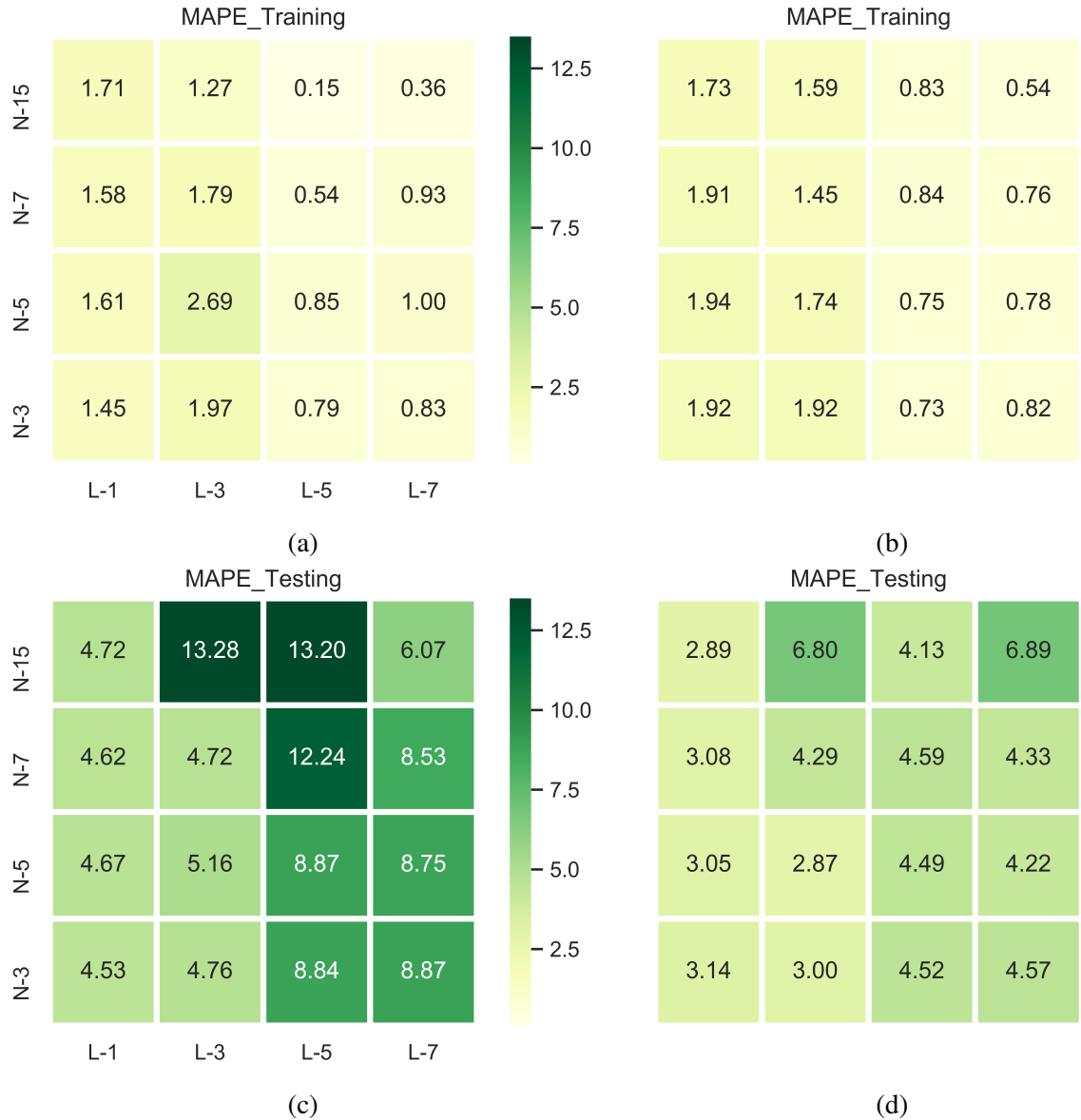
\begin{figure}
        \centering
 		\begin{subfigure}[b]{0.495\textwidth}
               \includegraphics[width=\textwidth,trim=85 18 50 20,clip]{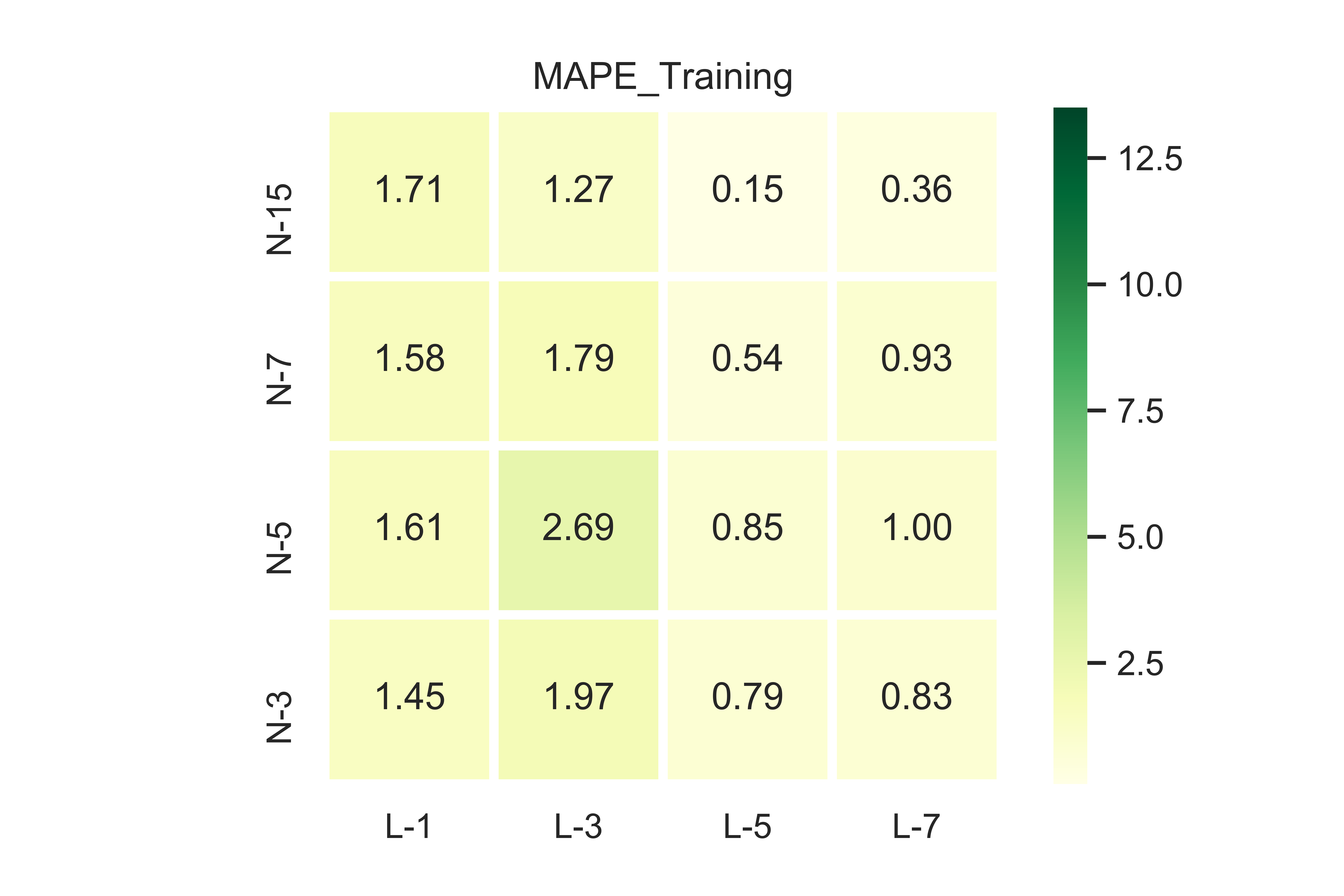}\begin{picture}(0,0)\put(-138,0){}\end{picture}
               \caption{}
        \end{subfigure}                                 
        \begin{subfigure}[b]{0.495\textwidth}
                \includegraphics[width=\textwidth,trim=85 18 50 20,clip]{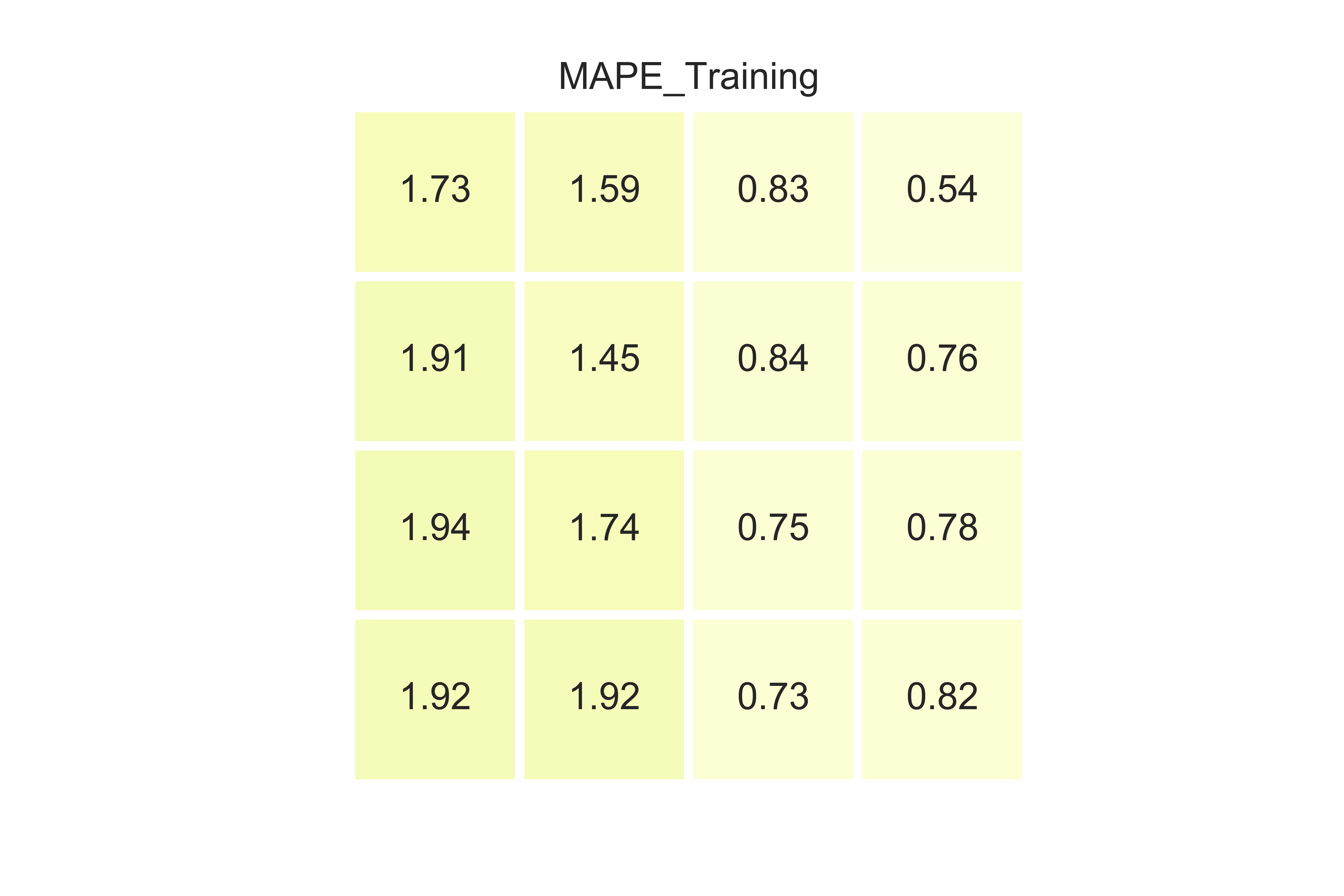}\begin{picture}(0,0)\put(-138,0){}\end{picture}
               \caption{} 
        \end{subfigure}
 		\begin{subfigure}[b]{0.495\textwidth}
               \includegraphics[width=\textwidth,trim=85 18 50 20,clip]{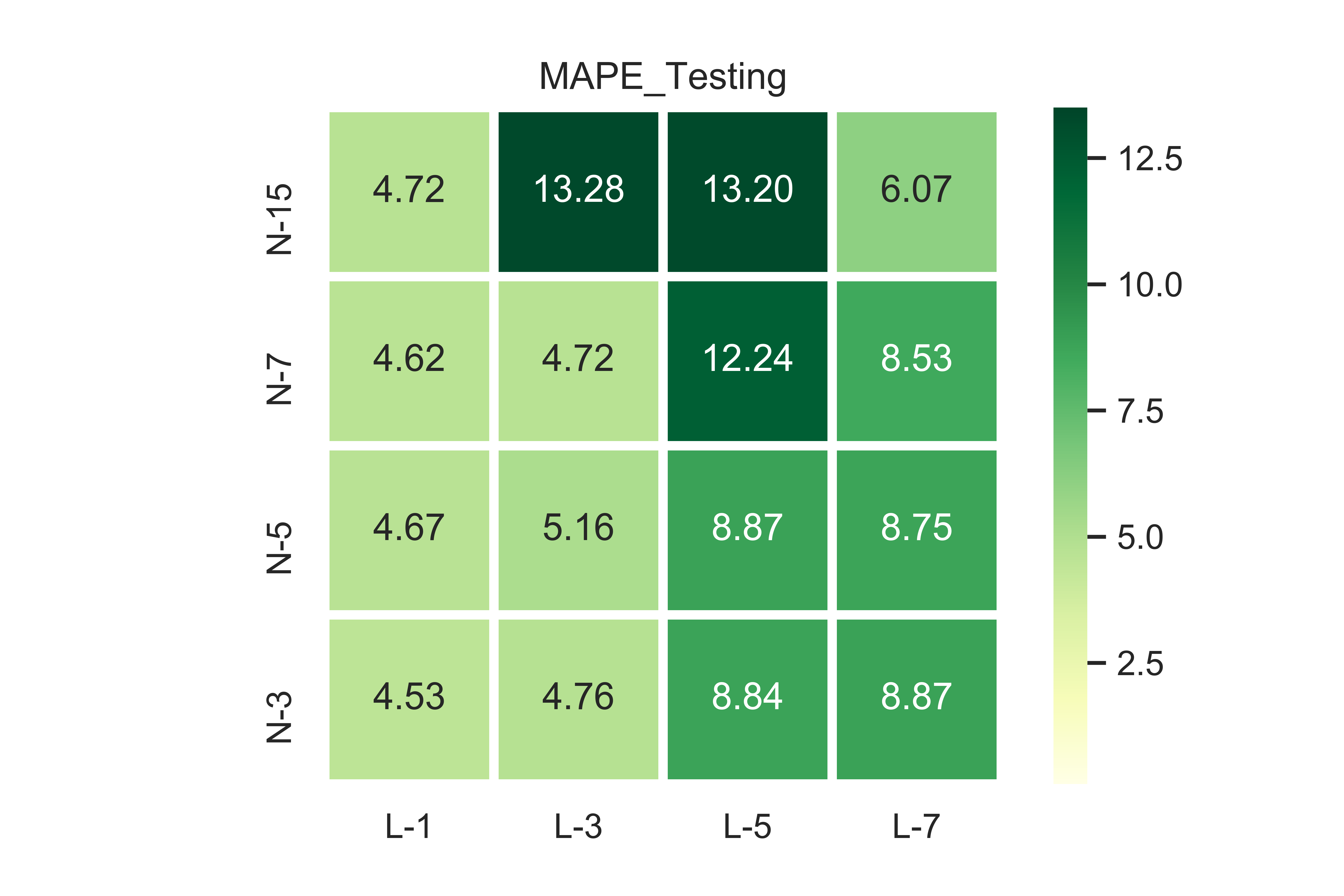}\begin{picture}(0,0)\put(-138,0){}\end{picture}
                \caption{}
        \end{subfigure}                                 
        \begin{subfigure}[b]{0.495\textwidth}
                \includegraphics[width=\textwidth,trim=85 18 50 20,clip]{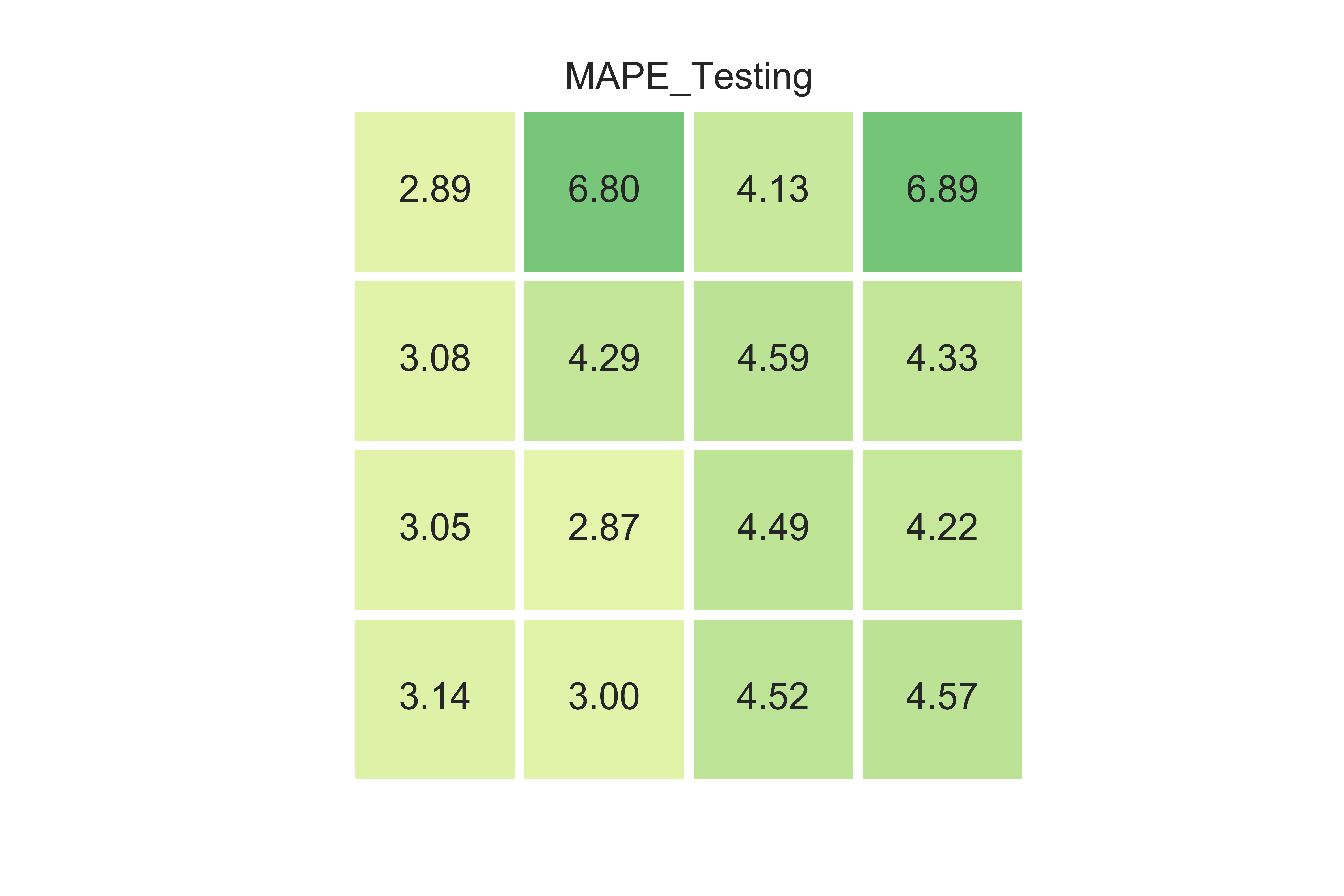}\begin{picture}(0,0)\put(-138,0){}\end{picture}
               \caption{} 
        \end{subfigure}         
        \caption{\label{fig_11} Training and testing MAPE for Case--2, (a) and (c) SSG, (b) and (d) LRR} 
        \end{figure}     

\subsection{\label{sec:Result_3} Case--3: Limited training data in the shear--dominated region}
\begin{figure}
        \centering
 		\begin{subfigure}[b]{0.495\textwidth}
               \includegraphics[width=\textwidth,trim=85 18 50 20,clip]{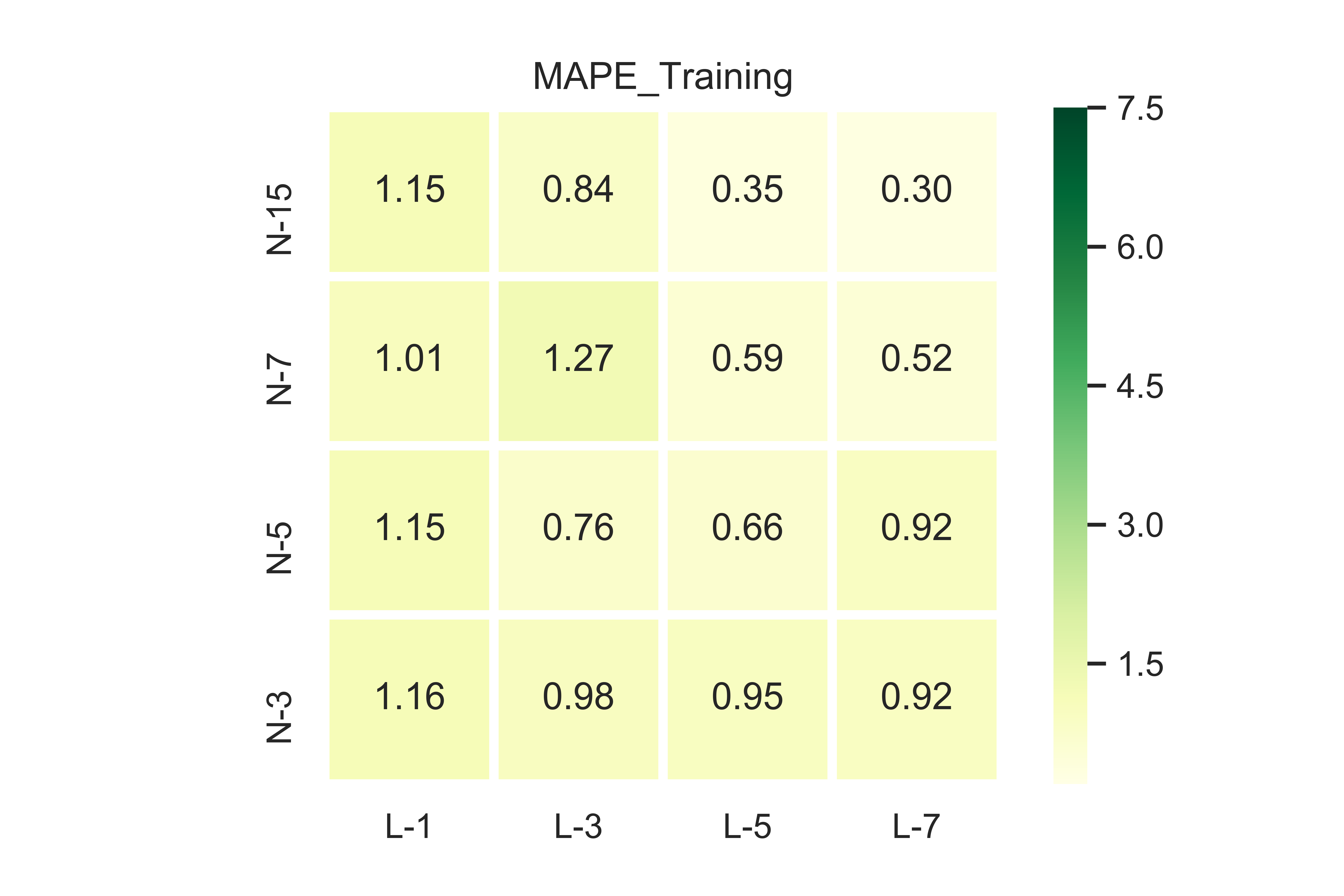}\begin{picture}(0,0)\put(-138,0){}\end{picture}
               \caption{}
        \end{subfigure}                                 
        \begin{subfigure}[b]{0.495\textwidth}
                \includegraphics[width=\textwidth,trim=85 18 50 20,clip]{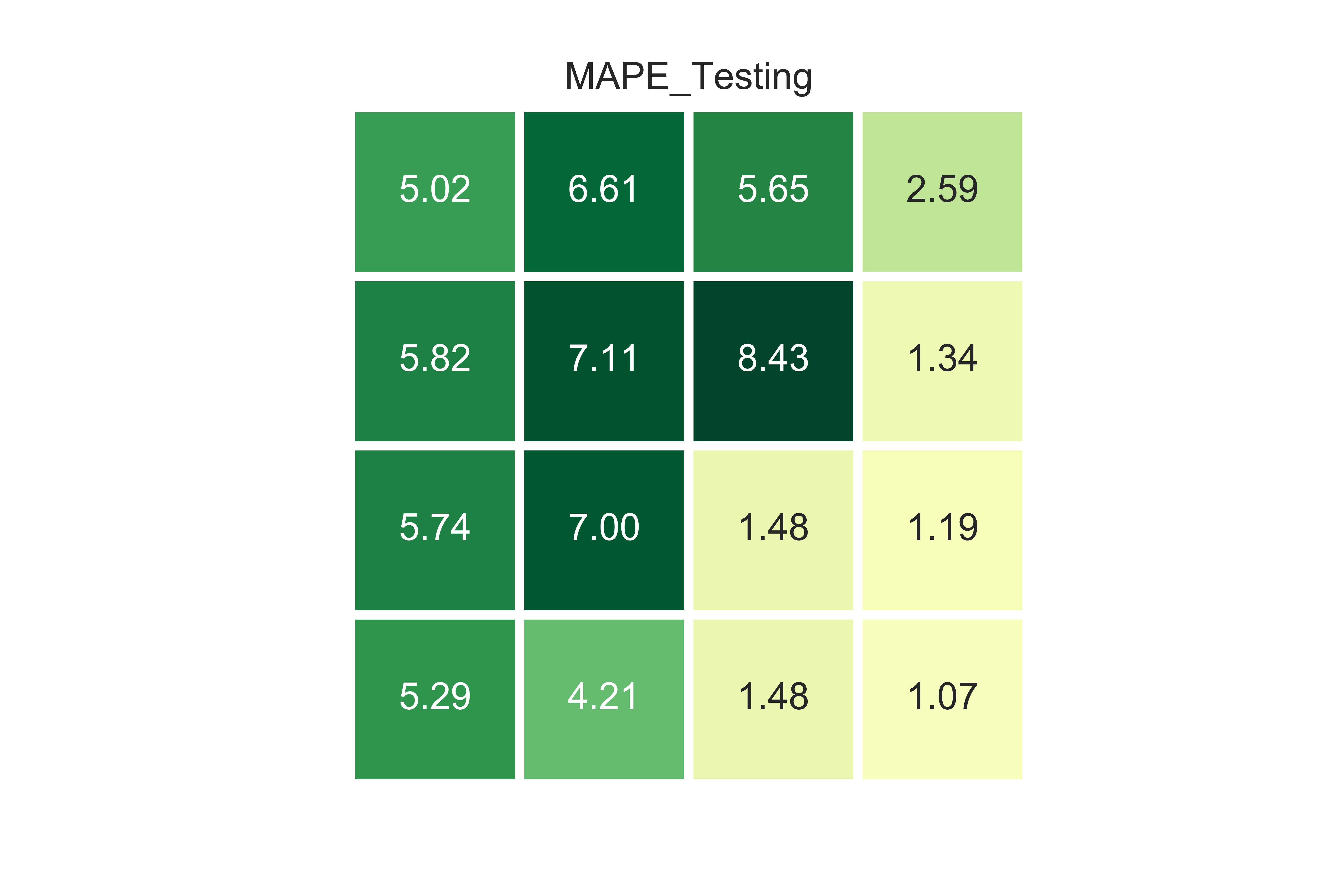}\begin{picture}(0,0)\put(-138,0){}\end{picture}
               \caption{} 
        \end{subfigure}
        \caption{\label{fig_13} MAPE of NNs with different architectures for Case--3 (a) training, (b) testing } 
        \end{figure}

\begin{figure}
        \centering
 		\captionsetup{justification=centering} 
 		\begin{subfigure}[b]{0.325\textwidth}
               \includegraphics[width=\textwidth,trim=2 2 2 2,clip]{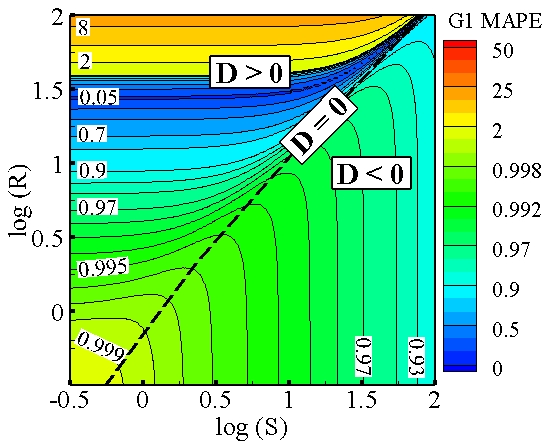}\begin{picture}(0,0)\put(-138,0){}\end{picture}
                \caption{}
        \end{subfigure}                                 
        \begin{subfigure}[b]{0.325\textwidth}
                \includegraphics[width=\textwidth,trim=1 1 1 2,clip]{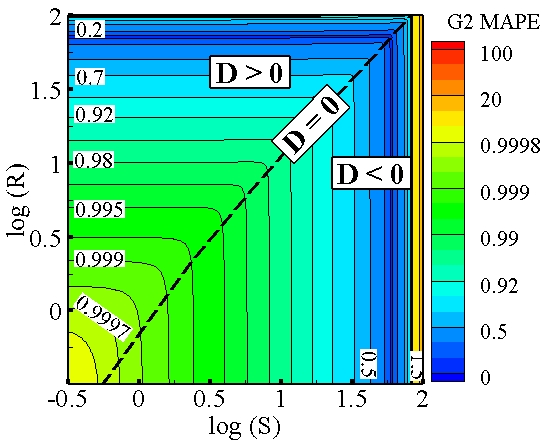}\begin{picture}(0,0)\put(-138,0){}\end{picture}
               \caption{} 
        \end{subfigure}
        \begin{subfigure}[b]{0.325\textwidth}
                \includegraphics[width=\textwidth,trim=1 1 1 2,clip]{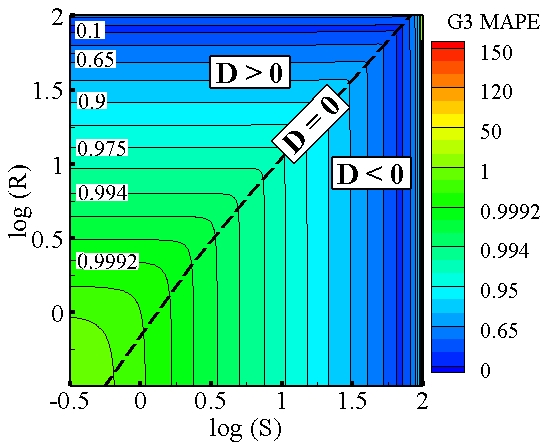}\begin{picture}(0,0)\put(-138,0){}\end{picture}
               \caption{} 
        \end{subfigure}    
 		\begin{subfigure}[b]{0.325\textwidth}
               \includegraphics[width=\textwidth,trim=2 2 2 2,clip]{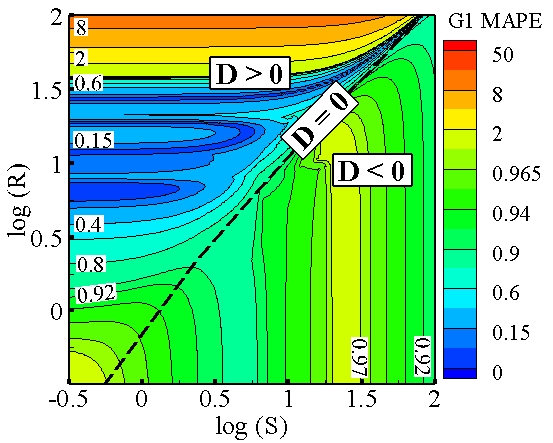}\begin{picture}(0,0)\put(-138,0){}\end{picture}
                \caption{}
        \end{subfigure}                                 
        \begin{subfigure}[b]{0.325\textwidth}
                \includegraphics[width=\textwidth,trim=1 1 1 2,clip]{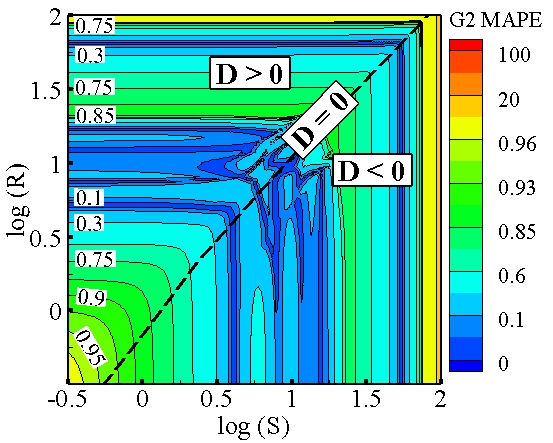}\begin{picture}(0,0)\put(-138,0){}\end{picture}
               \caption{} 
               \end{subfigure}
        \begin{subfigure}[b]{0.325\textwidth}
                \includegraphics[width=\textwidth,trim=1 1 1 2,clip]{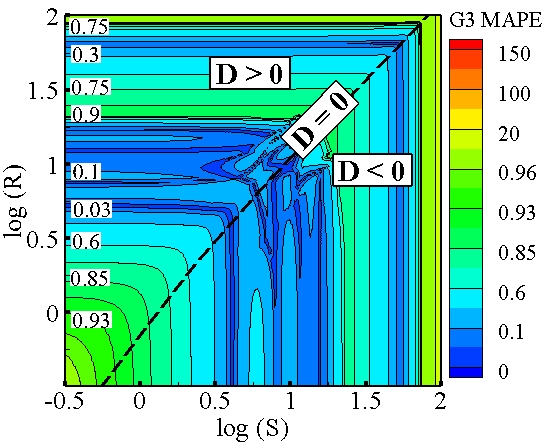}\begin{picture}(0,0)\put(-138,0){}\end{picture}
               \caption{} 
        \end{subfigure}    
 		\begin{subfigure}[b]{0.325\textwidth}
               \includegraphics[width=\textwidth,trim=2 2 2 2,clip]{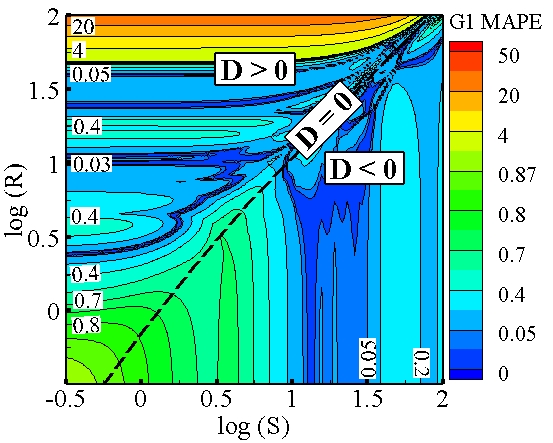}\begin{picture}(0,0)\put(-138,0){}\end{picture}
                \caption{}
        \end{subfigure}                                 
        \begin{subfigure}[b]{0.325\textwidth}
                \includegraphics[width=\textwidth,trim=1 1 1 2,clip]{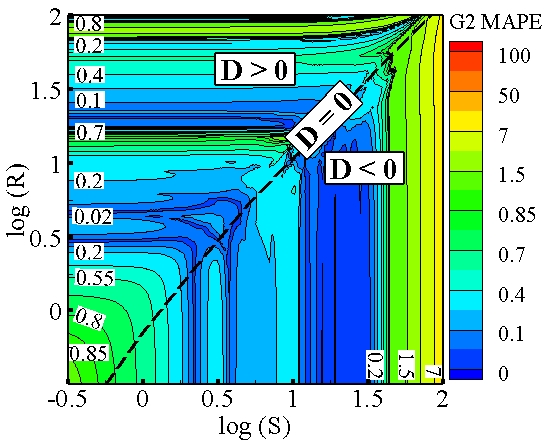}\begin{picture}(0,0)\put(-138,0){}\end{picture}
               \caption{} 
        \end{subfigure}
        \begin{subfigure}[b]{0.325\textwidth}
                \includegraphics[width=\textwidth,trim=1 1 1 2,clip]{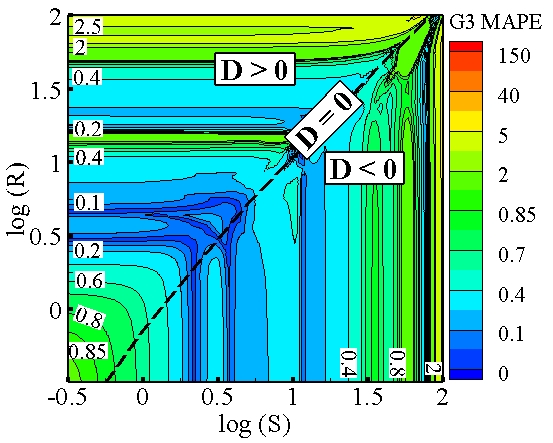}\begin{picture}(0,0)\put(-138,0){}\end{picture}
               \caption{} 
        \end{subfigure}           
        \caption{\label{fig_14} MAPE contours for Case--3, (a)-(c) 7L--5N, (d)-(f) 7L--7N, (g)-(i) 7L--15N} 
        \end{figure}

In this scenario the training dataset covers a limited but important region of the parameter space near the bifurcation line $D=0$. The data points in rest of the parameters space are used for testing the ML models. For this case, the data manifold is generated by ARSM with SSG model. The ReLU activation function is employed for all computation neurons and all other hyperparameters are fixed as shown in Table~\ref{tab:3}. The performance of NNs with 16 architectures in the training and testing datasets are compared in Fig.~\ref{fig_13}. Similar to  Case--2, a $L_2$ norm regularization with $\lambda = 0.1$ is used during training to reduce the overfitting of the ML models in this case. Although, the training error reduces by increasing number of hidden layers from three to seven in networks with large width, the testing error oscillates. Similar to Case--2, selecting a reasonably--sized generalizable NN is not straightforward when training data is limited to a narrow range in the shear--dominated region. Local MAPE contours in the entire parameter space for DNNs with different number of neurons are compared in Fig.~\ref{fig_14}. It is seen that the DNN with large width have relatively smaller errors in both training and testing regions. 

The main inferences from the three case studies can be summarized. We systematically examined the suitability of DNNs for RANS turbulence closure modeling. We studied the existence of a generalizable, moderate-sized NN given sufficiently many training data in the entire parameter space. The results show that NNs require more degrees of freedom (than the true proxy--physics model) to accurately approximate the true polynomial model with four coefficients even with entire data points of the parameter space. This implies that DNNs are not efficient approximations of data manifolds with non--linearity and bifurcation effects such as turbulence even in interpolation. We also observed that the approximation capability of the NNs significantly varies by activation function of neurons. Comparing the performance of the NNs with different loss functions (MAPE and RMSE) showed that networks with MAPE loss function have better performance for the generated datasets in this study. For the interpolation case, it has been shown that the reasonably--sized NNs trained with data manifold with no bifurcation (SZL model) have the smallest errors. However, the coexistence of non--linearity and bifurcation in the data manifold produced the largest level of testing errors in the NN solutions even in these simple proxy--physics models. As the non--linearity increases in the data manifolds with bifurcation, the approximation capability of the NNs reduces in both interpolation and extrapolation cases. Furthermore, in the extrapolation cases it is not straightforward to find an optimal architecture.

\section{\label{sec:Conc} Conclusion\protect}
Generalizability of ML--assisted RANS turbulence model to unseen flows still faces many challenges~\cite{beck2021perspective} due to flow--dependent non--linearity and bifurcations of the constitutive relations. Further, there is little consensus and great deal of uncertainty regarding the choice of NN hyperparameters and training techniques. Yet, these choices can significantly affect the predictive capability and generalizability of ML turbulence models. We seek to understand the optimal choice of hyperparameters, training process elements (type of loss function) and necessary number of neurons of DNNs required to allow a sufficiently accurate approximation at the RANS closure modeling level. Standard fully--connected NNs are trained in a supervised manner and their approximation capabilities are systematically investigated by considering the effects of: \emph{(i)} intrinsic complexity of the solution manifold; \emph{(ii)} sampling procedure (interpolation vs. extrapolation) and \emph{(iii)} optimization procedure.

A key novelty of this work is the adoption of simplified proxy--physics turbulence surrogates that incorporate some of the important features of real homogeneous flows to generate the sufficient training data to assess generalizability (interpolation vs. extrapolation)  characteristics. An important advantage of this approach is that training data for all flows in the parameter space can be generated easily. In contrast, DNS or experiments would be prohibitively expensive and may not even be feasible for all flows in the parameter space. Successful generalizability of ML models in this proxy--physics turbulence system is a necessary but not a sufficient condition for ML--model generalizability in actual turbulent flows. Nevertheless, this study provides valuable insight into the generalizability characteristics of different network architectures and hyperparameters in turbulence--like phenomena.

Three turbulence surrogates of different degrees of complexity are chosen: \emph{(i)} a non--linear constitutive relation with no bifurcation in the parameter regime \emph{(ii)} mildly non--linear constitutive relation with bifurcation in the regime of interest; and \emph{(iii)} moderately non--linear constitutive model with bifurcation. When the constitutive relation does not have bifurcation in the entire parameter space, the accuracy of the ML model is quite reasonable. However, when the surrogate data exhibits bifurcation, the combination of non--linearity and bifurcation produced the largest level of testing errors even in these simple proxy--physics models. Moreover, testing error increase with the level of non--linearity in the proxy--physics surrogate with bifurcation in the parameter space.

The conclusions of the study are two--fold. First, even for interpolation, the NN--based models require very large networks (more degrees of freedom than true proxy--physics model) to reduce errors to a reasonable level. Secondly, in a practical calculation, it is difficult \emph{a priori} to guarantee that the model will be queried only in the trained domain. Therefore, it is critical to establish that the NN--model can handle certain level of extrapolation. To assess the model ability for extrapolation/generalization, we train the model in limited parameter space and test outside of this regime. It is shown that even for this simple proxy--physics system, the NN--model performance is inadequate (Cases 2, 3). Further, we identify and distinguish the challenges to generalizability arising out of non--linearity and bifurcation. We believe that these findings are important and yet not clearly understood in current literature. Therefore, studies such as these are necessary for a balanced assessment of NN--based RANS turbulence models as predictive tools for use in unseen flows.

As mentioned before, true turbulence phenomena is much more complicated with many more degrees of freedom (many more physical parameters) and multiple bifurcations in the overall behavior. For example, mean flow three--dimensionality, large--scale instabilities, streamline curvature, stratification, compressibility and other effects encountered in typical engineering flows will lead to significantly more complex constitutive relations with non--equilibrium effects. It is very likely that sufficiently large samples of accurate (direct numerical simulations or experimental) time dependent data over all possible regimes of turbulence encompassing all these effects will not be available in the near future. Even if such data were available, it is unclear if an optimal set of hyperparameters can be found using currently available methods. Until such a time when \emph{(i)} sufficiently large volume of unsteady data is available over the entire parameter regime of turbulence; and \emph{(ii)} analytical methods for determining optimal network architecture are available, the results of the current study suggest that ML--enhanced RANS models will be limited in its ability to perform predictive computations of real engineering flows. It is important to note that when data is indeed available over the entire parameter range, traditional closure development methods may also lead to significantly improved models. Thus, the relative advantages of ML methods over traditional methods must be reassessed at that time.

\section*{\label{sec:App} APPENDIX\protect}
\begin{figure}
        \centering
 		\begin{subfigure}[b]{0.495\textwidth}
               \includegraphics[width=\textwidth,trim=85 18 50 20,clip]{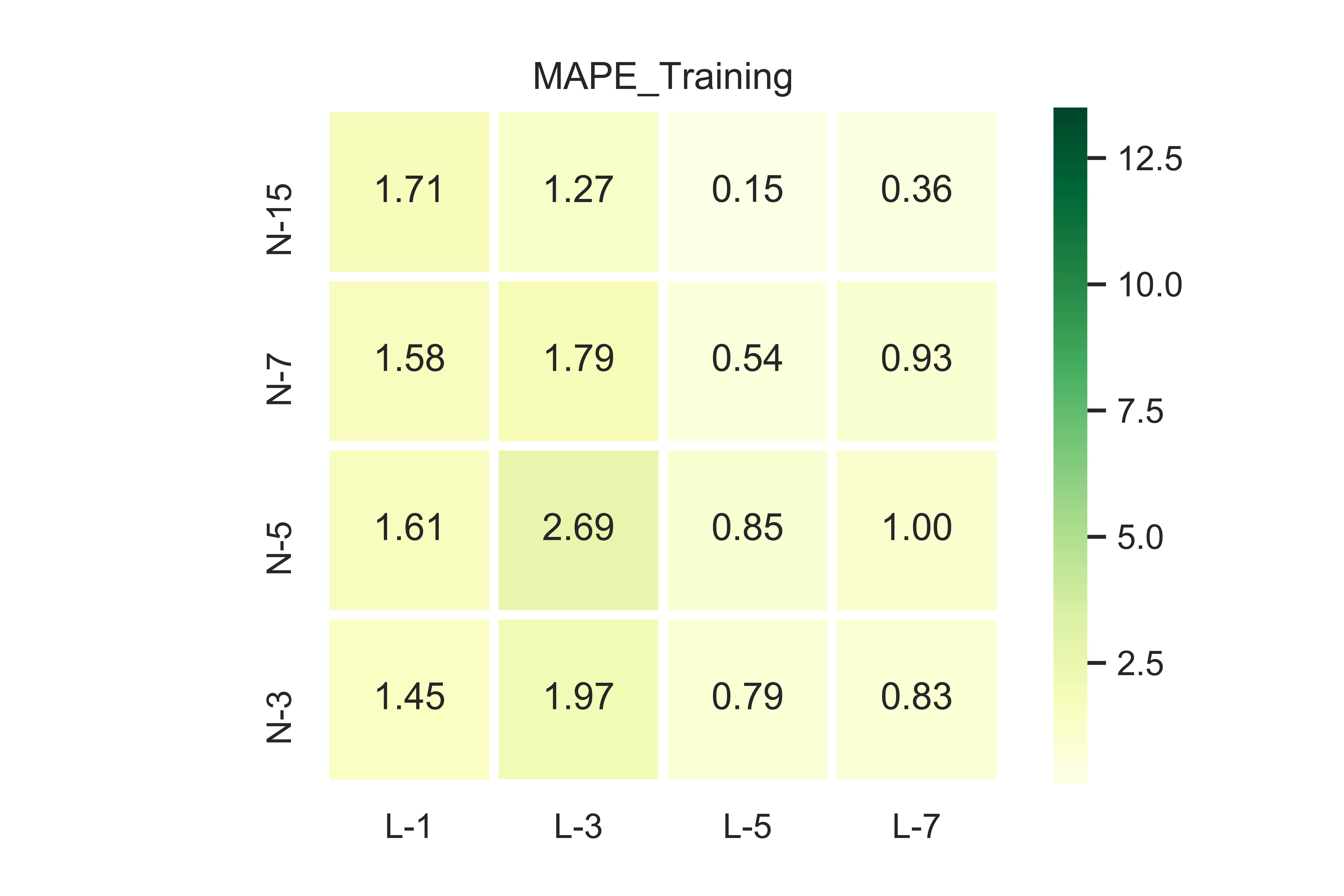}\begin{picture}(0,0)\put(-138,0){}\end{picture}
               \caption{}
        \end{subfigure}                                 
        \begin{subfigure}[b]{0.495\textwidth}
                \includegraphics[width=\textwidth,trim=85 18 50 20,clip]{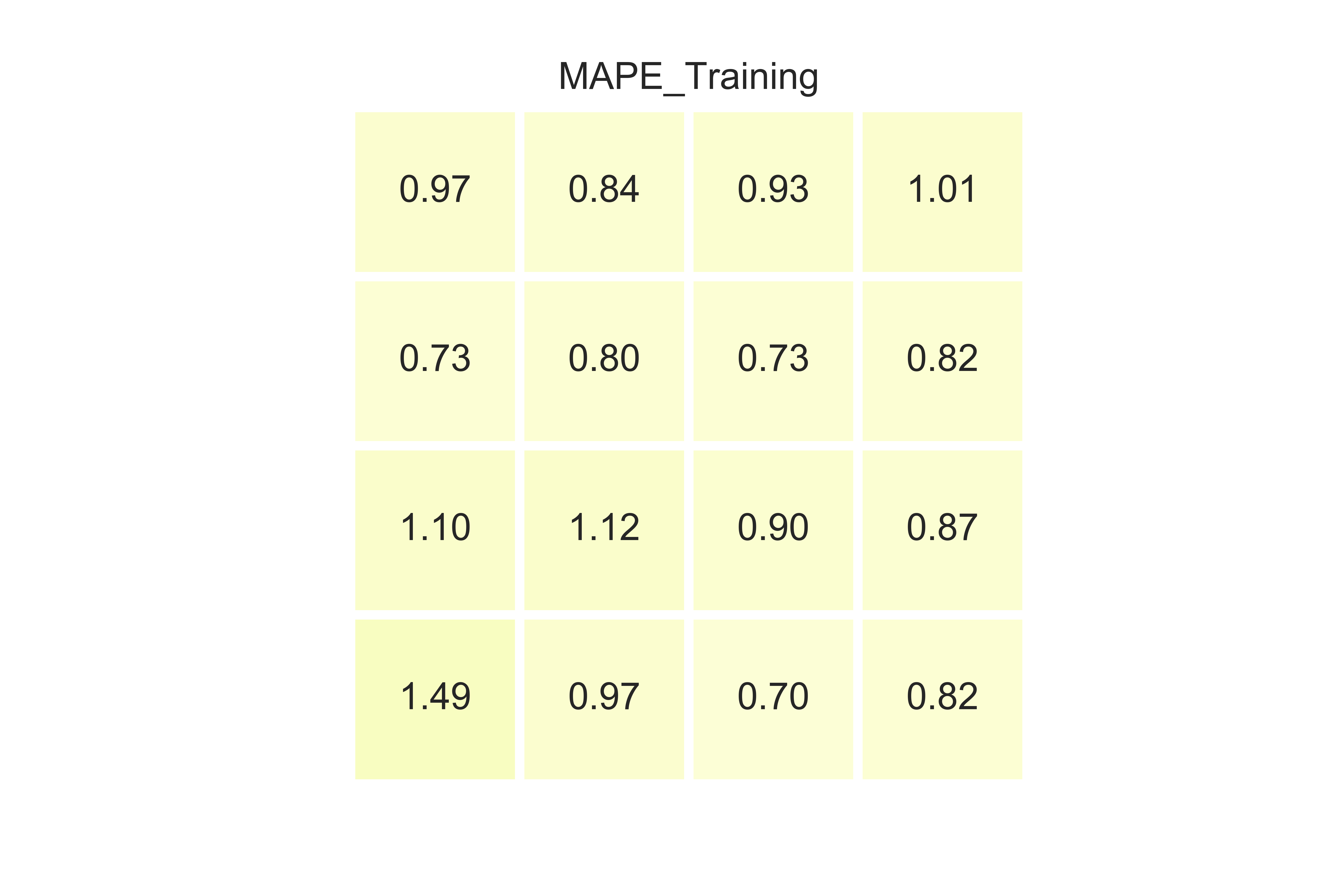}\begin{picture}(0,0)\put(-138,0){}\end{picture}
               \caption{}
        \end{subfigure}    
 		\begin{subfigure}[b]{0.495\textwidth}
               \includegraphics[width=\textwidth,trim=85 18 50 20,clip]{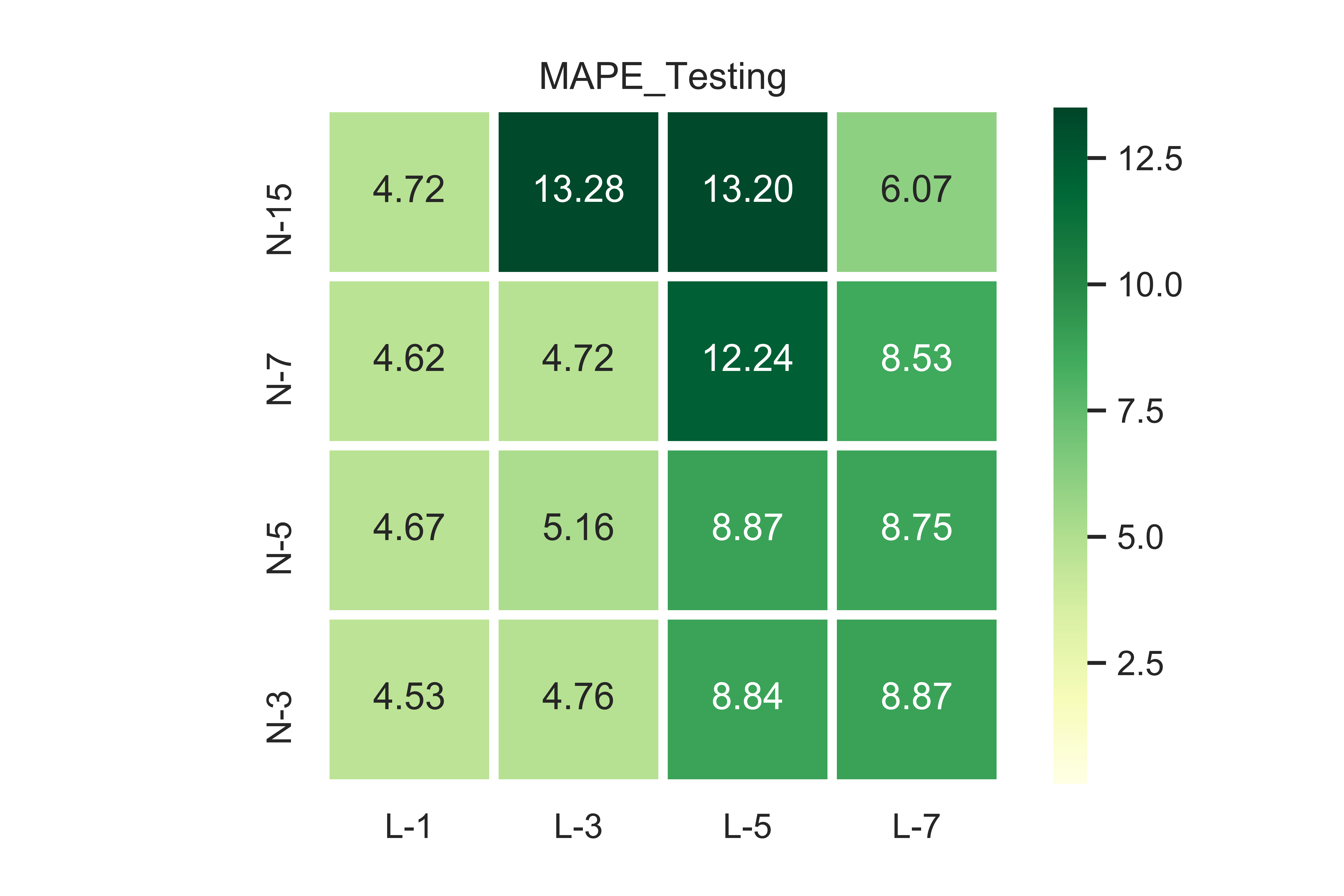}\begin{picture}(0,0)\put(-138,0){}\end{picture}
                \caption{}
        \end{subfigure}                                 
        \begin{subfigure}[b]{0.495\textwidth}
                \includegraphics[width=\textwidth,trim=85 18 50 20,clip]{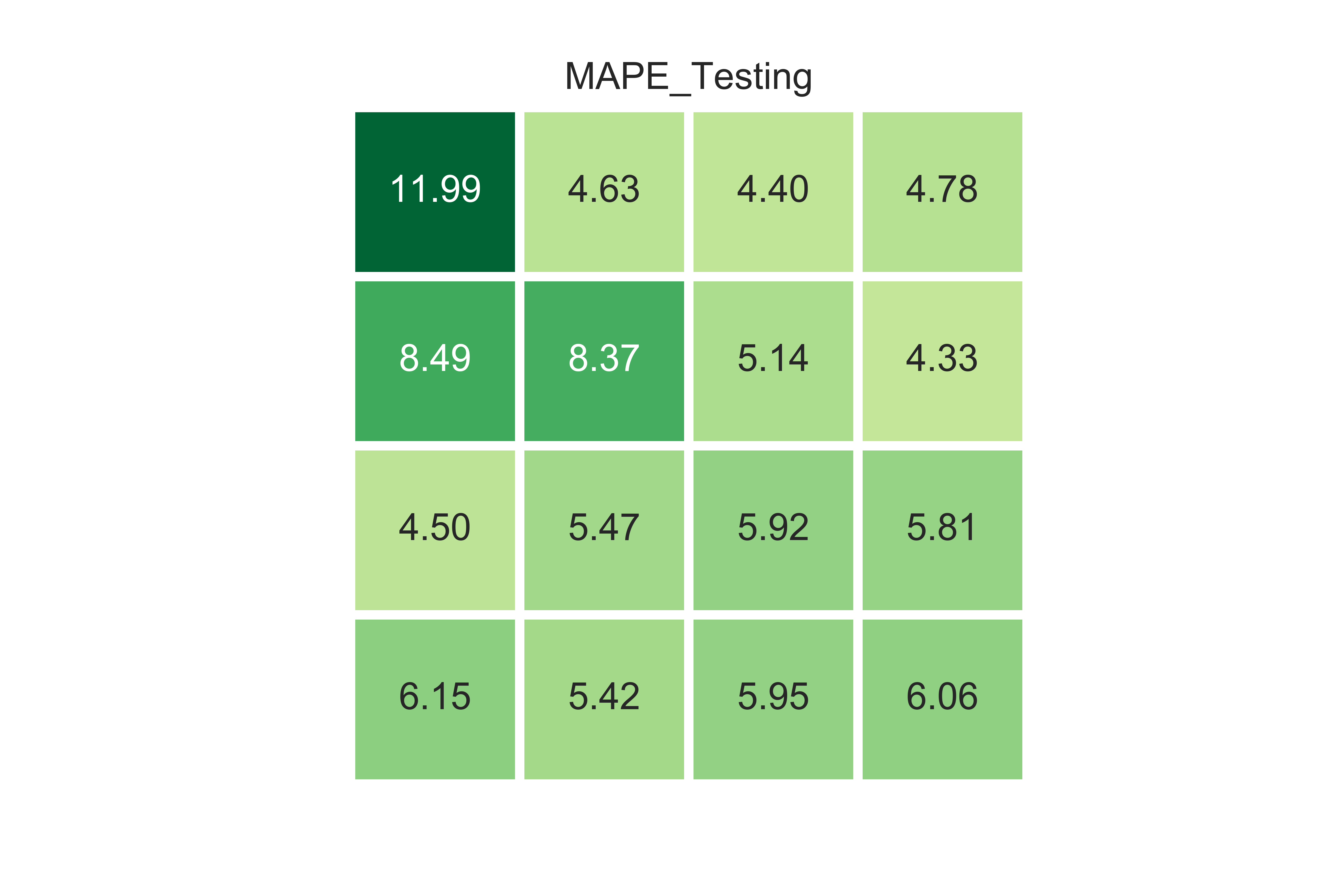}\begin{picture}(0,0)\put(-138,0){}\end{picture}
               \caption{} 
        \end{subfigure}    
        \caption{\label{fig_15} Training and testing MAPE for Case--2, (a) and (c) ReLU, (b) and (d) Sigmoid} 
        \end{figure}
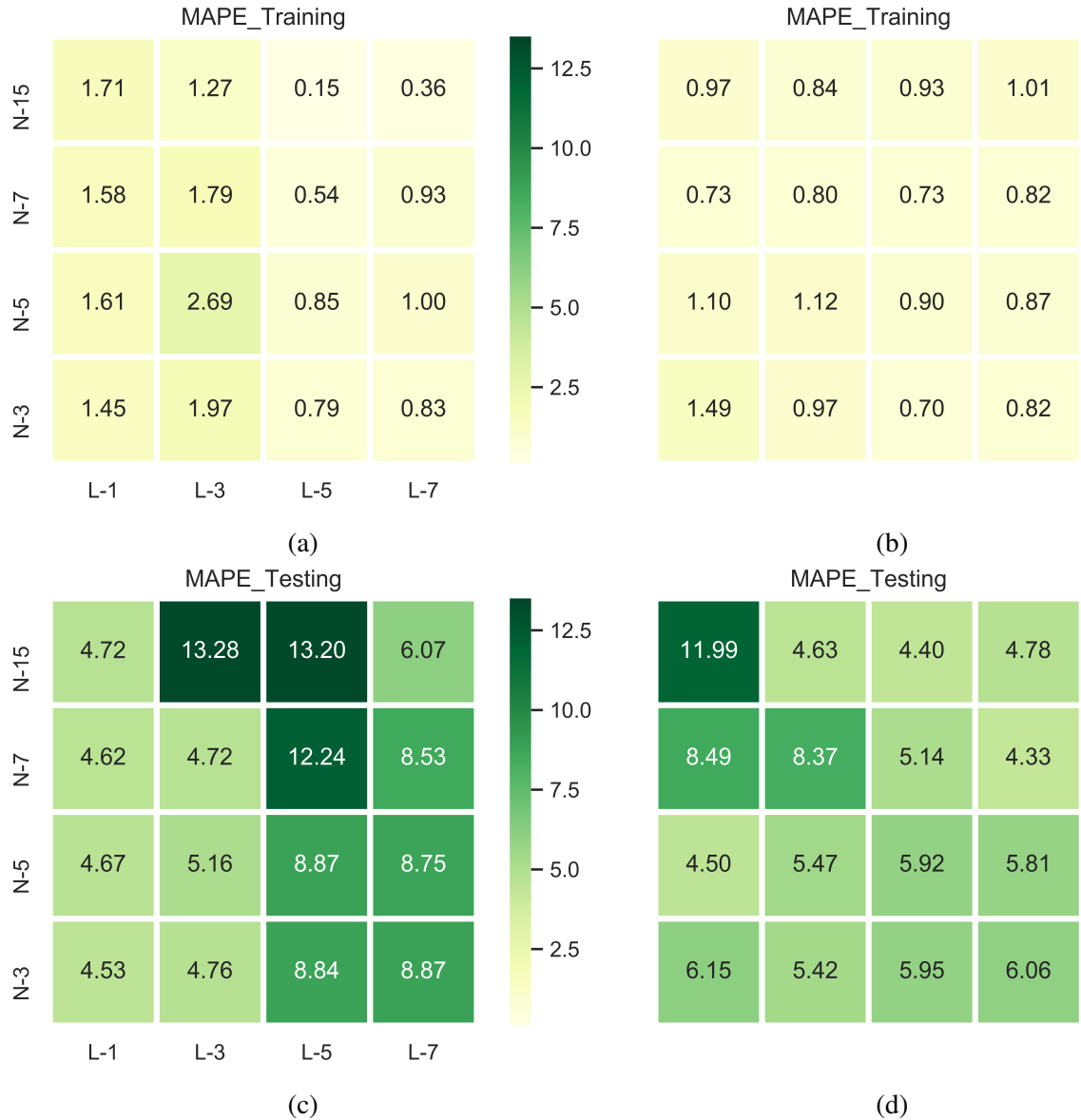

\begin{table}
  \caption{Performance of models trained with different activation functions }
  \label{tab:5}
  \begin{tabularx}{0.95\textwidth}{XXXXX}
    \hline\hline
    Type & MAPE--training & MAPE--testing \\
    \hline
    ReLU & 0.93 & 8.53 \\
    Elu  & 0.90 & 7.37 \\
    Leaky-ReLU & 0.5 & 11.60 \\
    Tanh & 0.75 & 12.03 \\
    Sigmoid & 0.85 & 4.39 \\
    \hline\hline
  \end{tabularx}
\end{table}
For the Case--2 in which training data is partially available in the strain--dominated region, a systematic grid search hyperparameter optimization is performed. For this case, the data manifold is generated by ARSM with SSG model. First, the performance of a 7L-7N NN with different activation functions (act) is studied. During the training a $L_2$ norm regularization with $\lambda = 0.1$ is used. All other hyperparameters are fixed as shown in Table~\ref{tab:3}. The training and testing MAPE errors for this experiment are shown in Table~\ref{tab:5}. It is seen that between the five activation functions, ReLU, Elu, and Sigmoid have reasonably small training and testing MAPE errors. By selecting two ReLU and Sigmoid activation functions, the performance of the NNs with different architectures are further investigated. MAPE of all the 16 network architectures in the training and testing datasets is shown in Fig.~\ref{fig_15}. Results show that for both activation functions, optimum network architecture should be selected from the deep networks with small training and testing MAPE errors. Additionally, it is seen that between two activation functions, larger networks with Sigmoid activation function have smaller testing errors in this case.

A 7L--7N NN with Sigmoid activation function are selected to study the effects of other hyperparameters on training and performance of the models. The results reported in Tables~\ref{tab:6}--\ref{tab:9} justifies the selection of optimum hyperparameters shown in Table~\ref{tab:3} in this study.

\begin{table}
  \caption{Performance of models trained with different learning rates, act=Sigmoid, bs=50, opt=Adam, $L_2 (\lambda$=0.1)}
  \label{tab:6}
  \begin{tabularx}{0.95\textwidth}{XXXXX}
    \hline\hline
    Learning rate & MAPE--training & MAPE--testing \\
    \hline
    $1.0\times10^{-2}$ & 2.48 & 3.98 \\
    $1.0\times10^{-3}$ & 0.85 & 4.39 \\
    $1.0\times10^{-4}$ & 0.77 & 10.81 \\
    $1.0\times10^{-5}$ & 0.74 & 10.98 \\
    \hline\hline
  \end{tabularx}
\end{table}

\begin{table}
  \caption{Performance of models trained with different batch sizes, act=Sigmoid, lr=0.001, opt=Adam, $L_2 (\lambda$=0.1)}
  \label{tab:7}
  \begin{tabularx}{0.95\textwidth}{XXXXX}
    \hline\hline
    Batch size & MAPE--training & MAPE--testing \\
    \hline
    25 & 0.88 & 4.41 \\
    50 & 0.85 & 4.39 \\
    100 & 0.91 & 9.63 \\
    1000 & 0.88 & 11.08 \\
    \hline\hline
  \end{tabularx}
\end{table}

\begin{table}
  \caption{Performance of models trained with different type of optimizers, act=Sigmoid, lr=0.001, bs=50, $L_2 (\lambda$=0.1)}
  \label{tab:8}
  \begin{tabularx}{0.95\textwidth}{XXXXX}
    \hline\hline
    Type & MAPE--training & MAPE--testing \\
    \hline
    Adam & 0.85 & 4.39 \\
    RMSProp & 1.92 & 5.43 \\
    \hline\hline
  \end{tabularx}
\end{table}

\begin{table}
  \caption{Performance of models trained with different regularization coefficients, act=Sigmoid, lr=0.001, bs=50}
  \label{tab:9}
  \begin{tabularx}{0.95\textwidth}{XXXXX}
    \hline\hline
    \multicolumn{3}{c}{$L_1$--norm}& \multicolumn{2}{c}{$L_2$--norm}  \\
    \hline
    Coefficient $\lambda$ & Training & Testing & Training & Testing \\
    \hline
    0 & 0.83 & 5.75 & 0.83 & 5.75 \\
    0.01 & 0.82 & 5.24 & 0.82 & 5.15 \\
    0.1 & 0.82 & 4.74 & 0.85 & 4.39 \\
    0.2 & 0.83 & 4.49 & 0.84 & 4.40 \\
    \hline\hline
  \end{tabularx}
\end{table}

\clearpage 
\section*{Acknowledgement}
The authors acknowledge support provided by Texas A\&M High Performance Research Computing center.
\section*{References}
\bibliographystyle{ieeetr}
\bibliography{main}


\end{document}